\documentclass[letterpaper,11pt]{article}

\usepackage{amsmath}
\usepackage{array}
\usepackage{arydshln}
\usepackage{extarrows}
\usepackage{jheppub}
\usepackage{natbib}
\usepackage{subcaption}
\usepackage{tikz}
\usepackage{tikzit}

\usetikzlibrary{arrows}

\newcommand{\calC}{{\cal C}}
\newcommand{\calH}{{\cal H}}
\newcommand{\calK}{{\cal K}}
\newcommand{\calW}{{\cal W}}

\usepackage{bm}
\newcommand{\bms}{{\bm S}}

\DeclareMathOperator{\Gr}{G}
\DeclareMathOperator*{\Res}{Res}
\DeclareMathOperator{\Trop}{Trop}

\title{Symbol Alphabets from Tensor Diagrams}

\author[1]{Lecheng Ren,}
\author[1,2]{Marcus Spradlin}
\author[1]{and Anastasia Volovich}

\affiliation[1]{Department of Physics,
Brown University,
Providence, RI 02912, USA}

\affiliation[2]{Brown Theoretical Physics Center,
Brown University,
Providence, RI 02912, USA}

\abstract{
We propose to use tensor diagrams and the Fomin-Pylyavskyy conjectures
to explore the connection between symbol alphabets of $n$-particle
amplitudes in planar $\mathcal{N}=4$ Yang-Mills theory and certain polytopes associated to the
Grassmannian $\Gr(4,n)$.  
We show how to assign a web (a planar tensor diagram) to each facet of these polytopes.
Webs with no inner
loops are associated to cluster variables (rational symbol letters).
For webs with a single
inner loop we propose and explicitly evaluate an associated web series
that contains
information about algebraic symbol letters.
In this manner we
reproduce the results of previous analyses of $n \le 8$, and
find that the polytope $\mathcal{C}^\dagger(4,9)$ encodes
all rational letters, and all square roots of the algebraic letters, of
known nine-particle amplitudes.
}

\begin{document}

\maketitle

\section{Introduction}

The perturbative $n$-particle scattering amplitudes of planar maximally supersymmetric Yang-Mills (SYM) theory are a remarkable collection of functions that reflect deep but still partially mysterious mathematical structure.  A prime example of this structure is the apparent connection~\cite{Golden:2013xva} between their physical singularities, which for certain amplitudes (those of polylogarithmic type) are encoded in symbol letters~\cite{Goncharov:2010jf} indicating where the amplitudes have logarithmic branch points, and the mathematics of cluster algebras~\cite{fomin2002cluster}.  All evidence available to date is consistent with the hypothesis that the symbol letters of all $6,7$-particle amplitudes (see~\cite{Caron-Huot:2020bkp} for a recent review) are cluster variables of the $\Gr(4,n)$ cluster algebra, whereas for $n\ge 8$ it is known that certain algebraic functions of cluster variables also appear as symbol letters (see Tab.~\ref{tab:symboldata})\footnote{The connection between cluster algebras and singularities of amplitudes could extend well beyond SYM theory~\cite{Chicherin:2020umh,He:2021esx}.}.

Recently this connection has been approached~\cite{Arkani-Hamed:2019rds,Drummond:2019cxm,Henke:2019hve,Herderschee:2021dez} via the study of certain fans one can naturally associate to the tropical positive Grassmannian $\Trop_{>0} \Gr(4,n)$ (generalizing the construction of~\cite{speyer2005tropical}) or (dually) certain ``$\Gr(4,n)$-polytopes'' one can associate to $\Gr(4,n)$ using the construction of~\cite{Arkani-Hamed:2019mrd}\footnote{The very same polytopes also appear in the study of $\mathbb{P}^{k-1}$-generalized scattering equations for biadjoint amplitudes~\cite{Cachazo:2019ngv,Cachazo:2019apa,Drummond:2019qjk,Cachazo:2020uup}.}.  This approach is appealing because for $n\ge 7$ the $\Gr(4,n)$-polytopes are different than cluster polytopes~\cite{fomin2003systems} in a way that makes them seem to more faithfully encode the symbol alphabets of SYM theory\footnote{Some finer aspects of this encoding for $n=7$ are discussed in~\cite{Henke:2019hve,Drummond:2020kqg}.}.  In particular, while the $\Gr(4,n)$ cluster algebra is infinite for $n \ge 8$, these polytopes are finite for any $n$, which matches the supposition~\cite{Prlina:2018ukf} that the symbol alphabet of $n$-particle amplitudes might be finite for any fixed $n$.  Moreover, while many facets of the relevant polytopes are naturally associated to certain finite subsets of cluster variables (many of which are known to appear as symbol letters of amplitudes), for $n\ge 8$ there are also ``exceptional'' facets that are not associated to any cluster variable.  Consequently the $\Gr(4,n)$-polytopes may have room to ``explain'' the algebraic symbol letters that appear in $n \ge 8$-particle amplitudes\footnote{It is an interesting open question whether any information about the more complicated analytic structure of non-polylogarithmic amplitudes (that are known to appear for $n \ge 10$~\cite{CaronHuot:2012ab,Bourjaily:2017bsb}) is encoded in polytopes.}.

There have been several different but related suggestions for how, precisely, algebraic letters are encoded in the structure of polytopes.  Here we follow the framework proposed in~\cite{Arkani-Hamed:2019rds}, where partial information about algebraic letters is encoded in certain formal power series associated to exceptional facets.  In that paper the authors studied a polytope called $\mathcal{C}^\dagger(4,8)$ (its definition is reviewed in Sec.~\ref{sec:polytopes}), finding that the cluster variables associated to its 272 non-exceptional facets include the 172 (non-frozen) cluster variables known to appear as symbol letters of $n=8$-particle amplitudes.  Moreover they conjectured a formula for the series associated to its 2 exceptional facets based on explicit computation of the first three nontrivial terms in the associated series, extending the results of~\cite{chang2020quantum} (whose method they used), and showed that it was consistent with the expected square root of four-mass box type.  Together, this was taken as evidence supporting the notion that $\mathcal{C}^\dagger(4,8)$ encodes information about the $n=8$ symbol alphabet (see also~\cite{Henke:2019hve}, and~\cite{Drummond:2019cxm,Herderschee:2021dez} which went further, fully reproducing all known algebraic letters).  Unfortunately the computational complexity of the series introduced in~\cite{Arkani-Hamed:2019rds} is such that computing higher order terms for $n=8$, or any nontrivial terms for $n>8$, seems vastly out of reach.

In this paper we bring new mathematical technology to bear on this problem: the ``web diagrams'' (called webs for short) and web invariants that star in the beautiful Fomin-Pylyavskyy (FP) conjectures~\cite{fomin2016tensor}.  We propose an algorithm for assigning webs to facets of $\Gr(k,n)$-polytopes.  FP conjectured that the invariant associated to a web is a cluster variable if the web can be rearranged into a tree using certain graphical moves (in which case it is called ``arborizable'').  Exceptional facets therefore correspond to non-arborizable webs.  To ``almost arborizable webs'' (those with only a single inner loop) we are able to associate a formal power series of webs whose invariants we can compute exactly, using a graphical recursion.  In this manner we are able to compute the complete series associated to the exceptional facets of $\mathcal{C}^\dagger(4,8)$, confirming the conjecture of~\cite{Arkani-Hamed:2019rds} (albeit in a slightly different basis; see Sec.~\ref{sec:clusteralgebraicseries}).

A prime motivation for this work was to construct and study the polytope $\mathcal{C}^\dagger(4,9)$ as a candidate for encoding information about the $n=9$ symbol letters known from~\cite{He:2020vob}.  We find that 3078 of its 3429 facets are associated with arborizable webs, and their invariants include all of the 522 (non-frozen) cluster variables known to appear in the $n=9$ symbol alphabet.  We find that 324 facets are associated to almost arborizable webs, and compute their web series exactly.  These encode the expected 9 square roots of four-mass box type that appear in~\cite{He:2020vob}, as well as 315 additional, more complicated square roots.  The apparent overabundance of facets suggests that $\mathcal{C}^\dagger(4,9)$ might be more complicated than needed to describe 9-particle amplitudes, although more complicated cluster variables and square roots could certainly be discovered in future computations at higher loop.

Some recent papers~\cite{Mago:2020kmp,He:2020uhb,Mago:2020nuv,UsToAppear} have explored the connection between symbol alphabets and plabic graphs~\cite{postnikov2006total,Arkani-Hamed:2016byb}, and have in particular shown that the complete symbol alphabets for $n=6,7,8$ (including the $n=8$ algebraic letters) can be ``derived'' by solving certain polynomial equations associated to plabic graphs.  It is intriguing to note that the webs appearing prominently in this paper bear superficial resemblance to plabic graphs (compare for example Fig.~\ref{fig:nonarb}(b) with Fig~5.1 of~\cite{Mago:2020kmp}), and it would be interesting to see if the approach to symbol alphabets taken here could be connected to that of~\cite{Mago:2020kmp,He:2020uhb,Mago:2020nuv}.  We note that a connection between webs and plabic graphs is known in the math literature; see Example~4.3 and Theorem~4.4 of~\cite{fraser2019dimers}.

The outline of this paper is as follows.  In Sec.~\ref{sec:outline} we briefly review basic information about $\Gr(k,n)$-polytopes, cluster series, and symbol letters, and in Sec.~\ref{sec:reviewfp} we review necessary details about webs and the FP conjectures.  Sec.~\ref{sec:Xmapdefinition} explains our main algorithm for associating webs to the facets of $\Gr(k,n)$-polytopes.  In Sec.~\ref{sec:series} we define and compute the web series for almost arborizable webs.  The results of our computations for $(k,n)$ up to $(3,10)$ and $(4,9)$, and the implications thereof for symbol alphabets, are summarized in Sec.~\ref{sec:results}.

{\bf Node Added.} After this paper was completed, we learned from G.~Muller of the paper~\cite{lamberti2020tensor} by L.~Lamberti, which considers $\mathfrak{sl}_3$ web series of the type we employ in Sec.~\ref{sec:series}.

\section{Outline of the Paper}
\label{sec:outline}

The central objects of study in this paper are certain $d = (k{-}1)(n{-}k{-}1)$-dimensional polytopes associated to the Grassmannian $\Gr(k,n)$.  In particular we are interested in the apparent connection (for $k=4$) between their facets and the symbol alphabet of $n$-particle amplitudes in SYM theory.  Our main result, which we apply to some polytopes associated to $\Gr(3,n \le 10)$ and $\Gr(4,n\le 9)$, is a new algorithm to address the problem:
\begin{quote}
\textit{How can one efficiently extract the symbol alphabet data associated to a given $\Gr(k,n)$-polytope?}
\end{quote}
The computational complexity of existing algorithms for answering this question is reviewed in Fig.~\ref{fig:outline} and Sec.~\ref{sec:gvectorappendix}.  In this section we very briefly review key features of these polytopes and the relations between the structures associated to their facets, enabling us to outline our new algorithm in Fig.~\ref{fig:outline}.

\subsection{\texorpdfstring{Warm-Up: The Polytope $\mathcal{C}(3,5)$}{Warm-Up: The Polytope C(3,5)}}
\label{sec:warmup}

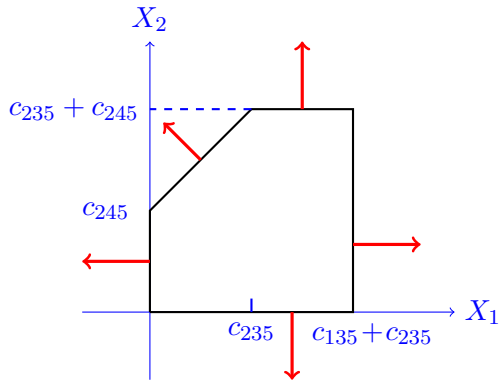
\begin{figure}[!htb]
    \centering
    \begin{tikzpicture}[scale=0.9]
        \draw[->,blue] (-1,0) -- (4.5,0) node[right]{$X_1$};
        \draw[->,blue] (0,-1) -- (0,4) node[above]{$X_2$};
        \draw[thick] (0,0) -- (3,0) node[below,blue]{$\quad  \ c_{135}\!+\!c_{235}$} -- (3,3) -- (1.5,3) -- (0,1.5) node[left, blue]{$c_{245}\ $} -- (0,0);
        \draw[thick,dashed,blue] (1.5,3) -- (0,3) node[left, blue]{$c_{235}+c_{245}$};
        \draw[thick,blue] (1.5,0) node[below,blue]{$c_{235}$} -- (1.5,.2);

        \draw[very thick,red,->] (0,0.75) -- (-1,0.75);
        \draw[very thick,red,->] (0.75,2.25) -- (0.2,3-0.2);
        \draw[very thick,red,->] (2.25,3) -- (2.25,4);
        \draw[very thick,red,->] (3,1) -- (4,1);
        \draw[very thick,red,->] (2.1,0) -- (2.1,-1);
    \end{tikzpicture}
    \caption{The polytope ${\cal C}(3,5)$, combinatorially equivalent to the exchange graph~\cite{fomin2002cluster} of the $\Gr(3,5)$ (or $A_2$) cluster algebra.}
    \label{qiupfhq9}
\end{figure}

In order to illustrate some key terminology we begin by considering the $2$-dimensional polytope called $\mathcal{C}(3,5)$ in~\cite{Arkani-Hamed:2019rds}.  It is equivalent to the associahedron~\cite{stasheff1,stasheff2} $K_4$ as constructed in~\cite{Arkani-Hamed:2017mur} and can be realized in $\mathbb{R}^2$ with coordinates $(X_1, X_2)$ by the inequalities
\begin{align}
\begin{split}
X_1 &\ge 0\,,\\
X_2 &\ge 0\,,\\
c_{135} + c_{235} - X_1 &\ge 0\,,\\
c_{235} + c_{245} - X_2 &\ge 0\,,\\
c_{245} + X_1 - X_2 &\ge 0\,,
\end{split}
\label{eq:c35example}
\end{align}
where the $c$'s are positive constants (see Fig.~\ref{qiupfhq9}).

There are three natural structures we can use to label each facet ${\mathcal{F}}$ of this polytope:
\begin{enumerate}
\item the function in~(\ref{eq:c35example}) that vanishes on ${\mathcal{F}}$ (and is positive inside the polytope),
\item the generator of the ray normal to ${\mathcal{F}}$ (we always choose the generator to be the first integer point along the \emph{outward} pointing normal ray),
\item or the $\Gr(3,5)$ cluster variable whose ${\bf g}$-vector~\cite{fomin2007cluster} is normal to ${\mathcal{F}}$.
\end{enumerate}
The correspondence between these structures for the five facets of $\mathcal{C}(3,5)$ is shown in Tab.~\ref{tab:one}.  Note that by convention we always fix the overall normalization of each kinematic function so that the coefficients of the $X$'s match the components of (the negative of) the corresponding generator.

\begin{table}
\begin{center}
\begin{tabular}{c|c|c}
kinematic function & generator & cluster variable \\
\hline\hline
$\color{brown} s_{125} = \color{black} c_{135} + c_{235} - X_1 $ & $(1,0)$ & $\langle 1\,2\,4 \rangle$\\
\hline
$\color{brown} s_{234} = \color{black} c_{235} + c_{245} - X_2$ & $(0,1)$ & $\langle 1\,3\,4 \rangle$\\
\hline
$\color{brown} s_{145} = \color{black} c_{245} + X_1 - X_2$ & $(-1,1)$ & $\langle 1\,3\,5 \rangle$\\
\hline
$\color{brown} s_{123} = \color{black} X_1$ & $(-1,0)$ & $\langle 2\,3\,5 \rangle$\\
\hline
$\color{brown} s_{345} = \color{black} X_2$ & $(0,-1)$ & $\langle 2\,4\,5 \rangle$\\
\end{tabular}
\end{center}
\caption{The correspondence between kinematic functions, generators, and cluster variables for the $\mathcal{C}(3,5)$ polytope.  Our conventions are summarized in Sec.~\ref{sec:conventions}, where we also review the definition and properties of the generalized Mandelstam variables $s_{ijk}$.}
\label{tab:one}
\end{table}

\subsection{Cluster Series}
\label{sec:clusteralgebraicseries}

The $\Gr(k,n)$ cluster algebra~\cite{fomin2002cluster,scott2006grassmannians} is finite if and only if $n{+}1 > d = (k{-}1)(n{-}k{-}1)$.  For polytopes associated to these algebras, it has been found (by explicit computation in all cases studied so far) that the generator of each normal ray is a ${\bf g}$-vector of the cluster algebra.  It is this fact that allowed us to fill in the third column in Tab.~\ref{tab:one}.  In contrast, \cite{Drummond:2019cxm,Arkani-Hamed:2019rds,Henke:2019hve} studied polytopes associated to the infinite algebra $\Gr(4,8)$ having a facet normal to
\begin{align}
(-1,1,0,1,0,-1,0,-1,1)\,,
\label{eq:firstray}
\end{align}
which is known to not be a ${\bf g}$-vector of $\Gr(4,8)$ (moreover, it is known to not even be inside the cluster fan~\cite{chang2020quantum}).  For this reason we call~(\ref{eq:firstray}) an \textit{exceptional generator}.  All exceptional generators of the polytopes studied in~\cite{Drummond:2019cxm,Arkani-Hamed:2019rds,Henke:2019hve} are related to~(\ref{eq:firstray}) by an element of the cluster modular group\footnote{In fact, (\ref{eq:firstray}) may be essentially unique in that all rays in $\mathbb{R}^9$ are either inside the $\Gr(4,8)$ cluster fan or related to the one generated by~(\ref{eq:firstray}) by an element of the cluster modular group~\cite{fraser2020braid}.  We thank C.~Kalousios for providing some data that supports this hypothesis.}.

Since it is not possible to assign a cluster variable to exceptional facets, \cite{Arkani-Hamed:2019rds} suggested instead to assign to each ray $\mathbb{R}^{+} {\bf y}$ the (formal) \textit{cluster series} (called a cluster algebraic function in~\cite{Arkani-Hamed:2019rds}) defined by
\begin{align}
f_{\bf y}(t) = \sum_{m \ge 0} \mathcal{B}(m {\bf y}) t^m\,,
\label{eq:clusteralgebraicseries}
\end{align}
where $\mathcal{B}({\bf y})$ is a cluster algebra basis element associated to the lattice point ${\bf y} \in \mathbb{Z}^d$.

In~\cite{Arkani-Hamed:2019rds} it was conjectured that the cluster series associated to the ray generated by~(\ref{eq:firstray}) takes the form
\begin{align}
\frac{1}{1 - A\,t + B\,t^2}
\label{eq:firstseries}
\end{align}
in the canonical basis~\cite{lusztig1990canonical}, where
\begin{align}
\begin{split}
A &= \langle 1\,2\,5\,6 \rangle \langle 3\,4\,7\,8\rangle - \langle 1\,2\,7\,8 \rangle \langle 3\,4\,5\,6 \rangle - \langle 1\,2\,3\,4 \rangle \langle 5\,6\,7\,8 \rangle\,, \\
B &= \langle 1\,2\,3\,4 \rangle \langle 3\,4\,5\,6 \rangle \langle 5\,6\,7\,8 \rangle \langle 1\,2\,7\,8 \rangle\,.
\end{split}
\label{eq:ABforfirstray}
\end{align}
This conjecture was checked through ${\cal{O}}(t^3)$ by explicit computation using the character formula of~\cite{chang2020quantum}.

In general $f_{\bf g}(t)$ may depend on the choice of basis for the cluster algebra, but we expect that certain important properties of $f_{\bf g}(t)$ are the same in any suitably reasonable basis.  In particular, we expect that it is a rational function of $t$ and that the locations of its poles (in $t$) are basis-independent and located on the positive $t$ axis when the series is evaluated at any point in the positive Grassmannian $\Gr_{>0}(k,n)$.

In Sec.~\ref{sec:series} of this paper we introduce a closely related \textit{web series}.  We conjecture that a web series exists for every ray, but we have not found the specific form of the series in general.  However, for certain rays (those corresponding to almost arborizable webs; see Sec.~\ref{sec:series}) we prove, to all orders in $t$, that the web series takes the form
\begin{align}
\frac{1 - B\,t^2}{1 - A\,t + B\,t^2}\,,
\label{eq:newseries}
\end{align}
with $A, B$ depending on the ray.  In particular we prove that the web series associated to the ray generated by~(\ref{eq:firstray}) takes the form~(\ref{eq:newseries}) with $A, B$ given by~(\ref{eq:ABforfirstray}).  Evidently our web series use a different basis than the one that gives the series~(\ref{eq:firstseries}), but is consistent with the abovementioned expectations (since it has the same poles, at $t = (A \pm \sqrt{A^2 - 4 B})/(2B)$).  We also prove that the web series has the form~(\ref{eq:newseries}), and evaluate the corresponding $A$'s and $B$'s, for 324 normal rays of the polytope $\mathcal{C}^\dagger(4,9)$ that might be relevant to the symbol alphabet of 9-particle scattering amplitudes in SYM theory.

Finally let us note that if ${\bf g}$ is a ${\bf g}$-vector, then $\mathcal{B}({\bf g})$ is the associated cluster variable and ${\mathcal{B}}(m {\bf g}) = {\mathcal{B}}({{\bf g}})^m$ for any choice of basis, so the cluster  series is basis-independent and geometric:
\begin{align}
f_{\bf g}(t) = \frac{1}{1 - t {\mathcal{B}}({\bf g})}\,.
\label{eq:geometric}
\end{align}
In such cases the information content of knowing $f_{\bf g}(t)$ is the same as that of knowing the cluster variable ${\mathcal{B}}({\bf g})$.  This provides a sense in which it is reasonable---for infinite algebras---to generalize the third column ``cluster variable'' of Tab.~\ref{tab:one} to ``cluster series'' as~\cite{Arkani-Hamed:2019rds} did, or to ``web series'', as we shall do.

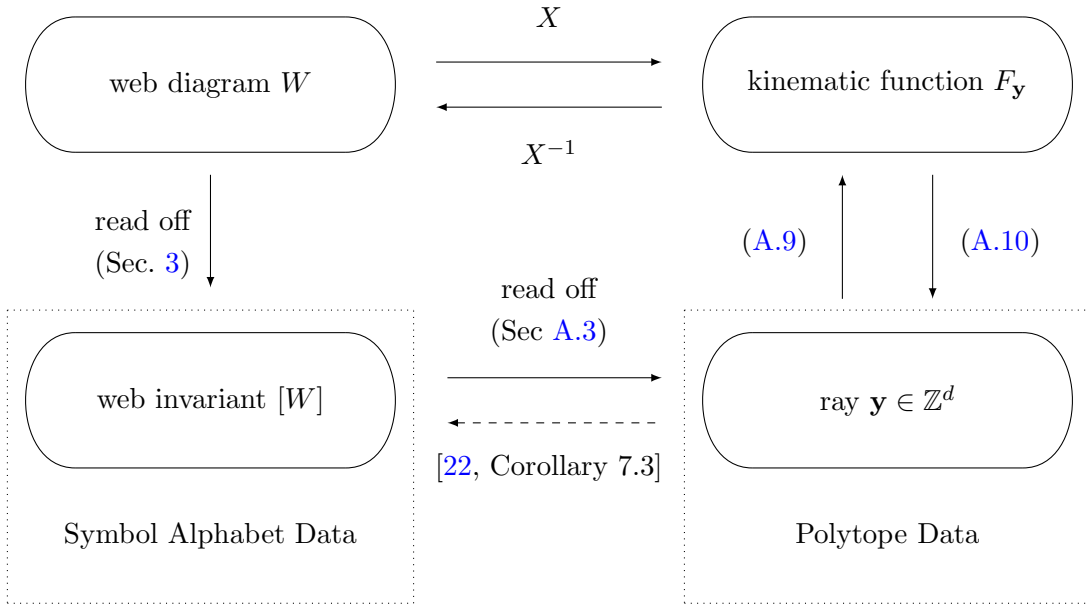
\begin{figure}
\begin{center}
\begin{tikzpicture}[scale=0.6]
	\begin{pgfonlayer}{nodelayer}
		\node [style=none] (0) at (-10.5, 5) {};
		\node [style=none] (1) at (-4.5, 5) {};
		\node [style=none] (2) at (-10.5, 2) {};
		\node [style=none] (3) at (-4.5, 2) {};
		\node [style=none] (7) at (-7.5, 3.5) {web diagram $W$};
		\node [style=none] (8) at (4.5, 5) {};
		\node [style=none] (9) at (10.5, 5) {};
		\node [style=none] (10) at (4.5, 2) {};
		\node [style=none] (11) at (10.5, 2) {};
		\node [style=none] (12) at (7.5, 3.5) {kinematic function $F_{\bf y}$};
		\node [style=none] (13) at (-10.5, -2) {};
		\node [style=none] (14) at (-4.5, -2) {};
		\node [style=none] (15) at (-10.5, -5) {};
		\node [style=none] (16) at (-4.5, -5) {};
		\node [style=none] (17) at (-7.5, -3.5) {web invariant $[W]$};
		\node [style=none] (18) at (4.5, -2) {};
		\node [style=none] (19) at (10.5, -2) {};
		\node [style=none] (20) at (4.5, -5) {};
		\node [style=none] (21) at (10.5, -5) {};
		\node [style=none] (22) at (7.5, -3.5) {ray ${\bf y} \in \mathbb{Z}^d$};
		\node [style=none] (23) at (-7.5, 1.5) {};
		\node [style=none] (24) at (-7.5, -1) {};
		\node [style=none] (25) at (6.5, 1.5) {};
		\node [style=none] (26) at (6.5, -1.25) {};
		\node [style=none] (27) at (8.5, 1.5) {};
		\node [style=none] (28) at (8.5, -1.25) {};
		\node [style=none] (29) at (-2.25, -3) {};
		\node [style=none] (30) at (2.5, -3) {};
		\node [style=none] (31) at (-2.5, 4) {};
		\node [style=none] (32) at (2.5, 4) {};
		\node [style=none] (33) at (-9, 0.5) {read off};
		\node [style=none] (34) at (5, 0) {(\ref{eq:fromy})};
		\node [style=none] (35) at (10, 0) {(\ref{eq:toy})};
		\node [style=none] (36) at (0, -2) {(Sec~\ref{sec:gvectorappendix})};
		\node [style=none] (37) at (0, 5) {$X$};
		\node [style=none] (38) at (0, -1) {read off};
		\node [style=none] (39) at (-9, -0.5) {(Sec.~\ref{sec:reviewfp})};
		\node [style=none] (40) at (-2.5, 3) {};
		\node [style=none] (41) at (2.5, 3) {};
		\node [style=none] (42) at (0, 2) {$X^{-1}$};
		\node [style=none] (43) at (-2.25, -4) {};
		\node [style=none] (44) at (2.5, -4) {};
		\node [style=none] (45) at (3, -1.5) {};
		\node [style=none] (46) at (12, -1.5) {};
		\node [style=none] (47) at (3, -8) {};
		\node [style=none] (48) at (12, -8) {};
		\node [style=none] (49) at (-12, -1.5) {};
		\node [style=none] (50) at (-3, -1.5) {};
		\node [style=none] (51) at (-3, -8) {};
		\node [style=none] (52) at (-12, -8) {};
		\node [style=none] (53) at (-7.5, -6.5) {Symbol Alphabet Data};
		\node [style=none] (54) at (7.5, -6.5) {Polytope Data};
		\node [style=none] (55) at (0, -5) {\cite[Corollary 7.3]{chang2020quantum}};
	\end{pgfonlayer}
	\begin{pgfonlayer}{edgelayer}
		\draw (0.center) to (1.center);
		\draw [bend left=90, looseness=1.25] (1.center) to (3.center);
		\draw [bend left=270, looseness=1.25] (0.center) to (2.center);
		\draw (2.center) to (3.center);
		\draw (8.center) to (9.center);
		\draw [bend left=90, looseness=1.25] (9.center) to (11.center);
		\draw [bend left=270, looseness=1.25] (8.center) to (10.center);
		\draw (10.center) to (11.center);
		\draw (13.center) to (14.center);
		\draw [bend left=90, looseness=1.25] (14.center) to (16.center);
		\draw [bend left=270, looseness=1.25] (13.center) to (15.center);
		\draw (15.center) to (16.center);
		\draw (18.center) to (19.center);
		\draw [bend left=90, looseness=1.25] (19.center) to (21.center);
		\draw [bend left=270, looseness=1.25] (18.center) to (20.center);
		\draw (20.center) to (21.center);
		\draw[arrows={-latex}] (31.center) to (32.center);
		\draw[arrows={-latex}] (29.center) to (30.center);
		\draw[arrows={latex-}] (24.center) to (23.center);
		\draw[arrows={-latex}] (26.center) to (25.center);
		\draw[arrows={latex-}] (28.center) to (27.center);
		\draw[arrows={latex-}] (40.center) to (41.center);
		\draw[dashed,arrows={latex-}] (43.center) to (44.center);
		\draw[dotted] (45.center) to (46.center);
		\draw[dotted] (47.center) to (48.center);
		\draw[dotted] (46.center) to (48.center);
		\draw[dotted] (45.center) to (47.center);
		\draw[dotted] (49.center) to (50.center);
		\draw[dotted] (49.center) to (52.center);
		\draw[dotted] (52.center) to (51.center);
		\draw[dotted] (51.center) to (50.center);
	\end{pgfonlayer}
\end{tikzpicture}
\end{center}
\caption[Paper Outline]{An outline of how our work fits into the literature.  Recent studies have uncovered an apparent connection between data associated to the facets (with normal ray ${\bf y}$) of certain $\Gr(4,n)$-polytopes (indicated on the second column) and the symbol alphabet of $n$-particle amplitudes in SYM theory (left column).  One route from the former to the latter~\cite{Arkani-Hamed:2019rds} (the dotted line) uses the character formula of~\cite{chang2020quantum} and requires the series~(\ref{eq:clusteralgebraicseries}) to be computed term by term, with the $m$-th term having computational complexity ${\mathcal{O}}((mw)!)$, where $w$ is an integer that depends on the facet.  In this paper we employ web invariants and the Fomin-Pylyavskyy conjectures to provide an alternate route, indicated by the arrow labeled $X^{-1}$.  Here $X$ is a map from web invariants to kinematic functions that we define in Sec.~\ref{sec:Xmapdefinition}.  Its inverse $X^{-1}$ can be computed in practice by scanning over a manifestly finite set of candidate preimages $W$ for any given $F_{\bf y}$.  Moreover, we provide a recursive (in $m$) all-order proof that for certain web invariants (those having a single closed loop), there is a web series that sums exactly to~(\ref{eq:newseries}), with $A, B$ depending on the facet.  This proof applies to the exceptional $\Gr(4,8)$ generators encountered in~\cite{Drummond:2019cxm,Arkani-Hamed:2019rds,Henke:2019hve}, and here we show that it also applies to the generators of 324 facets of the $\mathcal{C}^\dagger(4,9)$ polytope, for which we compute the exact web series.  Other alternate routes have been studied in~\cite{Drummond:2019cxm,Herderschee:2021dez}.}
\label{fig:outline}
\end{figure}

\begin{table}
\begin{center}
\begin{tabular}{l|c|c|c|c}
$n$-particle symbol data for $n=$ & 6 & 7 & 8 & 9 \\
\hline\hline
\# of rational letters & 9 & 42 & 172 & 522 \\
\hline
\# of algebraic letters & 0 & 0 & 18 & 99 \\
\hline
\# of distinct square roots & 0 & 0 & 2 & 9 \\
\multicolumn{1}{c}{}\\
$\mathcal{C}^\dagger(4,n)$ polytope data for $n=$ & 6 & 7 & 8 & 9 \\
\hline\hline
\# of facets normal to ${\bf g}$-vectors of $\Gr(4,n)$ & 9 & 42 & 272 & 3078 \\
\hline
\# of exceptional facets & 0 & 0 & 2 & 351
\end{tabular}
\end{center}
\caption{Summary of data about known symbol letters of $n$-particle amplitudes in SYM theory.  On the top line we omit the $n$ frozen variables of $\Gr(4,n)$.}
\label{tab:symboldata}
\end{table}

\subsection{Symbol Letters}

Finally let us briefly review the observed connection between the symbol alphabet of $n$-particle amplitudes in SYM theory and $\Gr(4,n)$-polytopes.  All known symbol letters (see Sec.~\ref{sec:summaryofsymbolletters} for more details) fall into two classes: \textit{rational} letters are cluster variables of $\Gr(4,n)$, and \textit{algebraic} letters have the form\footnote{All algebraic letters currently known have degree 2, but there is no reason to doubt that letters of arbitrarily high degree await discovery.} $\frac{a-\sqrt{b}}{a+\sqrt{b}}$ where $a, b$ are polynomials in Pl\"ucker coordinates.  Tab.~\ref{tab:symboldata} summarizes the number of each type of letter that is known to appear in various $n$-particle amplitudes, as well as the number of distinct square roots (each $\sqrt{b}$ appears in several algebraic letters, i.e.~paired with various different $a$'s).

Tab.~\ref{tab:symboldata} also summarizes data about the facets of the polytopes called $\mathcal{C}^\dagger(4,n)$ in~\cite{Arkani-Hamed:2019mrd}. There (and also in~\cite{Drummond:2019cxm,Henke:2019hve}) they were constructed and studied for $n \le 8$, and found to contain information about the $n$-particle symbol alphabet in the following sense.  First, the cluster variables associated to the ${\bf g}$-vector facets (second to last line) include all known rational symbol letters (top line).  For $n=8$ it was further observed that the square roots appearing in the poles of the (conjectured) series~(\ref{eq:firstseries}) associated to the two exceptional facets (bottom line), given by $\sqrt{A^2 - 4 B}$ in terms of~(\ref{eq:ABforfirstray}) and its image under a $\mathbb{Z}_8$ cyclic shift, agree precisely with the two distinct square roots known to appear in algebraic symbol letters of 8-particle amplitudes (third line).

In this paper we extend this analysis to $n=9$  using the algorithm summarized in Fig.~(\ref{fig:outline}).  We find that the polytope $\mathcal{C}^\dagger(4,9)$ has 3429 facets; 3078 are normal to ${\bf g}$-vectors of $\Gr(4,9)$ while the other 351 are exceptional.  The 3078 cluster variables associated to the former include the 522 (non-frozen) rational letters shown in Tab.~\ref{tab:symboldata}.  For 324 of the latter we prove that the web series has the form~(\ref{eq:newseries}); the 324 distinct square roots of the form $\sqrt{A^2 - 4 B}$ obtained in this way include the 9 counted in the third line of the table.  The remaining 27 exceptional facets remain more mysterious.  Although we expect that web series exist for them as well, we have not been able to find their explicit form.

As a further application of our technology we also define and study the polytopes $\mathcal{C}^\dagger(3,n)$.  Interestingly we find that they do not have any exceptional facets for $n \le 10$, which is as far as we have computed (see Sec.~\ref{sec:results}).

\section{Review of Tensor Diagrams and the Fomin-Pylyavskyy Conjectures}
\label{sec:reviewfp}

In this section we review some basic facts about tensor diagrams, which provide a graphical way to encode data about the cluster structure of the Grassmannian.  The connection between tensor diagrams and $\Gr(k,n)$ cluster algebras was first studied by Fomin and Pylyavskyy in~\cite{fomin2016tensor}.  The main elements of this connection are given by the Fomin-Pylyavskyy (FP) conjectures, which have been partly proved in~\cite{fraser2020braid}.  Our work relies on the FP conjectures in a manner discussed in Sec.~\ref{sec:results}.

An $\mathfrak{sl}_k$ \textit{tensor diagram} is a finite graph drawn inside a disk with $n$ marked points (labeled $1,\ldots,n$ clockwise around its boundary) satisfying the requirements:
\begin{enumerate}
\item all boundary vertices are colored black, and may have arbitrary valence,
\item each internal vertex may be either black or white, but must have valence $k$,
\item and each edge of the graph must connect a black vertex to a white vertex.
\end{enumerate}
A planar tensor diagram is called a \textit{web}, and a tensor diagram with no closed loops (of internal vertices) is called a \textit{tree}.  If we glue all of the vertices and edges of two or more webs into the same disk, we get a \textit{combination} of webs.

To each diagram $D$ we can associate a \textit{tensor invariant} $[D]$ constructed as follows.  First, we associate to each boundary vertex $i$ a $k$-component vector $Z_i^a$.  (For $k=4$ these are the familiar momentum twistor variables that encode massless $n$-particle kinematic data.)  Then to each white vertex we associate $\epsilon^{a_1 \cdots a_k}$, to each internal black vertex we associate $\epsilon_{a_1 \cdots a_k}$, and we contract all indices as indicated by the edges of the graph.  The resulting invariant is always a homogeneous polynomial in the Pl\"ucker coordinates
\begin{align}
\langle i_1\, i_2\, \ldots\, i_k \rangle = \det(Z_{i_1} Z_{i_2} \cdots Z_{i_k})
\end{align}
on $\Gr(k,n)$.  This definition suffices for our purposes, but it is not precise because when $k$ is even it leaves the overall sign of $[D]$ undetermined thanks to $\epsilon^{a_2 \cdots a_k a_1} = - \epsilon^{a_1 a_2 \cdots a_k}$.  For a proper definition of tensor invariants, including a detailed discussion of how to fix this sign, we refer the reader to~\cite{cautis2014webs,fraser2019dimers}.  In practice we will determine the ``correct'' overall sign for any invariant by requiring that it evaluates to a positive number when the $k \times n$ matrix $Z_i^a$ is an element of the positive Grassmannian $\Gr_{>0}(k,n)$.

If $W$ is a web satisfying certain additional conditions\footnote{\label{nonelliptic}For $k=3$ $W$ must be \emph{non-elliptic}, which means every pair of vertices is connected by at most one edge and each face formed by interior vertices has at least six sides~\cite{fomin2016tensor}.  For $k=4$ $W$ can have at most double edges and must have no 2-cycles~\cite{fraser2020braid}.} we call $[W]$ a \textit{web invariant}.  If $W$ is a combination of two or more webs $W_1, W_2, \ldots$ then $[W]$ is the product of the web invariants $[W_1], [W_2], \ldots$.  If $[W]$ is not a product of two or more web invariants then we say that $[W]$ is \textit{indecomposable}.

As their name suggests, tensor invariants are invariant under certain graphical moves known as \textit{skein relations} (see Fig.~\ref{figskein34})~\cite{kuperberg1996spiders,kim2003graphical,cautis2014webs,fomin2016tensor}.  One important application of these relations is that they can sometimes be used to convert a web $W$ with closed loops into a tree diagram $D$ that is equivalent in the sense that $[D] = [W]$; if this is possible then the web $W$ is called \textit{arborizable} (note that $D$ may or may not be a web, i.e.~it may be non-planar).

The \textit{Fomin-Pylyavskyy conjectures}~\cite{fomin2016tensor} comprise several interesting connections between tensor invariants and cluster variables.  These have been proven up to $\Gr(3,9)$ and $\Gr(4,8)$ in~\cite{fraser2020braid}.  For our purposes the key conjecture is: \textit{the set of cluster (and frozen) variables coincides with the set of indecomposable arborizable web invariants}.  Henceforth we only consider indecomposable diagrams.

In order to better familiarize the reader with tensor diagrams let us now introduce some tricks for quickly reading off the invariants associated to certain diagrams.

\subsection{\texorpdfstring{$\mathfrak{sl}_2$ Tensor Diagrams}{sl2 Tensor Diagrams}}
\label{sec:sl2diagrams}

This case is rather trivial in an instructive way.  The only structures an $\mathfrak{sl}_2$ tensor diagram can have are strands that begin and end on boundary vertices, passing along the way through an odd number of internal vertices alternating between white and black.  All internal vertices except for a single white vertex on each strand can be removed by a skein relation ($\epsilon^{ab} \epsilon_{bc} = \delta^a_c$).  The web invariants have the form $[W_{ij}] = \langle i\,j\rangle$, corresponding to the web $W_{ij}$ with a single strand connecting boundary vertices $1 \le i<j \le n$.

\subsection{\texorpdfstring{$\mathfrak{sl}_3$ Tensor Diagrams}{sl3 Tensor Diagrams}}
\label{sec:sl3diagrams}

The invariant for any $\mathfrak{sl}_3$ tree diagram $D$ can be read out in a simple way~\cite{fomin2016tensor}.  First we choose any internal vertex of $D$ to be the \textit{central vertex} $v$ and assign a direction to each edge in such a way that it points from the boundary of the diagram towards $v$; this assignment is unambiguous if $D$ is a tree.  Then at each internal vertex except $v$, there must be two inward pointing edges and one outward pointing edge.  Having already assigned a vector $Z_i$ to each boundary vertex $i$, we now assign to each internal vertex $v' \ne v$ the cross-product of the two vectors or covectors associated to the two inward edges at $v'$; this is a vector if $v'$ is black and a covector if $v'$ is white.  Then the invariant of $D$ is equal to the determinant of the three (co)vectors assigned to the three incoming edges at the central vertex $v$.

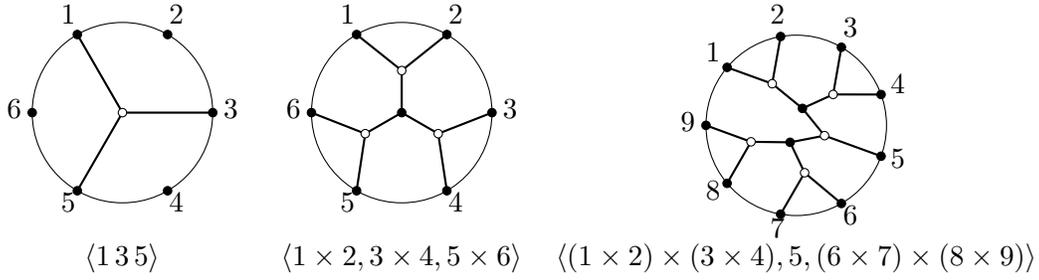
\begin{figure}[!ht]
     \centering
     \begin{tikzpicture}[baseline={([yshift=-2.2ex]current bounding box.center)},scale=0.8,xscale=-1]
     \def \cen{(0,0)};
     \def \rad{1.5};
     \draw {\cen} circle (\rad);

     \foreach \x in {1,2,3,4,5,6}{
          \filldraw[fill=black,draw=black] \cen+(60*\x:\rad) circle (2pt);
          \draw[white] \cen+({60* (\x)}:{\rad+0.3}) circle (0.1) node[black]{\tikz{\node at (0,0) {$\x$}}};
          \coordinate(v\x) at (60*\x:1.5);}

     \coordinate(A) at (0,0);
     \draw[thick] (A) -- (v1) (A) -- (v3) (A) -- (v5);
     \filldraw[fill=white,draw=black] (0,0) circle (2pt);

     \node at (0,-\rad-0.91) {$\langle 1\,3\,5\rangle$};

     \end{tikzpicture}
     \begin{tikzpicture}[baseline={([yshift=-2.2ex]current bounding box.center)},scale=0.8,xscale=-1]
     \def \cen{(0,0)};
     \def \rad{1.5};
     \draw {\cen} circle (\rad);

     \foreach \x in {1,2,3,4,5,6}{
          \filldraw[fill=black,draw=black] \cen+(60*\x:\rad) circle (2pt);
          \draw[white] \cen+({60* (\x)}:{\rad+0.3}) circle (0.1) node[black]{\tikz{\node at (0,0) {$\x$}}};
          \coordinate(v\x) at (60*\x:1.5);}

     \coordinate(A) at (0,0);
     \coordinate(B1) at (90:0.7);
     \coordinate(B2) at (210:0.7);
     \coordinate(B3) at (330:0.7);
     \draw[thick] (A) -- (B1) (A) -- (B3) (A) -- (B2);
     \draw[thick] (v1) -- (B1) -- (v2) (v3) -- (B2) -- (v4) (v5) -- (B3) -- (v6);

     \filldraw[fill=black,draw=black] (A) circle (2pt);
     \filldraw[fill=white,draw=black] (B1) circle (2pt);
     \filldraw[fill=white,draw=black] (B2) circle (2pt);
     \filldraw[fill=white,draw=black] (B3) circle (2pt);

     \node at (0,-\rad-0.91) {$\langle 1\times2, 3\times4, 5\times6\rangle$};

     \end{tikzpicture}
     \begin{tikzpicture}[baseline={([yshift=-2.2ex]current bounding box.center)},scale=0.8,xscale=-1]
     \def \cen{(0,0)};
     \def \rad{1.5};
     \draw {\cen} circle (\rad);

     \foreach \x in {1,2,3,4,5,6,7,8,9}{
          \filldraw[fill=black,draw=black] \cen+(40*\x:\rad) circle (2pt);
          \draw[white] \cen+({40* (\x)}:{\rad+0.3}) circle (0.1) node[black]{\tikz{\node at (0,0) {$\x$}}};
          \coordinate(v\x) at (40*\x:1.5);}

     \coordinate (A12) at (60:.8);
     \coordinate (A34) at (140:.8);
     \coordinate (A67) at (260:.8);
     \coordinate (A89) at (340:.8);
     \coordinate (B1234) at (110:.3);
     \coordinate (B6789) at (290:.3);
     \coordinate (C) at (200:.5);

     \draw[thick] (v1) -- (A12) -- (v2) (v3) -- (A34) -- (v4) (v6) -- (A67) -- (v7) (v8) -- (A89) -- (v9);
     \draw[thick] (A12) -- (B1234) -- (A34) (A67) -- (B6789) -- (A89);
     \draw[thick] (v5) -- (C) (B1234) -- (C) (B6789) -- (C);

     \foreach \x in {A12,A34,A67,A89,C}
     \filldraw[fill=white,draw=black] (\x) circle (2pt);

     \foreach \x in {B1234,B6789}
     \filldraw[fill=black,draw=black] (\x) circle (2pt);

     \node at (0,-\rad-.7) {$\langle (1\times2)\times (3\times4), 5, (6\times7)\times (8\times9) \rangle$};

     \end{tikzpicture}
     \caption{Examples of $\mathfrak{sl}_3$ webs and their corresponding invariants.}
     \label{figwebs}
\end{figure}

Some $\mathfrak{sl}_3$ webs and their corresponding invariants are shown in Fig.~\ref{figwebs}.  For example, let $D$ be the diagram shown in Fig.~\ref{figwebs}(b).  If we choose the black vertex in the middle to be the central vertex, then the covectors assigned to the top, bottom right, and bottom left white vertices are respectively $Z_1 \times Z_2$ (which we immediately abbreviate to $1 \times 2$), $3 \times 4$, and $5 \times 6$, yielding the tensor invariant $[D]=\langle 1 \times 2, 3 \times 4, 5 \times 6 \rangle$ as indicated in the figure.  We could also have chosen, say, the top white vertex as the center, in which case the invariant would have been computed as $[D]' = \langle 1, 2, (3 \times 4) \times (5 \times 6)\rangle$, but it is easy to check that $[D] = [D]'$.

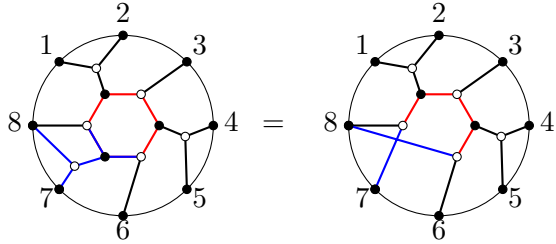
\begin{figure}
     \centering
     \begin{tikzpicture}[baseline={([yshift=-2.2ex]current bounding box.center)},scale=0.8,xscale=-1]

          \def \cen{(0,0)};
          \def \rad{1.5};
          \draw {\cen} circle (\rad);

          \foreach \x in {1,2,3,4,5,6,7,8}{
               \filldraw[fill=black,draw=black] \cen+(45*\x:\rad) circle (2pt);
               \draw[white] \cen+({45* (\x)}:{\rad+0.3}) circle (0.1) node[black]{\tikz{\node at (0,0) {$\x$}}};
               \coordinate(v\x) at (45*\x:1.5);}

          \foreach \x in {1,2,3,4,5,6}{
               \coordinate(A\x) at ({60*\x}:{.4*\rad});
               \draw[thick,red] ({60*\x}:{.4*\rad}) -- ({60+60*\x}:{.4*\rad});}

          \coordinate(B1) at (65:{.7*\rad});
          \coordinate(B2) at (190:{.7*\rad});
          \coordinate(B3) at (320:{.7*\rad});

          \draw[thick] (B1) -- (A1) (B2) -- (A3);
          \draw[thick] (A2) -- (v3) (A4) -- (v6) (A6) -- (v8);

          \draw[thick,blue] (A4) -- (A5) -- (A6) (A5) -- (B3) (v7) -- (B3) -- (v8);

          \draw[thick] (v1) -- (B1) -- (v2) (v4) -- (B2) -- (v5);

          \foreach \x in {1,3,5}{
               \filldraw[fill=black,draw=black] (A\x) circle (2pt);
               \filldraw[fill=white,draw=black] ({60+60*\x}:{.4*\rad}) circle (2pt);}
          \filldraw[fill=black,draw=black] (v7) circle (2pt);
          \filldraw[fill=black,draw=black] (v8) circle (2pt);
          \foreach \x in {1,2,3}
               \filldraw[fill=white,draw=black] (B\x) circle (2pt);

          \node at (-2.5,0) {$=$};
     \end{tikzpicture}
     \begin{tikzpicture}[baseline={([yshift=-2.2ex]current bounding box.center)},scale=0.8,xscale=-1]
          \def \cen{(0,0)};
          \def \rad{1.5};
          \draw {\cen} circle (\rad);

          \foreach \x in {1,2,3,4,5,6,7,8}{
               \filldraw[fill=black,draw=black] \cen+(45*\x:\rad) circle (2pt);
               \draw[white] \cen+({45* (\x)}:{\rad+0.3}) circle (0.1) node[black]{\tikz{\node at (0,0) {$\x$}}};
               \coordinate(v\x) at (45*\x:1.5);}

          \foreach \x in {1,2,3,4,6}{
               \coordinate(A\x) at ({60*\x}:{.4*\rad});}

          \draw[thick,red] (A6) -- (A1) -- (A2) -- (A3) -- (A4);

          \coordinate(B1) at (65:{.7*\rad});
          \coordinate(B2) at (190:{.7*\rad});

          \draw[thick] (B1) -- (A1) (B2) -- (A3);
          \draw[thick] (A2) -- (v3) (A4) -- (v6) (A6) -- (v8);
          \draw[thick] (v1) -- (B1) -- (v2) (v4) -- (B2) -- (v5);

          \draw[thick,blue] (A4) -- (v8) (A6) -- (v7);

          \foreach \x in {1,3}{
               \filldraw[fill=black,draw=black] (A\x) circle (2pt);
               \filldraw[fill=white,draw=black] ({60+60*\x}:{.4*\rad}) circle (2pt);}
          \filldraw[fill=white,draw=black] (A6) circle (2pt);

          \filldraw[fill=black,draw=black] (v7) circle (2pt);
          \filldraw[fill=black,draw=black] (v8) circle (2pt);
          \foreach \x in {1,2}
               \filldraw[fill=white,draw=black] (B\x) circle (2pt);

     \end{tikzpicture}

     \caption{An example of arborization, where we break the inner hexagon on the left by applying the first and seventh skein relations shown in Fig.~\ref{figskein34} to the blue edges.  (The blue and red colorings serve only to guide the eye.)}
     \label{figarb}
\end{figure}

This shortcut for computing a tensor invariant $[D]$ can also be used if $D$ is arborizable.  For example, by choosing the white vertex adjacent to $3$ as the center, the invariant associated to the diagram shown in Fig.~\ref{figarb} evaluates to
\begin{align}
\langle (7 \times 8) \times (1 \times 2), 3, (4 \times 5)  \times (6 \times 8) \rangle\,.
\label{strangeexample}
\end{align}

\subsection{\texorpdfstring{$\mathfrak{sl}_4$ Tensor Diagrams}{sl4 Tensor Diagrams}}

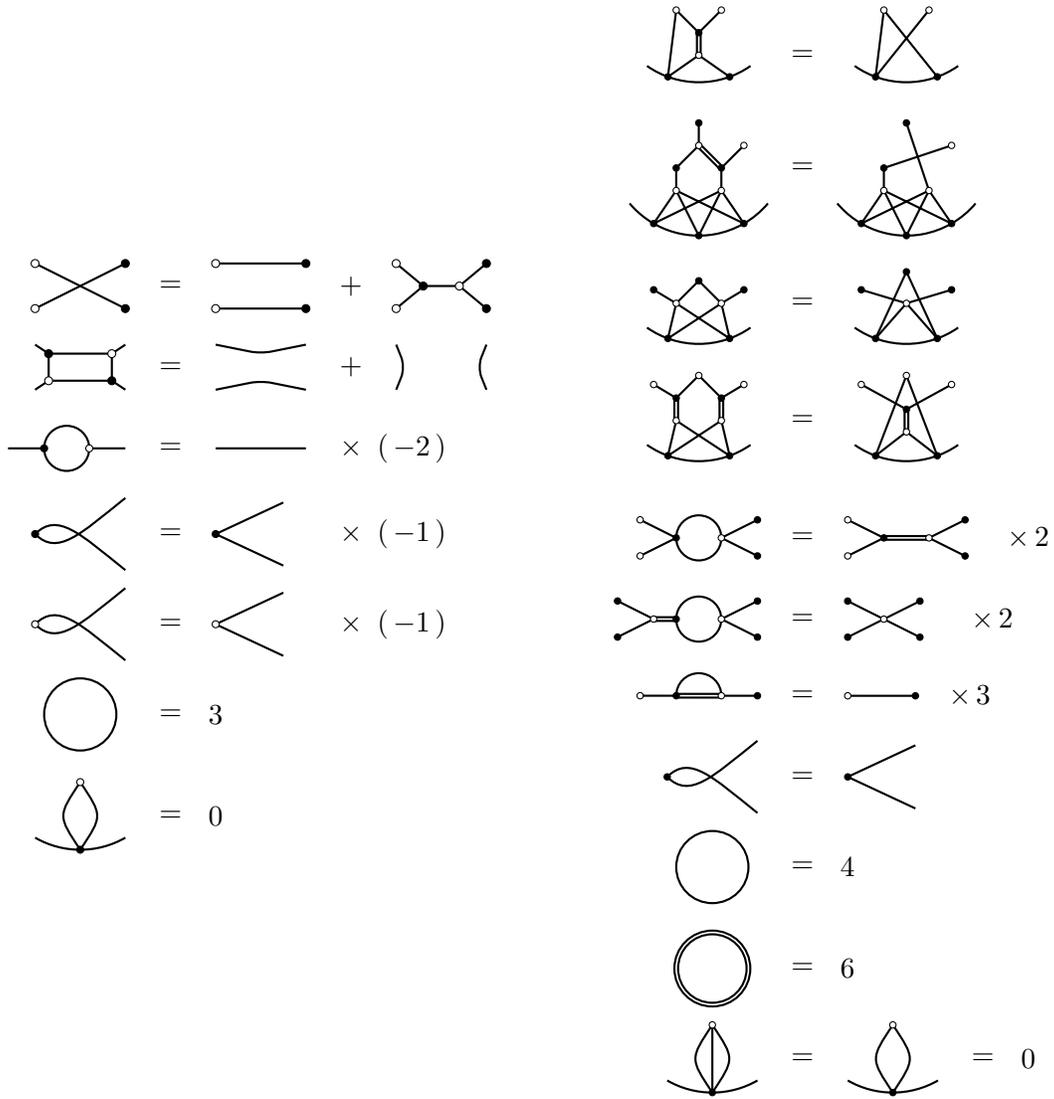
\begin{figure}
\centering
     \begin{subfigure}{0.49\textwidth}
	\centering
          \begin{tikzpicture}[scale=0.6]

          \draw[thick] (-3,5) node{\tikz{\filldraw[fill=white] (0,0) circle (1.5pt)}} -- (-1,4) node{\tikz{\filldraw[fill=black] circle (1.5pt)}};
          \draw[thick] (-3,4) node{\tikz{\filldraw[fill=white] (0,0) circle (1.5pt)}} -- (-1,5) node{\tikz{\filldraw[fill=black] circle (1.5pt)}};

          \node at (0,4.5) {$=$};

          \draw[thick] (1,5) node{\tikz{\filldraw[fill=white] (0,0) circle (1.5pt)}} -- (3,5) node{\tikz{\filldraw[fill=black] (0,0) circle (1.5pt)}};
          \draw[thick] (1,4) node{\tikz{\filldraw[fill=white] (0,0) circle (1.5pt)}} -- (3,4) node{\tikz{\filldraw[fill=black] (0,0) circle (1.5pt)}};

          \node at (4,4.5) {$+$};

          \draw[thick] (5.6,4.5) -- (6.4,4.5);
          \draw[thick] (5,5) node{\tikz{\filldraw[fill=white] (0,0) circle (1.5pt)}} -- (5.6,4.5) node{\tikz{\filldraw[fill=black] (0,0) circle (1.5pt)}} -- (5,4) node{\tikz{\filldraw[fill=white] (0,0) circle (1.5pt)}};
          \draw[thick] (7,4) node{\tikz{\filldraw[fill=black] (0,0) circle (1.5pt)}} -- (6.4,4.5) node{\tikz{\filldraw[fill=white] (0,0) circle (1.5pt)}} -- (7,5) node{\tikz{\filldraw[fill=black] (0,0) circle (1.5pt)}};

          \coordinate (vtext) at (0,2.7);
          \coordinate (v1) at (-3,3.2);
          \coordinate (v2) at (-1,3.2);
          \coordinate (v3) at (-1,2.2);
          \coordinate (v4) at (-3,2.2);
          \coordinate (v5) at (-2.7,3);
          \coordinate (v6) at (-1.3,3);
          \coordinate (v7) at (-1.3,2.4);
          \coordinate (v8) at (-2.7,2.4);

          \draw[thick] (v1) -- (v5) (v2) -- (v6) (v3) -- (v7) (v4) -- (v8);
          \draw[thick] (v8) -- (v5) node{\tikz{\filldraw (0,0) circle (1.5pt)}} -- (v6) node{\tikz{\filldraw[fill=white] (0,0) circle (1.5pt)}} -- (v7) node{\tikz{\filldraw (0,0) circle (1.5pt)}} -- (v8) node{\tikz{\filldraw[fill=white] (0,0) circle (1.5pt)}};

          \node at (vtext) {$=$};


          \draw[thick,rounded corners] (1,3.2) -- (2,3) -- (3,3.2);
          \draw[thick,rounded corners] (1,2.2) -- (2,2.4) -- (3,2.2);
          
          \node at (4,2.7) {$+$};
          
          \draw[thick,rounded corners] (5,3.2) -- (5.2,2.7) -- (5,2.2);
          \draw[thick,rounded corners] (7,3.2) -- (6.8,2.7) -- (7,2.2);
          
          \coordinate (v1) at (-3.6,0.9);
          \coordinate (v3) at ([shift={+(0.8,0)}] v1);
          \coordinate (vcirc) at ([shift={+(0.5,0)}] v3);
          \coordinate (v4) at ([shift={+(1,0)}] v3);
          \coordinate (v5) at ([shift={+(0.8,0)}] v4);
          \coordinate (vtext) at ([shift={+(1.8,0)}] v4);
          
          \draw[thick] (v1) -- (v3) (v5) -- (v4);
          \draw[thick] (vcirc) circle (0.5);
          
          \foreach \x in {4}
              \filldraw[fill=white] (v\x) circle (2.2pt);
          \foreach \x in {3}
              \filldraw[fill=black] (v\x) circle (2.2pt);
          
          \node at (vtext) {$=$};
          
          
          \coordinate (v1) at ([shift={+(1,0)}] vtext);
          \coordinate (v2) at ([shift={+(2,0)}] v1);
          \coordinate (vtext) at ([shift={+(3,0)}] v1);
          \coordinate (vtext2) at ([shift={+(1.35,0)}] vtext);
          
          \draw[thick] (v1) -- (v2);
          
          \node at (vtext) {$\times$};
          \node at (vtext2) {$(\,-2\,)$};
          
          \coordinate (v1) at (-3,-1);
          \coordinate (v2) at (-2.5,-1.5);
          \coordinate (v3) at (-2.5,-.5);
          \coordinate (v4) at (-2,-1);
          \coordinate (v5) at (-1,-.2);
          \coordinate (v6) at (-1,-1.8);
          \coordinate (vtext) at (0,-1);
          \coordinate (v7) at (1,-1);
          \coordinate (v8) at (2.5,-1.7);
          \coordinate (v9) at (2.5,-.3);
          \coordinate (vneg) at (4,-1);
          \coordinate (vneg1) at (5.35,-1);
          
          \draw[thick] (v1) .. controls (v2) and (v4) .. (v5);
          \draw[thick] (v1) .. controls (v3) and (v4) .. (v6);
          \draw[thick] (v8) -- (v7) -- (v9);
          
          \filldraw (v1) circle (2.2pt);
          \filldraw (v7) circle (2.2pt);
          
          \node at (vtext) {$=$};
          \node at (vneg) {$\times$};
          \node at (vneg1) {$(\,-1\,)$};
          
          \coordinate (v1) at (-3,-3);
          \coordinate (v2) at (-2.5,-3.5);
          \coordinate (v3) at (-2.5,-2.5);
          \coordinate (v4) at (-2,-3);
          \coordinate (v5) at (-1,-2.2);
          \coordinate (v6) at (-1,-3.8);
          \coordinate (vtext) at (0,-3);
          \coordinate (v7) at (1,-3);
          \coordinate (v8) at (2.5,-3.7);
          \coordinate (v9) at (2.5,-2.3);
          \coordinate (vneg) at (4,-3);
          \coordinate (vneg1) at (5.35,-3);
          
          \draw[thick] (v1) .. controls (v2) and (v4) .. (v5);
          \draw[thick] (v1) .. controls (v3) and (v4) .. (v6);
          \draw[thick] (v8) -- (v7) -- (v9);
          
          \filldraw[fill=white] (v1) circle (2.2pt);
          \filldraw[fill=white] (v7) circle (2.2pt);
          \node at (vtext) {$=$};
          \node at (vneg) {$\times$};
          \node at (vneg1) {$(\,-1\,)$};
          
          \draw[thick] (-2,-5) circle (0.8);
          \node at (0,-5) {$=$};
          \node at (1,-5) {$3$};
          \draw[thick] (-2,-6)+(240:2) arc (240:300:2);
          \draw[thick] (-2,-6.5) .. controls  (-1.5,-7.25) .. (-2,-8);
          \draw[thick] (-2,-6.5) .. controls  (-2.5,-7.25) .. (-2,-8);
          
          \filldraw[fill=white] (-2,-6.5) circle (2.2pt);
          \filldraw[fill=black] (-2,-8) circle (2.2pt);
          
          \node at (0,-7.25) {$=$};
          \node at (1,-7.25) {$0$};
          
          \end{tikzpicture}

     \end{subfigure}
     \begin{subfigure}{0.49\textwidth}
	\centering
     \begin{tikzpicture}[scale=0.6]

          \coordinate (vtext) at (0,8);
          \coordinate (v1) at ([shift={+(-2.3,0)}] vtext);
          \coordinate (v2) at ([shift={+(0,0.5)}] v1);
          \coordinate (v3) at ([shift={+(-0.5,0.5)}] v2);
          \coordinate (v4) at ([shift={+(0.5,0.5)}] v2);
          \coordinate (vcen) at ([shift={+(0,.9)}] v2);
          \coordinate (vleft) at ([shift={+(250:2)}] vcen);
          \coordinate (vright) at ([shift={+(290:2)}] vcen);
          \coordinate (vrim) at ([shift={+(235:2)}] vcen);

          \draw[thick] (vright) -- (v1) -- (vleft) -- (v3) -- (v2) -- (v4);
          \draw[thick,double] (v1) -- (v2);
          \draw[thick] (vrim) arc (235:305:2);

          \foreach \x in {v1,v3,v4}
               \filldraw[fill=white] (\x) circle (2pt);
          \foreach \x in {v2,vleft,vright}
               \filldraw (\x) circle (2pt);

          \node at (vtext) {$=$};

          \coordinate (v1) at ([shift={+(2.3,0)}] vtext);
          \coordinate (v2) at ([shift={+(0,0.5)}] v1);
          \coordinate (v3) at ([shift={+(-0.5,0.5)}] v2);
          \coordinate (v4) at ([shift={+(0.5,0.5)}] v2);
          \coordinate (vcen) at ([shift={+(0,.9)}] v2);
          \coordinate (vleft) at ([shift={+(250:2)}] vcen);
          \coordinate (vright) at ([shift={+(290:2)}] vcen);
          \coordinate (vrim) at ([shift={+(235:2)}] vcen);

          \draw[thick] (vright) -- (v3) -- (vleft) -- (v4);
          \draw[thick] (vrim) arc (235:305:2);

          \foreach \x in {v3,v4}
               \filldraw[fill=white] (\x) circle (2pt);
          \foreach \x in {vleft,vright}
               \filldraw (\x) circle (2pt);
          \coordinate (vtext) at (0,5.5);
          \coordinate (vtop) at ([shift={+(-2.3,1)}] vtext);
          \coordinate (v0) at ([shift={+(0,-0.5)}] vtop);
          \coordinate (v1) at ([shift={+(-0.5,-0.5)}] v0);
          \coordinate (v2) at ([shift={+(0.5,-0.5)}] v0);
          \coordinate (v20) at ([shift={+(0.5,0.5)}] v2);
          \coordinate (v3) at ([shift={+(0,-0.5)}] v1);
          \coordinate (v4) at ([shift={+(0,-0.5)}] v2);
          \coordinate (vleft) at ([shift={+(240:2)}] v0);
          \coordinate (vright) at ([shift={+(300:2)}] v0);
          \coordinate (vmed) at ([shift={+(270:2)}] v0);
          \coordinate (vrim) at ([shift={+(220:2)}] v0);

          \draw[thick] (v4) -- (vleft) -- (v3) (v4) -- (vright) -- (v3) (v4) -- (vmed) -- (v3);
          \draw[thick] (vtop) -- (v0) -- (v1) -- (v3) (v20) -- (v2) -- (v4);
          \draw[thick,double] (v0) -- (v2);

          \draw[thick] (vrim) arc (220:320:2);

          \foreach \x in {vtop,v1,v2,vleft,vmed,vright}
               \filldraw (\x) circle (2pt);
          \foreach \x in {v0,v3,v4,v20}
               \filldraw[fill=white] (\x) circle (2pt);

          \node at (vtext) {$=$};

          \coordinate (vtop) at ([shift={+(2.3,1)}] vtext);
          \coordinate (v0) at ([shift={+(0,-0.5)}] vtop);
          \coordinate (v1) at ([shift={+(-0.5,-0.5)}] v0);
          \coordinate (v2) at ([shift={+(0.5,-0.5)}] v0);
          \coordinate (v20) at ([shift={+(0.5,0.5)}] v2);
          \coordinate (v3) at ([shift={+(0,-0.5)}] v1);
          \coordinate (v4) at ([shift={+(0,-0.5)}] v2);
          \coordinate (vleft) at ([shift={+(240:2)}] v0);
          \coordinate (vright) at ([shift={+(300:2)}] v0);
          \coordinate (vmed) at ([shift={+(270:2)}] v0);
          \coordinate (vrim) at ([shift={+(220:2)}] v0);

          \draw[thick] (v4) -- (vleft) -- (v3) (v4) -- (vright) -- (v3) (v4) -- (vmed) -- (v3);
          \draw[thick] (vtop) -- (v4) (v20) -- (v1) -- (v3);

          \draw[thick] (vrim) arc (220:320:2);

          \foreach \x in {vtop,v1,vleft,vmed,vright}
               \filldraw (\x) circle (2pt);
          \foreach \x in {v3,v4,v20}
               \filldraw[fill=white] (\x) circle (2pt);
          
          \coordinate (vtext) at (0,2.5);
          \coordinate (v1) at ([shift={+(-2.3,0.5)}] vtext);
          \coordinate (v2) at ([shift={+(-0.5,-0.5)}] v1);
          \coordinate (v3) at ([shift={+(0.5,-0.5)}] v1);
          \coordinate (v4) at ([shift={+(-0.5,0.3)}] v2);
          \coordinate (v5) at ([shift={+(0.5,0.3)}] v3);
          \coordinate (vcen) at ([shift={+(0,0.6)}] v1);
          \coordinate (vleft) at ([shift={+(250:2)}] vcen);
          \coordinate (vright) at ([shift={+(290:2)}] vcen);
          \coordinate (vrim) at ([shift={+(235:2)}] vcen);

          \draw[thick] (v3) -- (vleft) -- (v2) -- (vright) -- (v3);
          \draw[thick] (v4) -- (v2) -- (v1) -- (v3) -- (v5);
          \draw[thick] (vrim) arc (235:305:2);

          \foreach \x in {v2,v3}
               \filldraw[fill=white] (\x) circle (2pt);
          \foreach \x in {v1,vleft,vright,v4,v5}
               \filldraw (\x) circle (2pt);

          \node at (vtext) {$=$};

          \coordinate (v1) at ([shift={+(2.3,0.7)}] vtext);
          \coordinate (v2) at ([shift={+(-0.5,-0.7)}] v1);
          \coordinate (v3) at ([shift={+(0.5,-0.7)}] v1);
          \coordinate (v4) at ([shift={+(-0.5,0.3)}] v2);
          \coordinate (v5) at ([shift={+(0.5,0.3)}] v3);
          \coordinate (v0) at ([shift={+(0,-0.7)}] v1);
          \coordinate (vcen) at ([shift={+(0,0.4)}] v1);
          \coordinate (vleft) at ([shift={+(250:2)}] vcen);
          \coordinate (vright) at ([shift={+(290:2)}] vcen);
          \coordinate (vrim) at ([shift={+(235:2)}] vcen);

          \draw[thick] (v4) -- (v0) -- (v5) (vleft) -- (v0) -- (vright);
          \draw[thick] (vleft) -- (v1) -- (vright);
          \draw[thick] (vrim) arc (235:305:2);

          \foreach \x in {v0}
               \filldraw[fill=white] (\x) circle (2pt);
          \foreach \x in {v1,vleft,vright,v4,v5}
               \filldraw (\x) circle (2pt);

          \coordinate (vtext) at (0,-0.1);
          \coordinate (v1) at ([shift={+(-2.3,1)}] vtext);
          \coordinate (v2) at ([shift={+(-0.5,-0.5)}] v1);
          \coordinate (v3) at ([shift={+(0.5,-0.5)}] v1);
          \coordinate (v4) at ([shift={+(-0.5,0.3)}] v2);
          \coordinate (v5) at ([shift={+(0.5,0.3)}] v3);
          \coordinate (v6) at ([shift={+(0,-0.5)}] v2);
          \coordinate (v7) at ([shift={+(0,-0.5)}] v3);
          \coordinate (vcen) at ([shift={+(0,0.1)}] v1);
          \coordinate (vleft) at ([shift={+(250:2)}] vcen);
          \coordinate (vright) at ([shift={+(290:2)}] vcen);
          \coordinate (vrim) at ([shift={+(235:2)}] vcen);

          \draw[thick] (v7) -- (vleft) -- (v6) -- (vright) -- (v7);
          \draw[thick] (v4) -- (v2) -- (v1) -- (v3) -- (v5);
          \draw[thick,double] (v2) -- (v6) (v3) -- (v7);
          \draw[thick] (vrim) arc (235:305:2);

          \foreach \x in {v1,v4,v5,v6,v7}
               \filldraw[fill=white] (\x) circle (2pt);
          \foreach \x in {vleft,vright,v2,v3}
               \filldraw (\x) circle (2pt);

          \node at (vtext) {$=$};

          \coordinate (v1) at ([shift={+(2.3,1)}] vtext);
          \coordinate (v10) at ([shift={+(0,-.75)}] v1);
          \coordinate (v11) at ([shift={+(0,-.5)}] v10);
          \coordinate (v2) at ([shift={+(-0.5,-0.5)}] v1);
          \coordinate (v3) at ([shift={+(0.5,-0.5)}] v1);
          \coordinate (v4) at ([shift={+(-1,-0.2)}] v1);
          \coordinate (v5) at ([shift={+(1,-0.2)}] v1);
          \coordinate (vcen) at ([shift={+(0,0.1)}] v1);
          \coordinate (vleft) at ([shift={+(250:2)}] vcen);
          \coordinate (vright) at ([shift={+(290:2)}] vcen);
          \coordinate (vrim) at ([shift={+(235:2)}] vcen);

          \draw[thick] (vleft) -- (v11) -- (vright) (v4) -- (v10) -- (v5);
          \draw[thick] (vleft) -- (v1) -- (vright);
          \draw[thick,double] (v10) -- (v11);
          \draw[thick] (vrim) arc (235:305:2);

          \foreach \x in {v1,v4,v5,v11}
               \filldraw[fill=white] (\x) circle (2pt);
          \foreach \x in {v10,vleft,vright}
               \filldraw (\x) circle (2pt);
          \coordinate (v1) at (-3.6,-2.3);
          \coordinate (v2) at ([shift={+(0,-0.8)}] v1);
          \coordinate (v3) at ([shift={+(0.8,0.4)}] v2);
          \coordinate (vcirc) at ([shift={+(0.5,0)}] v3);
          \coordinate (v4) at ([shift={+(1,0)}] v3);
          \coordinate (v5) at ([shift={+(0.8,0.4)}] v4);
          \coordinate (v6) at ([shift={+(0.8,-0.4)}] v4);
          \coordinate (vtext) at ([shift={+(1.8,0)}] v4);

          \draw[thick] (v1) -- (v3) -- (v2) (v5) -- (v4) -- (v6);
          \draw[thick] (vcirc) circle (0.5);

          \foreach \x in {1,2,4}
               \filldraw[fill=white] (v\x) circle (2pt);
          \foreach \x in {3,5,6}
               \filldraw[fill=black] (v\x) circle (2pt);

          \node at (vtext) {$=$};


          \coordinate (v1) at ([shift={+(1,0.4)}] vtext);
          \coordinate (v2) at ([shift={+(0,-0.8)}] v1);
          \coordinate (v3) at ([shift={+(0.8,0.4)}] v2);
          \coordinate (v4) at ([shift={+(1,0)}] v3);
          \coordinate (v5) at ([shift={+(0.8,0.4)}] v4);
          \coordinate (v6) at ([shift={+(0.8,-0.4)}] v4);
          \coordinate (vtext) at ([shift={+(2.2,0)}] v4);

          \draw[thick] (v1) -- (v3) -- (v2) (v5) -- (v4) -- (v6);
          \draw[thick,double] (v3) -- (v4);

          \foreach \x in {1,2,4}
               \filldraw[fill=white] (v\x) circle (2pt);
          \foreach \x in {3,5,6}
               \filldraw[fill=black] (v\x) circle (2pt);

          \node at (vtext) {$\times\,2$};

          \coordinate (v1) at (-4.1,-4.1);
          \coordinate (v2) at ([shift={+(0,-0.8)}] v1);
          \coordinate (v3) at ([shift={+(0.8,0.4)}] v2);
          \coordinate (v7) at ([shift={+(0.5,0)}] v3);;;
          \coordinate (vcirc) at ([shift={+(0.5,0)}] v7);
          \coordinate (v4) at ([shift={+(1,0)}] v7);
          \coordinate (v5) at ([shift={+(0.8,0.4)}] v4);
          \coordinate (v6) at ([shift={+(0.8,-0.4)}] v4);
          \coordinate (vtext) at ([shift={+(1.8,0)}] v4);

          \draw[thick] (v1) -- (v3) -- (v2) (v5) -- (v4) -- (v6);
          \draw[thick] (vcirc) circle (0.5);
          \draw[thick,double] (v3) -- (v7);

          \foreach \x in {3,4}
               \filldraw[fill=white] (v\x) circle (2pt);
          \foreach \x in {1,2,5,6,7}
               \filldraw[fill=black] (v\x) circle (2pt);

          \node at (vtext) {$=$};

          \coordinate (v1) at ([shift={+(1,0.4)}] vtext);
          \coordinate (v2) at ([shift={+(0,-0.8)}] v1);
          \coordinate (v3) at ([shift={+(0.8,0.4)}] v2);
          \coordinate (v4) at ([shift={+(0.8,0.4)}] v3);
          \coordinate (v5) at ([shift={+(0,-0.8)}] v4);
          \coordinate (vtext) at ([shift={+(2.4,0)}] v3);

          \draw[thick] (v1) -- (v3) -- (v2) (v4) -- (v3) -- (v5);

          \foreach \x in {3}
               \filldraw[fill=white] (v\x) circle (2pt);
          \foreach \x in {1,2,4,5}
               \filldraw[fill=black] (v\x) circle (2pt);

          \node at (vtext) {$\times\,2$};

          \coordinate (v1) at (-3.6,-6.2);
          \coordinate (v2) at ([shift={+(0.8,0)}] v1);
          \coordinate (v3) at ([shift={+(1,0)}] v2);
          \coordinate (v4) at ([shift={+(0.8,0)}] v3);
          \coordinate (vtext) at ([shift={+(1,0)}] v4);

          \draw[thick] (v1) -- (v2) (v3) -- (v4);
          \draw[thick,double] (v2) -- (v3);
          \draw[thick] (v3) arc (0:180:0.5);

          \foreach \x in {1,3}
               \filldraw[fill=white] (v\x) circle (2pt);
          \foreach \x in {2,4}
               \filldraw[fill=black] (v\x) circle (2pt);

          \node at (vtext) {$=$};

          \coordinate (v1) at ([shift={+(1,0)}] vtext);
          \coordinate (v2) at ([shift={+(1.5,0)}] v1);
          \coordinate (vtext) at ([shift={+(1.2,0)}] v2);

          \draw[thick] (v1) -- (v2);
          \filldraw[fill=white] (v1) circle (2pt);
          \filldraw[fill=black] (v2) circle (2pt);

          \node at (vtext) {$\times\,3$};

          \coordinate (v1) at (-3,-8);
          \coordinate (v2) at (-2.5,-8.5);
          \coordinate (v3) at (-2.5,-7.5);
          \coordinate (v4) at (-2,-8);
          \coordinate (v5) at (-1,-7.2);
          \coordinate (v6) at (-1,-8.8);
          \coordinate (vtext) at (0,-8);
          \coordinate (v7) at (1,-8);
          \coordinate (v8) at (2.5,-8.7);
          \coordinate (v9) at (2.5,-7.3);
          \coordinate (vneg) at (4,-8);

          \draw[thick] (v1) .. controls (v2) and (v4) .. (v5);
          \draw[thick] (v1) .. controls (v3) and (v4) .. (v6);
          \draw[thick] (v8) -- (v7) -- (v9);

          \filldraw (v1) circle (2pt);
          \filldraw (v7) circle (2pt);

          \node at (vtext) {$=$};
          \draw[thick] (-2,-10) circle (0.8);
          \node at (0,-10) {$=$};
          \node at (1,-10) {$4$};

          \draw[thick,double] (-2,-12.25) circle (0.8);
          \node at (0,-12.25) {$=$};
          \node at (1,-12.25) {$6$};

          \draw[thick] (-2,-13)+(240:2) arc (240:300:2);
          \draw[thick] (-2,-13.5) -- (-2,-15);
          \draw[thick] (-2,-13.5) .. controls  (-2.5,-14.25) .. (-2,-15);
          \draw[thick] (-2,-13.5) .. controls  (-1.5,-14.25) .. (-2,-15);

          \filldraw[fill=white] (-2,-13.5) circle (2pt);
          \filldraw[fill=black] (-2,-15) circle (2pt);

          \node at (0,-14.25) {$=$};

          \draw[thick] (2,-13)+(240:2) arc (240:300:2);
          \draw[thick] (2,-13.5) .. controls  (2.5,-14.25) .. (2,-15);
          \draw[thick] (2,-13.5) .. controls  (1.5,-14.25) .. (2,-15);

          \filldraw[fill=white] (2,-13.5) circle (2pt);
          \filldraw[fill=black] (2,-15) circle (2pt);

          \node at (4,-14.25) {$=$};
          \node at (5,-14.25) {$0$};
     \end{tikzpicture}
     \end{subfigure}
     \caption{The skein relations for $\mathfrak{sl}_3$ and $\mathfrak{sl}_4$ tensor diagrams, adapted from~\cite{kuperberg1996spiders,fomin2016tensor}.  As mentioned in the text, the latter graphical relations are equivalences from skein relations between tensor invariants mod overall sign.  Additionally, the $\mathfrak{sl}_4$ relations in the sixth and eighth lines hold with all colors exchanged. These relations are all derivable from identities for contractions of Levi-Civita tensors; for example the top relation on the left shows $\delta_b^d \delta_c^e = \delta_c^d \delta_b^e + \epsilon_{abc}\epsilon^{ade}$.}
     \label{figskein34}
\end{figure}

Again we emphasize that we have only explained how to compute tensor invariants mod sign when $k$ is even.  In order to formulate the skein relations for $\mathfrak{sl}_4$ tensor diagrams, it would be necessary to be careful about the detailed sign convention explained in~\cite{kim2003graphical,cautis2014webs}.  In Fig.~\ref{figskein34} we show the equivalence relations (mod sign) that we require for the calculations in this paper.

Freed from having to worry about the sign, we can compute invariants for $\mathfrak{sl}_4$ trees in a similar manner to those of $\mathfrak{sl}_3$.  If two vertices are connected by a pair of edges then we call the pair a \textit{double edge}.  After choosing a central vertex $v$ and assigning to each edge a direction pointing towards $v$, every other internal vertex $v' \ne v$ has either a double or single outgoing edge, and the other two or three edges are incoming.  Extending the $k=3$ analysis in the obvious way, we now assign to $v'$ the 1- or 2-index co- or contravariant tensor constructed by contracting $\epsilon^{abcd}$ (if $v'$ is white) or $\epsilon_{abcd}$ (if $v'$ is black) with the tensors associated to the incoming edges (multiplied by $\frac{1}{2}$ if there is a double edge).  Finally, at the central vertex $v$ all edges are incoming; contracting their indices with the appropriate $\epsilon$ computes the diagram's invariant (up to overall sign).

For example, if in the diagram
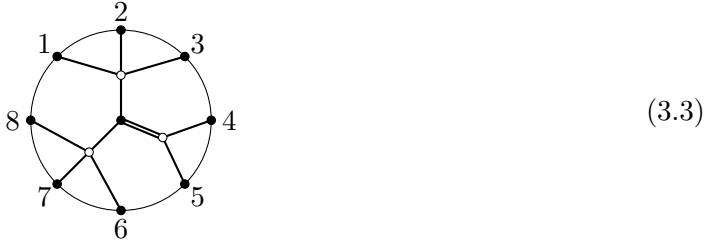
\begin{equation}
     \begin{tikzpicture}[baseline={(current bounding box.center)},scale=0.8,xscale=-1]
          \def \cen{(0,0)};
          \def \rad{1.5};
          \draw {\cen} circle (\rad);
          
          \foreach \x in {1,2,3,4,5,6,7,8}{
               \filldraw[fill=black,draw=black] \cen+(45*\x:\rad) circle (2pt);
               \draw[white] \cen+({45* (\x)}:{\rad+0.3}) circle (0.1) node[black]{\tikz{\node at (0,0) {$\x$}}};
               \coordinate(v\x) at (45*\x:1.5);}

          \coordinate (A123) at (90:0.75);
          \coordinate (B45) at ({180+22.5}:0.75);
          \coordinate (A678) at ({270+45}:0.75);
          \coordinate (C) at (0,0);

          \draw[thick] (v1) -- (A123) -- (v2) (A123) -- (v3);
          \draw[thick] (v7) -- (A678) -- (v8) (A678) -- (v6);
          \draw[thick] (v4) -- (B45) -- (v5);
          \draw[thick] (A678) --(C) -- (A123);
          \draw[thick,double] (B45) -- (C);

          \filldraw[fill=white] (A123) circle (2pt);
          \filldraw[fill=white] (B45) circle (2pt);
          \filldraw[fill=white] (A678) circle (2pt);
          \filldraw (C) circle (2pt);
     \end{tikzpicture}
\end{equation}
we choose the internal black vertex as the center, then the top, bottom right, and bottom left vertices are assigned $(123)$, $(45)$, and $(678)$, respectively, where we use the shorthand
\begin{align}
(ij) = \frac{1}{2} \epsilon_{abcd} Z_i^a Z_j^b\,, \qquad
(ijk) = \epsilon_{abcd} Z_i^a Z_j^b Z_k^c\,.
\label{eq:shorthand}
\end{align}
Contracting indices at the central vertex computes the diagram's invariant, which can be expressed as $\langle 45 (123) \cap (678) \rangle$
using the notation
\begin{align}
\langle ab(cde) \cap (fgh) \rangle = \langle acde \rangle \langle bfgh \rangle - \langle bcde \rangle \langle afgh\rangle\,.
\label{eq:capdef}
\end{align}

\bigskip

\subsection{Non-Arborizable Web Invariants}

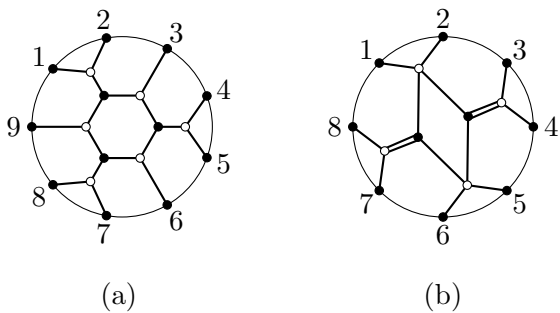
\begin{figure}[!htb]
\centering
\begin{tabular}
{
>{\centering\arraybackslash} m{0.25\textwidth}
>{\centering\arraybackslash} m{0.25\textwidth}
}
{
     \begin{tikzpicture}[scale=0.8,xscale=-1]
          \def \cen{(0,0)};
          \def \rad{1.5};
          \draw {\cen} circle (\rad);
          
          \foreach \x in {1,2,3,4,5,6,7,8,9}{
               \filldraw[fill=black,draw=black] \cen+(40*\x:\rad) circle (2pt);
               \draw[white] \cen+({40* (\x)}:{\rad+0.3}) circle (0.1) node[black]{\tikz{\node at (0,0) {$\x$}}};
               \coordinate(v\x) at (40*\x:1.5);}
          
          \foreach \x in {1,2,3,4,5,6}{
               \coordinate(A\x) at ({60*\x}:{.4*\rad});
               \draw[thick] ({60*\x}:{.4*\rad}) -- ({60+60*\x}:{.4*\rad});}
          
          \coordinate(B1) at (60:{.7*\rad});
          \coordinate(B2) at (180:{.7*\rad});
          \coordinate(B3) at (300:{.7*\rad});
          
          \draw[thick] (B1) -- (A1) (B2) -- (A3) (B3) -- (A5);
          \draw[thick] (A2) -- (v3) (A4) -- (v6) (A6) -- (v9);
          
          \draw[thick] (v1) -- (B1) -- (v2) (v4) -- (B2) -- (v5)  (v7) -- (B3) -- (v8);

          \foreach \x in {1,3,5}{
               \filldraw[fill=black,draw=black] (A\x) circle (2pt);
               \filldraw[fill=white,draw=black] ({60+60*\x}:{.4*\rad}) circle (2pt);}
          \foreach \x in {1,2,3}
               \filldraw[fill=white,draw=black] (B\x) circle (2pt);
     \end{tikzpicture}
}
&
{
     \begin{tikzpicture}[scale=.8,xscale=-1]
          \def \cen{(0,0)};
          \def \rad{1.5};
          \draw {\cen} circle (\rad);
          
          \foreach \x in {1,2,3,4,5,6,7,8}{
               \filldraw[fill=black,draw=black] \cen+(45*\x:\rad) circle (2pt);
               \draw[white] \cen+({45* (\x)}:{\rad+0.3}) circle (0.1) node[black]{\tikz{\node at (0,0) {$\x$}}};
               \coordinate(v\x) at (45*\x:1.5);}
          
          \foreach \x in {1,2,3,4}
               \coordinate(A\x) at ({135/2-90+90*\x}:{\rad*0.7});
          
          \foreach \x in {1,2}{
               \coordinate(B\x) at ({135/2-90+180*\x}:{\rad*0.3});
               \filldraw (B\x) circle (2pt);
          }
          
          \draw[thick] (A1) -- (B1) -- (A3) -- (B2) -- cycle;
          \draw[thick] (v1) -- (A1) -- (v2) (v3) -- (A2) -- (v4) (v5) -- (A3) -- (v6) (v7) -- (A4) -- (v8);
          \foreach \x in {135/2+90,135/2-90}{
               \draw[thick] ({\x+6}:{\rad*0.3}) -- ({\x+2}:{\rad*0.7});
               \draw[thick] ({\x-6}:{\rad*0.3}) -- ({\x-2}:{\rad*0.7});}
          
          \foreach \x in {1,2,3,4}
               \filldraw[fill=white] (A\x) circle (2pt);
          \end{tikzpicture}
}
\\
(a) & (b)
\end{tabular}
	  \caption{Two of the simplest non-arborizable webs, for (a) $\Gr(3,9)$ and (b) $\Gr(4,8)$.}
	  \label{fig:nonarb}
\end{figure}

According to the Fomin-Pylyavskyy conjectures, every cluster monomial (a product of compatible cluster variables) in $\Gr(k,n)$ is an $n$-point $\mathfrak{sl}_k$ web invariant.  However, the converse is not true because of the existence of non-arborizable webs.  Their invariants are multiplicatively independent of cluster variables and so indicate that bases for cluster algebras must (in general) have elements beyond cluster monomials (see Sec.~III.A of~\cite{Arkani-Hamed:2019rds} for an explicit example discussed in the physics literature).  The simplest non-arborizable $\mathfrak{sl}_3$ webs appear at $n=9$ and the simplest $\mathfrak{sl}_4$ webs appear at $n=8$.  These include for example~\cite[Figure 31]{fomin2016tensor} and~\cite[(8.2)]{chang2020quantum}, shown in Fig.~\ref{fig:nonarb}.

\section{From Tensor Diagrams to Kinematic Functions}
\label{sec:Xmapdefinition}

In this section we study a map $X$ that associates a kinematic function $F = X([D])$ to certain tensor invariants $[D]$.  A key property we want the map to have is that if $[D]$ is a cluster variable, then $X([D])$ should be the kinematic function naturally associated to $[D]$ (in the same sense of association as between the first and third columns of Tab.~\ref{tab:one}).

More specifically, and more generally, $X$ is defined as follows: if $[D]$ is a tensor invariant whose ${\bf g}$-vector (defined as reviewed in Sec.~\ref{sec:gvectorappendix}) is ${\bf y} \in \mathbb{Z}^d$, and if ${\bf y}$ is the first integer point along the ray $\mathbb{R}^+ {\bf g}$ (in which case we say that ${\bf y}$ and $[D]$ are \emph{primitive}), then $X([D])$ is the kinematic function $F_{\bf y}$ computed according to~(\ref{eq:fromy}).  These steps trace counterclockwise around Fig.~\ref{fig:outline}, when applied to a diagram that is not necessarily a web.  In the rest of this section we present \emph{conjectural} formulas that compute $X([D])$ for $k \le 4$ directly, as opposed to tracing around the figure.  We have confirmed that our formulas obey the defining property for all kinematic functions and web invariants that we have encountered in this work (summarized in Tab.~\ref{tab:summary}), and conjecture them to be valid in general.  In Sec.~\ref{sec:series} we discuss web series, which extend this discussion to arbitrary integer points along $\mathbb{R}^+ {\bf y}$.

It is sufficient to focus our attention on indecomposable invariants; more generally we have $X([D_1] \cdot [D_2] \cdot \cdots) = X([D_1]) + X([D_2]) + \cdots$.  In order to make our results easily accessible to users of two different sets of conventions we study two distinct versions of the map: $X_L$ and $X_R$.  The former is attached to the conventions reviewed in Secs.~\ref{sec:webvariables}--\ref{sec:polytopes} (and used in the example of Sec.~\ref{sec:warmup}), while the latter is attached to the ``Langlands dual'' convention where all arrows in Fig.~\ref{figinigkn} are reversed (see Sec.~\ref{sec:langlands}).

For $k=2$ it is possible to write down an explicit general formula for the $X$-map, exploiting the fact that every indecomposable $\mathfrak{sl}_2$ tensor invariant has the form $\langle i\,j\rangle$ for $i < j$ (see Sec.~\ref{sec:sl2diagrams}).  The associated kinematic function, for the two choices of convention, is simply
\begin{align}
\begin{split}
X_L(\langle a\,b\rangle) &= \sum_{a+1 \le i < j \le b} s_{ij}\,,\\
X_R(\langle a\,b\rangle) &= \sum_{a \le i < j \le b-1} s_{ij}\,.
\end{split}
\end{align}

\subsection{\texorpdfstring{$\mathfrak{sl}_3$ Arborizable Invariants}{sl3 Arborizable Invariants}}
\label{sec:threetreesandwebs}

For $k>2$ we conjecture a recursive formula for the $X$-map, first for tree diagrams.  In order to apply the recursion, $[D]$ must be written in the form obtained by reading it off from some diagram $D$ as described in Sec.~\ref{sec:sl3diagrams}.  If we are handed $[D]$ in some random form (as a polynomial in Pl\"ucker coordinates), we would first need to draw some tree diagram $D$ whose invariant is $[D]$.  Next, $D$ must be put into \emph{canonical order}, which means that all lines and vertices are placed so as to minimize the number of crossed lines but \emph{without} producing or annihilating any lines or vertices.  Specifically, all crossing structures of the type shown in the fourth and fifth lines of the $\mathfrak{sl}_3$ skein relations (Fig.~\ref{figskein34}, left panel) should be cleared, but without using the moves shown on the first and second lines.

The recursion is seeded by the simplest possible $\mathfrak{sl}_3$ tensor diagrams: those with only a single internal vertex (see for example the left panel of Fig.~\ref{figwebs}), for which we have
\begin{align}
\begin{split}
\label{sl3seed}
X_L(\langle a\,b\,c\rangle) = S_{a+1,\ldots,b(b+1,\ldots,c)}\,, \\
X_R(\langle a\,b\,c\rangle) = S_{(a,\ldots,b-1)b,\ldots,c-1}\,,
\end{split}
\end{align}
using notation reviewed in Sec.~\ref{sec:notation}.

For more general diagrams, our definition of the $X$-map is motivated by the $\Gr(3,6)$ ``bipyramid relation'' of~\cite{speyer2005tropical} (see for example (6.10) of~\cite{Drummond:2019qjk} for more specificity) which in our notation reads
\begin{align}
S_{(12)34} + S_{12(34)} = S_{1234} + S_{3456} + S_{1256}\,.
\end{align}
Working for a moment with the ``left'' conventions, we can use~(\ref{sl3seed}) to rewrite all but the first term as $X_L$ images of cluster variables:
\begin{align}
S_{(12)34} + X_L(\langle 6\,2\,4\rangle) = X_L(\langle 6\,3\,4\rangle) + X_L(\langle 2\,5\,6\rangle) + X_L(\langle 4\,1\,2\rangle)\,.
\end{align}
We don't yet know what to do with the first term, but one can calculate that the ${\bf g}$-vector of $S_{(12)34}$ is the same as that of the cluster variable $\langle 5\times 6,1 \times 2,3 \times 4\rangle$, which motivates us to \emph{define}
\begin{align}
X_L(\langle 5\times 6,1 \times 2,3 \times 4\rangle) := X_L(\langle 6\,3\,4\rangle) + X_L(\langle 2\,5\,6\rangle) + X_L(\langle 4\,1\,2\rangle) - X_L(\langle 6\,2\,4\rangle)\,.
\label{eq:toextend}
\end{align}
Our recursive definition of $X_L$ (and similar for $X_R$) is motivated by the desire to extend~(\ref{eq:toextend}) to more general cases without having to rely on computing ${\bf g}$-vectors at intermediate stages.

Therefore we now consider more complicated tensor diagrams whose invariants involve cross-products of the form $\langle{a\times b,\,c\times d,\, e\times f\,}\rangle$.  Here $a,b,\ldots,f$ are either all vectors or all covectors (the expression makes no sense otherwise), and each could itself be a string of nested cross-products such as $a = (1 \times 2) \times (3 \times 4)$.  The generalization of~(\ref{eq:toextend}) is
\begin{align}
\begin{split}
\label{eq:sl3vector}
X_L(\langle{a\times b,c\times d,e\times f}\rangle) &= X_L(\langle{a,b,d}\rangle) + X_L(\langle{c,d,f}\rangle) + X_L(\langle{e,f,b}\rangle) - X_L(\langle{b,d,f}\rangle)\,,\\
X_R(\langle{a\times b,c\times d,e\times f}\rangle) &= X_R(\langle{a,c,d}\rangle) + X_R(\langle{c,e,f}\rangle) + X_R(\langle{e,a,b}\rangle) - X_R(\langle{a,c,e}\rangle)
\end{split}
\end{align}
for vectors and
\begin{align}
\begin{split}
\label{eq:sl3covector}
X_L(\langle{a\times b,c\times d,e\times f}\rangle) &= X_L(\langle{a,c,d}\rangle) + X_L(\langle{c,e,f}\rangle) + X_L(\langle{e,a,b}\rangle) - X_L(\langle{a,c,e}\rangle)\,,\\
X_R(\langle{a\times b,\,c\times d,\, e\times f\,}\rangle) &= X_R(\langle{a,b,d}\rangle) + X_R(\langle{c,d,f}\rangle) + X_R(\langle{e,f,b}\rangle) - X_R(\langle{b,d,f}\rangle)
\end{split}
\end{align}
for covectors.

The formulas~(\ref{sl3seed}), (\ref{eq:sl3vector}) and~(\ref{eq:sl3covector}) provide a recursive formula for $X_L$ and $X_R$ for all $\mathfrak{sl}_3$ tree diagrams, but it may not immediately be clear how to apply the recursion to a diagram like the one in the right panel of Fig.~\ref{figwebs} because its invariant has no manifest cross-product in the middle entry.  However, it is always possible to rewrite an invariant in an equivalent way that exposes a cross-product in each entry; in this case via the identity
\begin{align}
\langle (1 \times 2)\times (3\times 4), 5, (6 \times 7) \times (8\times 9)\rangle = \langle 1 \times 2, 3 \times 4, 5 \times ( (6 \times 7) \times (8 \times 9)) \rangle\,,
\end{align}
to which one can now apply~(\ref{eq:sl3covector}).  In this example one has to perform further rearrangements at the next step in the recursion; the important point is that it is always possible to do so.

As emphasized above, although we have written the above recursion in a seemingly general way in terms of nested brackets, it will give inconsistent results unless $[D]$ is expressed as the invariant read off from a canonically ordered diagram.  One obvious manifestation of the inconsistency for improper ordering is the fact that we clearly have
\begin{align}
\langle 1 \times 2, 4 \times 3, 6 \times 5\rangle = \langle 1 \times 2, 3 \times 4, 5 \times 6 \rangle\,,
\label{eq:co}
\end{align}
but~(\ref{eq:sl3vector}) and~(\ref{eq:sl3covector}) are not invariant under $c \leftrightarrow d, e \leftrightarrow f$; the recursion must be applied only to the right-hand side of~(\ref{eq:co}).

\subsection{\texorpdfstring{$\mathfrak{sl}_3$ Non-Arborizable Invariants}{sl3 Non-Arborizable Invariants}}
\label{sec:unroll3}

Now we propose recursion relations for computing $X([D])$ when $[D]$ is a non-arborizable invariant.  The basic step is to ``unroll'' each internal loop by appropriately cutting one of its edges.  Consider first the case when $[D]$ is the invariant of a diagram having a single internal loop.  Then, for any choice of edge on the internal loop (shown in red) we define
\begin{align}
\begin{split}
     X_L\left(
     \raisebox{-.7cm}{
     \begin{tikzpicture}[scale=0.6]
          \def \cen{(0,0)};
          \def \rad{1.5};
          \foreach \x in {1,2}{
               \filldraw[fill=black,draw=black] \cen+(40*\x:\rad) circle (2pt);
               \coordinate(v\x) at (40*\x:1.5);}
          \foreach \x in {3,9}{
               \filldraw[fill=black,draw=black] \cen+(40*\x:{0.8*\rad}) circle (2pt);
               \coordinate(v\x) at (40*\x:{0.8*\rad});}
          \draw[white] \cen+({40* (1)}:{\rad+0.45}) circle (0.1) node[black]{\tikz{\node at (0,0) {$c$}}};
          \draw[white] \cen+({40* (2)}:{\rad+0.45}) circle (0.1) node[black]{\tikz{\node at (0,0) {$b$}}};
          \draw[white] \cen+({40* (3)}:{.8*\rad+0.45}) circle (0.1) node[black]{\tikz{\node at (0,0) {$a$}}};
          \foreach \x in {1,2,3,4,5,6}{
               \coordinate(A\x) at ({60*\x}:{.4*\rad});}
          \draw[thick] (A2) -- (A3) (A6) -- (A1);
          \draw[very thick,red] (A1) -- (A2);
          \coordinate(B1) at (60:{.7*\rad});
          \coordinate(B2) at (180:{.8*\rad});
          \coordinate(B3) at (300:{.8*\rad});
          \draw[thick] (B1) -- (A1) (B2) -- (A3);
          \draw[thick] (A2) -- (v3) (A6) -- (v9);
          \draw[thick] (v1) -- (B1) -- (v2) ;
          \draw[very thick,dashed,blue] (-{0.4*\rad},0) arc (180:360:{0.4*\rad});
          \foreach \x in {1,3} {
               \filldraw[fill=black,draw=black] (A\x) circle (2pt);
               \filldraw[fill=white,draw=black] ({-60+60*\x}:{.4*\rad}) circle (2pt);}
          \foreach \x in {1,2}
               \filldraw[fill=white,draw=black] (B\x) circle (2pt);
     \end{tikzpicture}}\right) = X_L\left(
     \raisebox{-.7cm}{
     \begin{tikzpicture}[scale=0.6]
          \def \cen{(0,0)};
          \def \rad{1.5};
          \coordinate (v11) at (21:{\rad*1.2});
          \coordinate (v12) at (45:{1.2*\rad});
          \coordinate (v21) at (60:{1.1*\rad});
          \coordinate (v22) at (85:{0.9*\rad});
          \coordinate (v31) at ({-.5*\rad+0.3},{.3*\rad+0.6});
          \coordinate (v32) at ({-.5*\rad-0.3},{.3*\rad+0.6});
          \foreach \x in {9}{
               \filldraw[fill=black,draw=black] \cen+(40*\x:{0.8*\rad}) circle (2pt);
               \coordinate(v\x) at (40*\x:{0.8*\rad});}
          \draw[white] \cen+(21:{\rad*1.2+0.4}) circle (0.1) node[black]{\tikz{\node at (0,0) {$c$}}};
          \draw[white] \cen+(45:{1.2*\rad+0.4}) circle (0.1) node[black]{\tikz{\node at (0,0) {$b$}}};
          \draw[white] \cen+(60:{1.1*\rad+0.4}) circle (0.1) node[black]{\tikz{\node at (0,0) {$a$}}};
          \draw[white] \cen+(85:{0.9*\rad+0.4}) circle (0.1) node[black]{\tikz{\node at (0,0) {$m$}}};
          \draw[white] ({-.5*\rad+0.3},{.3*\rad+0.6})+(60:{0.25}) circle (0.1) node[black]{\tikz{\node at (0,0) {$c$}}};
          \draw[white] ({-.5*\rad-0.3},{.3*\rad+0.6})+(120:{0.25}) circle (0.1) node[black]{\tikz{\node at (0,0) {$a$}}};
          \coordinate(A1) at ({.4*\rad},{.3*\rad});
          \coordinate(A2) at ({-.5*\rad},{.4*\rad});
          \foreach \x in {3,4,5,6}{
               \coordinate(A\x) at ({60*\x}:{.4*\rad});}
          \draw[thick] (A2) -- (A3) (A6) -- (A1);
          \coordinate(B1) at (35:{.8*\rad});
          \coordinate(B15) at ({.3*\rad},{.6*\rad});
          \coordinate(B2) at (180:{.8*\rad});
          \coordinate(B3) at (300:{.8*\rad});
          \draw[thick] (B1) -- (A1) (B2) -- (A3);
          \draw[thick] (A6) -- (v9);
          \draw[thick] (v11) -- (B1) -- (v12) (A2) -- (v32);
          \draw[thick,red] (v31) -- (A2) (v21) -- (B15) -- (v22) (B15) -- (A1);
          \draw[very thick,dashed,blue] (-{0.4*\rad},0) arc (180:360:{0.4*\rad});
          \foreach \x in {11,12,32}
               \filldraw (v\x) circle (2pt);
          \foreach \x in {21,22,31}
               \filldraw[draw=red,fill=red] (v\x) circle (2pt);
          \filldraw[fill=black,draw=black] (A1) circle (2pt);
          \filldraw[fill=white,draw=black] (A2) circle (2pt);
          \foreach \x in {3}{
               \filldraw[fill=black,draw=black] (A\x) circle (2pt);
               \filldraw[fill=white,draw=black] ({-180+60*\x}:{.4*\rad}) circle (2pt);}
          \foreach \x in {1,2}
               \filldraw[fill=white,draw=black] (B\x) circle (2pt);
          \filldraw[fill=white,draw=red] (B15) circle (2pt);
     \end{tikzpicture}}\right) - X_L\big(\langle{m,a,c}\rangle)\,,
\\
     X_R\left(
     \raisebox{-.7cm}{
     \begin{tikzpicture}[scale=0.6,xscale=-1]
          \def \cen{(0,0)};
          \def \rad{1.5};
          \foreach \x in {1,2}{
               \filldraw[fill=black,draw=black] \cen+(40*\x:\rad) circle (2pt);
               \coordinate(v\x) at (40*\x:1.5);}
          \foreach \x in {3,9}{
               \filldraw[fill=black,draw=black] \cen+(40*\x:{0.8*\rad}) circle (2pt);
               \coordinate(v\x) at (40*\x:{0.8*\rad});}
          \draw[white] \cen+({40* (1)}:{\rad+0.45}) circle (0.1) node[black]{\tikz{\node at (0,0) {$a$}}};
          \draw[white] \cen+({40* (2)}:{\rad+0.45}) circle (0.1) node[black]{\tikz{\node at (0,0) {$b$}}};
          \draw[white] \cen+({40* (3)}:{.8*\rad+0.45}) circle (0.1) node[black]{\tikz{\node at (0,0) {$c$}}};
          \foreach \x in {1,2,3,4,5,6}{
               \coordinate(A\x) at ({60*\x}:{.4*\rad});}
          \draw[thick] (A2) -- (A3) (A6) -- (A1);
          \draw[very thick,red] (A1) -- (A2);
          \coordinate(B1) at (60:{.7*\rad});
          \coordinate(B2) at (180:{.8*\rad});
          \coordinate(B3) at (300:{.8*\rad});
          \draw[thick] (B1) -- (A1) (B2) -- (A3);
          \draw[thick] (A2) -- (v3) (A6) -- (v9);
          \draw[thick] (v1) -- (B1) -- (v2) ;
          \draw[very thick,dashed,blue] (-{0.4*\rad},0) arc (180:360:{0.4*\rad});
          \foreach \x in {1,3} {
               \filldraw[fill=black,draw=black] (A\x) circle (2pt);
               \filldraw[fill=white,draw=black] ({-60+60*\x}:{.4*\rad}) circle (2pt);}
          \foreach \x in {1,2}
               \filldraw[fill=white,draw=black] (B\x) circle (2pt);
     \end{tikzpicture}}\right) = X_R\left(
     \raisebox{-.7cm}{
     \begin{tikzpicture}[scale=0.6,xscale=-1]
          \def \cen{(0,0)};
          \def \rad{1.5};
          \coordinate (v11) at (21:{\rad*1.2});
          \coordinate (v12) at (45:{1.2*\rad});
          \coordinate (v21) at (60:{1.1*\rad});
          \coordinate (v22) at (85:{0.9*\rad});
          \coordinate (v31) at ({-.5*\rad+0.3},{.3*\rad+0.6});
          \coordinate (v32) at ({-.5*\rad-0.3},{.3*\rad+0.6});
          \foreach \x in {9}{
               \filldraw[fill=black,draw=black] \cen+(40*\x:{0.8*\rad}) circle (2pt);
               \coordinate(v\x) at (40*\x:{0.8*\rad});}
          \draw[white] \cen+(21:{\rad*1.2+0.4}) circle (0.1) node[black]{\tikz{\node at (0,0) {$a$}}};
          \draw[white] \cen+(45:{1.2*\rad+0.4}) circle (0.1) node[black]{\tikz{\node at (0,0) {$b$}}};
          \draw[white] \cen+(60:{1.1*\rad+0.4}) circle (0.1) node[black]{\tikz{\node at (0,0) {$c$}}};
          \draw[white] \cen+(85:{0.9*\rad+0.4}) circle (0.1) node[black]{\tikz{\node at (0,0) {$m$}}};
          \draw[white] ({-.5*\rad+0.3},{.3*\rad+0.6})+(60:{0.25}) circle (0.1) node[black]{\tikz{\node at (0,0) {$a$}}};
          \draw[white] ({-.5*\rad-0.3},{.3*\rad+0.6})+(120:{0.25}) circle (0.1) node[black]{\tikz{\node at (0,0) {$c$}}};
          \coordinate(A1) at ({.4*\rad},{.3*\rad});
          \coordinate(A2) at ({-.5*\rad},{.4*\rad});
          \foreach \x in {3,4,5,6}{
               \coordinate(A\x) at ({60*\x}:{.4*\rad});}
          \draw[thick] (A2) -- (A3) (A6) -- (A1);
          \coordinate(B1) at (35:{.8*\rad});
          \coordinate(B15) at ({.3*\rad},{.6*\rad});
          \coordinate(B2) at (180:{.8*\rad});
          \coordinate(B3) at (300:{.8*\rad});
          \draw[thick] (B1) -- (A1) (B2) -- (A3);
          \draw[thick] (A6) -- (v9);
          \draw[thick] (v11) -- (B1) -- (v12) (A2) -- (v32);
          \draw[thick,red] (v31) -- (A2) (v21) -- (B15) -- (v22) (B15) -- (A1);
          \draw[very thick,dashed,blue] (-{0.4*\rad},0) arc (180:360:{0.4*\rad});
          \foreach \x in {11,12,32}
               \filldraw (v\x) circle (2pt);
          \foreach \x in {21,22,31}
               \filldraw[draw=red,fill=red] (v\x) circle (2pt);
          \filldraw[fill=black,draw=black] (A1) circle (2pt);
          \filldraw[fill=white,draw=black] (A2) circle (2pt);
          \foreach \x in {3}{
               \filldraw[fill=black,draw=black] (A\x) circle (2pt);
               \filldraw[fill=white,draw=black] ({-180+60*\x}:{.4*\rad}) circle (2pt);}
          \foreach \x in {1,2}
               \filldraw[fill=white,draw=black] (B\x) circle (2pt);
          \filldraw[fill=white,draw=red] (B15) circle (2pt);
     \end{tikzpicture}}\right) - X_R\big(\langle{a,c,m}\rangle)\,,
\end{split}
\label{eq:sl3unroll}
\end{align}
where we have removed the red edge of the loop and replaced it with the new red edges shown, the blue dashed lines denote an arbitrary (even) number of additional vertices on the inner loop, and $m$ is an arbitrary reference point.  One can check that all terms involving the reference point $m$ cancel out when the recursion is applied and the right-hand sides are expanded out.  We emphasize again that~(\ref{eq:sl3unroll}) is only valid when the diagram is drawn in canonical order, with minimal number of crossings.

To see an example of the recursion at work, let us compute the kinematic functions associated to Fig.~\ref{fig:nonarb}(a).  Applying~(\ref{eq:sl3unroll}) and then~(\ref{eq:sl3vector}), (\ref{eq:sl3covector}) we have
\begin{equation}
     \begin{aligned}
          &X_L\left(\!\!\!
          \raisebox{-1.35cm}{
          \begin{tikzpicture}[scale=0.6,xscale=-1]
               \def \cen{(0,0)};
               \def \rad{1.5};
               \draw {\cen} circle (\rad); 

               \foreach \x in {1,2,3,4,5,6,7,8,9}{
                    \filldraw[fill=black,draw=black] \cen+(40*\x:\rad) circle (2pt);
                    \draw[white] \cen+({40* (\x)}:{\rad+0.35}) circle (0.1) node[black]{\tikz{\node at (0,0) {$\x$}}};
                    \coordinate(v\x) at (40*\x:1.5);}

               \foreach \x in {1,2,3,4,5,6}{
                    \coordinate(A\x) at ({60*\x}:{.4*\rad});}

               \draw[thick] (A2) -- (A1) -- (A6) -- (A5) -- (A4) -- (A3);
               \draw[thick,red] (A3) -- (A2);

               \coordinate(B1) at (60:{.7*\rad});
               \coordinate(B2) at (180:{.7*\rad});
               \coordinate(B3) at (300:{.7*\rad});

               \draw[thick] (B1) -- (A1) (B2) -- (A3) (B3) -- (A5);
               \draw[thick] (A2) -- (v3) (A4) -- (v6) (A6) -- (v9);

               \draw[thick] (v1) -- (B1) -- (v2) (v4) -- (B2) -- (v5)  (v7) -- (B3) -- (v8);

               \foreach \x in {1,3,5}{
                    \filldraw[fill=black,draw=black] (A\x) circle (2pt);
                    \filldraw[fill=white,draw=black] ({60+60*\x}:{.4*\rad}) circle (2pt);}
               \foreach \x in {1,2,3}
                    \filldraw[fill=white,draw=black] (B\x) circle (2pt);
          \end{tikzpicture}
          }\!\!\!\right) \\&= X_L\Big(\Big\langle{\big((m\times 3)\times(4\times5)\big)\times6,\, 7\times8, \, 9\times\big((1\times2)\times(3\times5)\big)}\Big\rangle\Big) - X_L(\langle{m,3,5}\rangle)\\
          &= X_L(\langle{m\times3,\,4\times5,\, 6\times8}\rangle) + X_L(\langle{7\times8,\,1\times2,\,3\times5}\rangle) + X_L(\langle{6\times9,\,1\times2,\,3\times5}\rangle) \\
          & \qquad - X_L(\langle{6\times8,\,1\times2,\,3\times5}\rangle) - X_L(\langle{m,3,5}\rangle) \\
          &= X_L(\langle{1,2,5}\rangle) + X_L(\langle{4,5,8}\rangle) + X_L(\langle{7,8,2}\rangle) + X_L(\langle{2,6,9}\rangle) + X_L(\langle{3,5,9}\rangle) \\
          & \qquad + X_L(\langle{3,6,8}\rangle)  - X_L(\langle{2,5,9}\rangle) - X_L(\langle{2,6,8}\rangle) - X_L(\langle{3,5,8}\rangle)\,,
     \end{aligned}
\end{equation}
and similarly
\begin{equation}
     \begin{aligned}
          &X_R\left(\!\!\!
          \raisebox{-1.35cm}{
          \begin{tikzpicture}[scale=0.6,xscale=-1]
               \def \cen{(0,0)};
               \def \rad{1.5};
               \draw {\cen} circle (\rad); 

               \foreach \x in {1,2,3,4,5,6,7,8,9}{
                    \filldraw[fill=black,draw=black] \cen+(40*\x:\rad) circle (2pt);
                    \draw[white] \cen+({40* (\x)}:{\rad+0.35}) circle (0.1) node[black]{\tikz{\node at (0,0) {$\x$}}};
                    \coordinate(v\x) at (40*\x:1.5);}

               \foreach \x in {1,2,3,4,5,6}{
                    \coordinate(A\x) at ({60*\x}:{.4*\rad});}

               \draw[thick] (A1) -- (A6) -- (A5) -- (A4) -- (A3) -- (A2);
               \draw[thick,red] (A1) -- (A2);

               \coordinate(B1) at (60:{.7*\rad});
               \coordinate(B2) at (180:{.7*\rad});
               \coordinate(B3) at (300:{.7*\rad});

               \draw[thick] (B1) -- (A1) (B2) -- (A3) (B3) -- (A5);
               \draw[thick] (A2) -- (v3) (A4) -- (v6) (A6) -- (v9);

               \draw[thick] (v1) -- (B1) -- (v2) (v4) -- (B2) -- (v5)  (v7) -- (B3) -- (v8);

               \foreach \x in {1,3,5}{
                    \filldraw[fill=black,draw=black] (A\x) circle (2pt);
                    \filldraw[fill=white,draw=black] ({60+60*\x}:{.4*\rad}) circle (2pt);}
               \foreach \x in {1,2,3}
                    \filldraw[fill=white,draw=black] (B\x) circle (2pt);
          \end{tikzpicture}
          }\!\!\!\right)\\
	  &=   X_R\Big(\Big\langle{\big((1\times 3)\times(4\times5)\big)\times6,\, 7\times8, \, 9\times\big((1\times2)\times(3\times m)\big)}\Big\rangle\Big) - X_R(\langle{1,3,m}\rangle)\\
          &=  X_R(\langle{1\times3,\,4\times5,\, 7\times8}\rangle) + X_R(\langle{7\times9,\,1\times2,\,3\times   m}\rangle) + X_R(\langle{1\times3,\,4\times5,\,6\times 9}\rangle) \\
          & \qquad - X_R(\langle{1\times3,\,4\times5,\,7\times 9}\rangle) - X_L(\langle{1,3,m}\rangle) \\
          &= X_R(\langle{7,1,2}\rangle) + X_R(\langle{1,4,5}\rangle) + X_R(\langle{4,7,8}\rangle) + X_R(\langle{4,6,9}\rangle) + X_R(\langle{3,7,9}\rangle) \\
          & \qquad  + X_R(\langle{1,3,6}\rangle) - X_R(\langle{4,7,9}\rangle) - X_R(\langle{1,3,7}\rangle) - X_R(\langle{1,4,6}\rangle)\,.
     \end{aligned}
\end{equation}

A diagram with $\ell > 1$ inner loops can be treated similarly, by recursively unrolling each loop via introduction of a new reference point.  All $\ell$ reference points will disappear in the final result. 

Of course, some diagrams with internal loops are equivalent to trees by the skein relations and it is important to check that the recursion relations we have given respect this equivalence.  To see this we must look at the two types of arborization processes.  First we consider
\begin{align}\label{eq:arb1}
     \begin{tikzpicture}[scale=0.8,baseline={(current bounding box.center)}]
          \def \cen{(0,0)};
          \def \rad{1.5};
          \foreach \x in {1,2}{
               \filldraw[fill=black,draw=black] \cen+(50*\x:\rad) circle (2pt);
               \coordinate(v\x) at (50*\x:1.5);}
          \foreach \x in {9}{
               \filldraw[fill=black,draw=black] \cen+(40*\x:{0.8*\rad}) circle (2pt);
               \coordinate(v\x) at (40*\x:{0.8*\rad});}
          \draw[white] \cen+({50* (1)}:{\rad+0.45}) circle (0.1) node[black]{\tikz{\node at (0,0) {$b$}}};
          \draw[white] \cen+({50* (2)}:{\rad+0.45}) circle (0.1) node[black]{\tikz{\node at (0,0) {$a$}}};
          \draw[white] \cen+({50* (0)}:{0.8*\rad+0.45}) circle (0.1) node[black]{\tikz{\node at (0,0) {$c$}}};
          \foreach \x in {1,2,3,4,5,6}{
               \coordinate(A\x) at ({60*\x}:{.4*\rad});}
          \draw[thick] (A2) -- (A3) (A6) -- (A1);
          \draw[very thick,blue] (A1) -- (A2) -- (A3) (A6) -- (A1);
          \coordinate(B1) at (75:{.7*\rad});
          \coordinate(B2) at (180:{.8*\rad});
          \coordinate(B3) at (300:{.8*\rad});
          \draw[thick] (B2) -- (A3) (A6) -- (v9);
          \draw[very thick,blue] (A2) -- (v2) (B1) -- (A1);
          \draw[thick] (v1) -- (B1) -- (v2) ;
          \draw[very thick,dashed,blue] (-{0.4*\rad},0) arc (180:360:{0.4*\rad});
          \foreach \x in {1,3} {
               \filldraw[fill=black,draw=black] (A\x) circle (2pt);
               \filldraw[fill=white,draw=black] ({-60+60*\x}:{.4*\rad}) circle (2pt);}
          \foreach \x in {1,2}
               \filldraw[fill=white,draw=black] (B\x) circle (2pt);
          \node at (1.7*\rad,0.7) {\quad$\longrightarrow$};
     \end{tikzpicture}\quad
     \begin{tikzpicture}[scale=0.8,baseline={(current bounding box.center)}]
          \def \cen{(0,0)};
          \def \rad{1.5};
          \foreach \x in {1,2}{
               \filldraw[fill=black,draw=black] \cen+(50*\x:\rad) circle (2pt);
               \coordinate(v\x) at (50*\x:1.5);}
          \foreach \x in {9}{
               \filldraw[fill=black,draw=black] \cen+(40*\x:{0.8*\rad}) circle (2pt);
               \coordinate(v\x) at (40*\x:{0.8*\rad});}
          \draw[white] \cen+({50* (1)}:{\rad+0.45}) circle (0.1) node[black]{\tikz{\node at (0,0) {$b$}}};
          \draw[white] \cen+({50* (2)}:{\rad+0.45}) circle (0.1) node[black]{\tikz{\node at (0,0) {$a$}}};
          \draw[white] \cen+({50* (0)}:{0.8*\rad+0.45}) circle (0.1) node[black]{\tikz{\node at (0,0) {$c$}}};
          \foreach \x in {1,2,3,4,5,6}{
               \coordinate(A\x) at ({60*\x}:{.4*\rad});}
          \coordinate(B1) at (75:{.7*\rad});
          \coordinate(B2) at (180:{.8*\rad});
          \coordinate(B3) at (300:{.8*\rad});
          \draw[thick] (B2) -- (A3) (A6) -- (v9);
          \draw[very thick,blue] (A3) -- (B1) (A6) -- (v2);
          \draw[thick] (v1) -- (B1) -- (v2) ;
          \draw[very thick,dashed,blue] (-{0.4*\rad},0) arc (180:360:{0.4*\rad});
          \foreach \x in {3} {
               \filldraw[fill=black,draw=black] (A\x) circle (2pt);
               \filldraw[fill=white,draw=black] ({-180+60*\x}:{.4*\rad}) circle (2pt);}
          \foreach \x in {1,2}
               \filldraw[fill=white,draw=black] (B\x) circle (2pt);
     \end{tikzpicture}
     \end{align}
Applying $X_L$ to the left-hand side and using~(\ref{eq:sl3unroll}) gives
\begin{equation}
\label{eq:XLunroll}
     X_L\left(\!\!\!
     \raisebox{-0.85cm}{
     \begin{tikzpicture}[scale=0.8]
          \def \cen{(0,0)};
          \def \rad{1.5};

          \foreach \x in {1,2}{
               \filldraw[fill=black,draw=black] \cen+(50*\x:\rad) circle (2pt);
               \coordinate(v\x) at (50*\x:1.5);}

          \foreach \x in {9}{
               \filldraw[fill=black,draw=black] \cen+(40*\x:{0.8*\rad}) circle (2pt);
               \coordinate(v\x) at (40*\x:{0.8*\rad});}

          \draw[white] \cen+({50* (1)}:{\rad+0.3}) circle (0.1) node[black]{\tikz{\node at (0,0) {$b$}}};
          \draw[white] \cen+({50* (2)}:{\rad+0.3}) circle (0.1) node[black]{\tikz{\node at (0,0) {$a$}}};
          \draw[white] \cen+({50* (0)}:{0.8*\rad+0.25}) circle (0.1) node[black]{\tikz{\node at (0,0) {$c$}}};

          \foreach \x in {1,2,3,4,5,6}{
               \coordinate(A\x) at ({60*\x}:{.4*\rad});}
          \draw[thick] (A2) -- (A3) (A6) -- (A1);

          \draw[very thick,blue] (A2) -- (A3) (A6) -- (A1);
          \draw[very thick,red] (A1) -- (A2);

          \coordinate(B1) at (75:{.7*\rad});
          \coordinate(B2) at (180:{.8*\rad});
          \coordinate(B3) at (300:{.8*\rad});

          \draw[thick] (B2) -- (A3) (A6) -- (v9);
          \draw[very thick,blue] (A2) -- (v2) (B1) -- (A1);

          \draw[thick] (v1) -- (B1) -- (v2) ;

          \draw[very thick,dashed,blue] (-{0.4*\rad},0) arc (180:360:{0.4*\rad});

          \foreach \x in {1,3} {
               \filldraw[fill=black,draw=black] (A\x) circle (2pt);
               \filldraw[fill=white,draw=black] ({-60+60*\x}:{.4*\rad}) circle (2pt);}
          \foreach \x in {1,2}
               \filldraw[fill=white,draw=black] (B\x) circle (2pt);
     \end{tikzpicture}}
     \right) =
     X_L\left(\!\!\!
     \raisebox{-0.85cm}{
     \begin{tikzpicture}[scale=0.8]
          \def \cen{(0,0)};
          \def \rad{1.5};

          \coordinate (v11) at (21:{\rad*1.2});
          \coordinate (v12) at (50:{1.2*\rad});
          \coordinate (v22) at (80:{1*\rad});
          \coordinate (v31) at ({-.5*\rad+0.3},{.3*\rad+0.6});
          \coordinate (v32) at ({-.5*\rad-0.3},{.3*\rad+0.6});

          \foreach \x in {9}{
               \filldraw[fill=black,draw=black] \cen+(40*\x:{0.8*\rad}) circle (2pt);
               \coordinate(v\x) at (40*\x:{0.8*\rad});}

          \draw[white] \cen+(21:{\rad*1.2+0.3}) circle (0.1) node[black]{\tikz{\node at (0,0) {$b$}}};
          \draw[white] \cen+(45:{1.2*\rad+0.3}) circle (0.1) node[black]{\tikz{\node at (0,0) {$a$}}};
          \draw[white] \cen+(0:\rad) circle (0.1) node[black]{\tikz{\node at ({50* (0)}:{0.8*\rad+0.25}) {$c$}}};
          \draw[white] \cen+(85:{0.9*\rad+0.4}) circle (0.1) node[black]{\tikz{\node at (0,0) {$m$}}};
          \draw[white] ({-.5*\rad+0.3},{.3*\rad+0.6})+(60:{0.25}) circle (0.1) node[black]{\tikz{\node at (0,0) {$b$}}};
          \draw[white] ({-.5*\rad-0.3},{.3*\rad+0.6})+(120:{0.25}) circle (0.1) node[black]{\tikz{\node at (0,0) {$a$}}};

          \coordinate(A1) at ({.4*\rad},{.3*\rad});
          \coordinate(A2) at ({-.5*\rad},{.4*\rad});

          \foreach \x in {3,4,5,6}{
               \coordinate(A\x) at ({60*\x}:{.4*\rad});}
          \draw[very thick,blue] (A2) -- (A3) (A6) -- (A1);
          
          \coordinate(B1) at (35:{.8*\rad});
          \coordinate(B15) at ({.35*\rad},{.65*\rad});
          \coordinate(B2) at (180:{.8*\rad});
          \coordinate(B3) at (300:{.8*\rad});

          \draw[very thick,blue] (B1) -- (A1);
          \draw[thick] (A6) -- (v9) (B2) -- (A3);

          \draw[thick] (v11) -- (B1) -- (v12);
          \draw[very thick,blue] (A2) -- (v32);

          \draw[very thick,red] (v31) -- (A2) (v12) -- (B15) -- (v22) (B15) -- (A1);

          \draw[very thick,dashed,blue] (-{0.4*\rad},0) arc (180:360:{0.4*\rad});

          \foreach \x in {11,32}
               \filldraw (v\x) circle (2pt);
          \foreach \x in {12,22,31}
               \filldraw[draw=red,fill=red] (v\x) circle (2pt);
          \filldraw[fill=black,draw=black] (A1) circle (2pt);
          \filldraw[fill=white,draw=black] (A2) circle (2pt);
          \foreach \x in {3}{
               \filldraw[fill=black,draw=black] (A\x) circle (2pt);
               \filldraw[fill=white,draw=black] ({-180+60*\x}:{.4*\rad}) circle (2pt);}
          \foreach \x in {1,2}
               \filldraw[fill=white,draw=black] (B\x) circle (2pt);
          \filldraw[fill=white,draw=red] (B15) circle (2pt);
     \end{tikzpicture}}\right) - X_L\big(\langle{m,a,b}\rangle).
\end{equation}
The first term on the right has the form $X_L(\langle m\times a,a\times b,c\times\cdots\rangle)$ where ``$\cdots$'' stands for everything along the dotted blue semicircle and to its left in the figure.  Then using~(\ref{eq:sl3vector}) we have
\begin{align}
\begin{split}
(\ref{eq:XLunroll}) &= X_L(\langle m,a,b\rangle) + X_L(\langle a,b,\cdots\rangle) + X_L(\langle c,\cdots,a \rangle) - X_L(\langle a,b,\cdots \rangle) - X_L(\langle m,a,b \rangle)\\
          &= X_L(\langle a,c,\cdots\rangle)\,,
	  \end{split}
\end{align}
but this is the same as $X_L$ applied to the right-hand side of~(\ref{eq:arb1}), as required.

Next consider applying $X_L$ to the left-hand side of
\begin{align}
\label{eq:arb2}
     \begin{tikzpicture}[baseline={(current bounding box.center)},scale=0.8,xscale=-1]
          \def \cen{(0,0)};
          \def \rad{1.5};
          \foreach \x in {1,2}{
               \filldraw[fill=black,draw=black] \cen+({40*\x+20}:\rad) circle (2pt);
               \coordinate(v\x) at ({40*\x+20}:1.5);}
          \foreach \x in {9}{
               \filldraw[fill=black,draw=black] \cen+(40*\x:{0.8*\rad}) circle (2pt);
               \coordinate(v\x) at (40*\x:{0.8*\rad});}
          \draw[white] \cen+({40* (1)+20}:{\rad+0.45}) circle (0.1) node[black]{\tikz{\node at (0,0) {$b$}}};
          \draw[white] \cen+({40* (2)+20}:{\rad+0.45}) circle (0.1) node[black]{\tikz{\node at (0,0) {$c$}}};\draw[white] \cen+({50* (0)}:{0.8*\rad+0.45}) circle (0.1) node[black]{\tikz{\node at (0,0) {$a$}}};
          \foreach \x in {1,2,3,4,5,6}{
               \coordinate(A\x) at ({60*\x}:{.4*\rad});}
          \draw[thick] (A2) -- (A3) (A6) -- (A1);
          \draw[very thick,blue] (A1) -- (A2) -- (A3) (A6) -- (A1);
          \coordinate(B1) at (75:{.7*\rad});
          \coordinate(B2) at (180:{.8*\rad});
          \coordinate(B3) at (300:{.8*\rad});
          \draw[thick] (B2) -- (A3) (A6) -- (v9);
          \draw[very thick,blue] (A2) -- (v2) (B1) -- (A1);
          \draw[thick] (v1) -- (B1) -- (v2) ;
          \draw[very thick,dashed,blue] (-{0.4*\rad},0) arc (180:360:{0.4*\rad});
          \foreach \x in {1,3} {
               \filldraw[fill=black,draw=black] (A\x) circle (2pt);
               \filldraw[fill=white,draw=black] ({-60+60*\x}:{.4*\rad}) circle (2pt);}
          \foreach \x in {1,2}
               \filldraw[fill=white,draw=black] (B\x) circle (2pt);
          \node at (-1.0*\rad,0.7) {\qquad\quad$\longrightarrow$};
     \end{tikzpicture}\quad
     \begin{tikzpicture}[baseline={(current bounding box.center)},scale=0.8,xscale=-1]
          \def \cen{(0,0)};
          \def \rad{1.5};
          \foreach \x in {1,2}{
               \filldraw[fill=black,draw=black] \cen+({40*\x+20}:\rad) circle (2pt);
               \coordinate(v\x) at ({40*\x+20}:1.5);}
          \foreach \x in {9}{
               \filldraw[fill=black,draw=black] \cen+(40*\x:{0.8*\rad}) circle (2pt);
               \coordinate(v\x) at (40*\x:{0.8*\rad});}
          \draw[white] \cen+({40* (1)+20}:{\rad+0.45}) circle (0.1) node[black]{\tikz{\node at (0,0) {$b$}}};
          \draw[white] \cen+({50* (2)}:{\rad+0.45}) circle (0.1) node[black]{\tikz{\node at (0,0) {$c$}}};
          \draw[white] \cen+({50* (0)}:{0.8*\rad+0.45}) circle (0.1) node[black]{\tikz{\node at (0,0) {$a$}}};
          \foreach \x in {1,2,3,4,5,6}{
               \coordinate(A\x) at ({60*\x}:{.4*\rad});}
          \coordinate(B1) at (75:{.7*\rad});
          \coordinate(B2) at (180:{.8*\rad});
          \coordinate(B3) at (300:{.8*\rad});
          \draw[thick] (B2) -- (A3) (A6) -- (v9);
          \draw[very thick,blue] (A3) -- (B1) (A6) -- (v2);
          \draw[thick] (v1) -- (B1) -- (v2) ;
          \draw[very thick,dashed,blue] (-{0.4*\rad},0) arc (180:360:{0.4*\rad});
          \foreach \x in {3} {
               \filldraw[fill=black,draw=black] (A\x) circle (2pt);
               \filldraw[fill=white,draw=black] ({-180+60*\x}:{.4*\rad}) circle (2pt);}
          \foreach \x in {1,2}
               \filldraw[fill=white,draw=black] (B\x) circle (2pt);
     \end{tikzpicture}
\end{align}
which by~(\ref{eq:sl3unroll}) gives
\begin{align}
\begin{split}
     & X_L\left(
     \raisebox{-.85cm}{
     \begin{tikzpicture}[scale=0.8]
          \def \cen{(0,0)};
          \def \rad{1.5};
          \foreach \x in {1,2}{
               \filldraw[fill=black,draw=black] \cen+({50*\x-50}:{(0.6+\x*0.2)*\rad}) circle (2pt);
               \coordinate(v\x) at ({50*\x-50}:{(0.6+\x*0.2)*\rad});}
          \foreach \x in {3,9}{
               \filldraw[fill=black,draw=black] \cen+(40*\x:{0.8*\rad}) circle (2pt);
               \coordinate(v\x) at (40*\x:{0.8*\rad});}
          \draw[white] \cen+({50* (0)}:{0.8*\rad+0.45}) circle (0.1) node[black]{\tikz{\node at (0,0) {$c$}}};
          \draw[white] \cen+({50* (1)}:{\rad+0.45}) circle (0.1) node[black]{\tikz{\node at (0,0) {$b$}}};
          \draw[white] \cen+({40* (3)}:{.8*\rad+0.45}) circle (0.1) node[black]{\tikz{\node at (0,0) {$a$}}};
          \foreach \x in {1,2,3,4,5,6}{
               \coordinate(A\x) at ({60*\x}:{.4*\rad});}
          \draw[thick] (A2) -- (A3);
          \draw[very thick,red] (A1) -- (A2);
          \coordinate(B1) at (35:{.7*\rad});
          \coordinate(B2) at (180:{.8*\rad});
          \coordinate(B3) at (300:{.8*\rad});
          \draw[very thick, blue] (B1) -- (A1)  (v1) -- (A6) -- (A1);
          \draw[thick] (A2) -- (v3) (B2) -- (A3);
          \draw[thick] (v1) -- (B1) -- (v2) ;
          \draw[very thick,dashed,blue] (-{0.4*\rad},0) arc (180:360:{0.4*\rad});
          \foreach \x in {1,3} {
               \filldraw[fill=black,draw=black] (A\x) circle (2pt);
               \filldraw[fill=white,draw=black] ({-60+60*\x}:{.4*\rad}) circle (2pt);}
          \foreach \x in {1,2}
               \filldraw[fill=white,draw=black] (B\x) circle (2pt);
     \end{tikzpicture}}\right) = X_L\left(
     \raisebox{-.85cm}{
     \begin{tikzpicture}[scale=0.8]
          \def \cen{(0,0)};
          \def \rad{1.5};
          \coordinate (v11) at (0:{\rad*1});
          \coordinate (v12) at (38:{1.2*\rad});
          \coordinate (v21) at (55:{1.1*\rad});
          \coordinate (v22) at (85:{0.9*\rad});
          \coordinate (v31) at ({-.5*\rad+0.3},{.3*\rad+0.6});
          \coordinate (v32) at ({-.5*\rad-0.3},{.3*\rad+0.6});
          \foreach \x in {9}{
               \filldraw[fill=black,draw=black] \cen+(40*\x:{1*\rad}) circle (2pt);
               \coordinate(v\x) at (40*\x:{1*\rad});}
          \draw[white] \cen+(0:{\rad*1+0.4}) circle (0.1) node[black]{\tikz{\node at (0,0) {$c$}}};
          \draw[white] \cen+(38:{1.2*\rad+0.4}) circle (0.1) node[black]{\tikz{\node at (0,0) {$b$}}};
          \draw[white] \cen+(55:{1.1*\rad+0.4}) circle (0.1) node[black]{\tikz{\node at (0,0) {$a$}}};
          \draw[white] \cen+(85:{0.9*\rad+0.4}) circle (0.1) node[black]{\tikz{\node at (0,0) {$m$}}};
          \draw[white] ({-.5*\rad+0.3},{.3*\rad+0.6})+(60:{0.25}) circle (0.1) node[black]{\tikz{\node at (0,0) {$c$}}};
          \draw[white] ({-.5*\rad-0.3},{.3*\rad+0.6})+(120:{0.25}) circle (0.1) node[black]{\tikz{\node at (0,0) {$a$}}};
          \coordinate(A1) at ({.4*\rad},{.3*\rad});
          \coordinate(A2) at ({-.5*\rad},{.4*\rad});
          \foreach \x in {3,4,5,6}{
               \coordinate(A\x) at ({60*\x}:{.4*\rad});}
          \draw[thick] (A2) -- (A3);
          \coordinate(B1) at (25:{.8*\rad});
          \coordinate(B15) at ({.3*\rad},{.6*\rad});
          \coordinate(B2) at (180:{.8*\rad});
          \coordinate(B3) at (300:{.8*\rad});
          \draw[very thick, blue] (B1) -- (A1) (v9) -- (A6) -- (A1);
          \draw[thick] (B2) -- (A3);
          \draw[thick] (v11) -- (B1) -- (v12) (A2) -- (v32);
          \draw[thick,red] (v31) -- (A2) (v21) -- (B15) -- (v22) (B15) -- (A1);
          \draw[very thick,dashed,blue] (-{0.4*\rad},0) arc (180:360:{0.4*\rad});
          \foreach \x in {11,12,32}
               \filldraw (v\x) circle (2pt);
          \foreach \x in {21,22,31}
               \filldraw[draw=red,fill=red] (v\x) circle (2pt);
          \filldraw[fill=black,draw=black] (A1) circle (2pt);
          \filldraw[fill=white,draw=black] (A2) circle (2pt);
          \foreach \x in {3}{
               \filldraw[fill=black,draw=black] (A\x) circle (2pt);
               \filldraw[fill=white,draw=black] ({-180+60*\x}:{.4*\rad}) circle (2pt);}
          \foreach \x in {1,2}
               \filldraw[fill=white,draw=black] (B\x) circle (2pt);
          \filldraw[fill=white,draw=red] (B15) circle (2pt);
     \end{tikzpicture}}\right) - X_L\big(\langle{m,a,c}\rangle) \\
          &\qquad =X_L(\langle m\times a, b\times c, c\times\cdots \rangle) - X_L(\langle m,a,c) \rangle \\
          &\qquad = X_L(\langle m,a,c \rangle) + X_L(\langle b,c,\cdots \rangle) + X_L(\langle c,\cdots,a \rangle) - X_L(\langle a,c,\cdots \rangle) - X_L(\langle m,a,c) \\
          &\qquad = X_L(\langle b,c,\cdots \rangle)\,,
     \end{split}
\end{align}
which again is the required answer: $X_L$ applied to the right-hand side of~(\ref{eq:arb2}).

We omit the proof for $X_R$, which is essentially the same, and instead illustrate with the example shown in Fig.~\ref{figarb}:
\begin{equation}\label{eq:figarb}
     \begin{tikzpicture}[baseline={([yshift=-2.2ex]current bounding box.center)},scale=0.8,xscale=-1]

          \def \cen{(0,0)};
          \def \rad{1.5};
          \draw {\cen} circle (\rad); 

          \foreach \x in {1,2,3,4,5,6,7,8}{
               \filldraw[fill=black,draw=black] \cen+(45*\x:\rad) circle (2pt);
               \draw[white] \cen+({45* (\x)}:{\rad+0.3}) circle (0.1) node[black]{\tikz{\node at (0,0) {$\x$}}};
               \coordinate(v\x) at (45*\x:1.5);}

          \foreach \x in {1,2,3,4,5,6}{
               \coordinate(A\x) at ({60*\x}:{.4*\rad});
               \draw[thick,black] ({60*\x}:{.4*\rad}) -- ({60+60*\x}:{.4*\rad});}

          \foreach \x in {1}{
               \coordinate(A\x) at ({60*\x}:{.4*\rad});
               \draw[thick,red] ({60*\x}:{.4*\rad}) -- ({60+60*\x}:{.4*\rad});}

          \coordinate(B1) at (65:{.7*\rad});
          \coordinate(B2) at (190:{.7*\rad});
          \coordinate(B3) at (320:{.7*\rad});

          \draw[thick] (B1) -- (A1) (B2) -- (A3);
          \draw[thick] (A2) -- (v3) (A4) -- (v6) (A6) -- (v8);

          \draw[thick,black] (A4) -- (A5) -- (A6) (A5) -- (B3) (v7) -- (B3) -- (v8);

          \draw[thick] (v1) -- (B1) -- (v2) (v4) -- (B2) -- (v5);

          \foreach \x in {1,3,5}{
               \filldraw[fill=black,draw=black] (A\x) circle (2pt);
               \filldraw[fill=white,draw=black] ({60+60*\x}:{.4*\rad}) circle (2pt);}
          \filldraw[fill=black,draw=black] (v7) circle (2pt);
          \filldraw[fill=black,draw=black] (v8) circle (2pt);
          \foreach \x in {1,2,3}
               \filldraw[fill=white,draw=black] (B\x) circle (2pt);

          \node at (-2.5,0) {$=$};
     \end{tikzpicture}
     \begin{tikzpicture}[baseline={([yshift=-2.2ex]current bounding box.center)},scale=0.8,xscale=-1]
          \def \cen{(0,0)};
          \def \rad{1.5};
          \draw {\cen} circle (\rad); 

          \foreach \x in {1,2,3,4,5,6,7,8}{
               \filldraw[fill=black,draw=black] \cen+(45*\x:\rad) circle (2pt);
               \draw[white] \cen+({45* (\x)}:{\rad+0.3}) circle (0.1) node[black]{\tikz{\node at (0,0) {$\x$}}};
               \coordinate(v\x) at (45*\x:1.5);}

          \foreach \x in {1,2,3,4,6}{
               \coordinate(A\x) at ({60*\x}:{.4*\rad});}

          \draw[thick,black] (A6) -- (A1) -- (A2) -- (A3) -- (A4);

          \coordinate(B1) at (65:{.7*\rad});
          \coordinate(B2) at (190:{.7*\rad});

          \draw[thick] (B1) -- (A1) (B2) -- (A3);
          \draw[thick] (A2) -- (v3) (A4) -- (v6) (A6) -- (v8);
          \draw[thick] (v1) -- (B1) -- (v2) (v4) -- (B2) -- (v5);

          \draw[thick,black] (A4) -- (v8) (A6) -- (v7);

          \foreach \x in {1,3}{
               \filldraw[fill=black,draw=black] (A\x) circle (2pt);
               \filldraw[fill=white,draw=black] ({60+60*\x}:{.4*\rad}) circle (2pt);}
          \filldraw[fill=white,draw=black] (A6) circle (2pt);

          \filldraw[fill=black,draw=black] (v7) circle (2pt);
          \filldraw[fill=black,draw=black] (v8) circle (2pt);
          \foreach \x in {1,2}
               \filldraw[fill=white,draw=black] (B\x) circle (2pt);

          \node at (-6.3,0) {~$=\langle(7\times8)\times(1\times2),3,(4\times5)\times(6\times8)\rangle$};
     \end{tikzpicture}
\end{equation}
Using~(\ref{eq:sl3unroll}) to unroll the red edge on the inner loop of the left diagram gives
\begin{align}
     \begin{split}
          &X_R(\langle ((1 \times 3)\times(4 \times 5))\times 6,\, 7 \times 8,\, 8\times((1 \times 2) \times (3 \times m)) \rangle) - X_R(\langle 1,3,m \rangle)\\
          &\qquad = X_R(\langle{7,1,2}\rangle) + X_R(\langle{1,4,5}\rangle) + X_R(\langle{4,6,8}\rangle) + X_R(\langle{3,7,8}\rangle) \\
          & \qquad\qquad\qquad  + X_R(\langle{1,3,6}\rangle)- X_R(\langle{1,3,7}\rangle) - X_R(\langle{1,4,6}\rangle).
     \end{split}
\label{eq:skeincheck}
\end{align}
On the other hand, the right diagram in~(\ref{eq:figarb}) is a tree; according to~(\ref{eq:sl3covector}) its image under the $X_R$ map is
\begin{align}\label{eq:figarb3}
     \begin{split}
          &X_R(\langle(7\times8)\times(1\times2),3,(4\times5)\times(6\times8)\rangle)\\
          &\qquad  = X_R(\langle{7,1,2}\rangle) + X_R(\langle{1,4,5}\rangle) + X_R(\langle{4,6,8}\rangle) + X_R(\langle{3,7,8}\rangle) + X_R(\langle{1,3,6}\rangle) \\
          & \qquad\qquad  - X_R(\langle{1,3,7}\rangle) - X_R(\langle{1,4,6}\rangle)\,,
     \end{split}
\end{align}
in agreement with~(\ref{eq:skeincheck}).

\subsection{\texorpdfstring{$\mathfrak{sl}_4$ Arborizable Invariants}{sl4 Arborizable Invariants}}

For $k=4$ the recursion for calculating the $X$-map is seeded by
\begin{align}
\begin{split}
X_L(\langle a\,b\,c\,d \rangle) &= {\bf S}_{a+1,\ldots,b[b+1,\ldots,c(c+1,\ldots,d)]}\,,\\
X_R(\langle a\,b\,c\,d \rangle) &= {\bf S}_{[(a,\ldots,b-1)b,\ldots,c-1]c,\ldots,d-1}\,,
\end{split}
\label{eq:sl4seed}
\end{align}
using notation reviewed in Sec.~\ref{sec:notation}.

There are two basic types of structures that can appear in place of simple vectors $a,b,c,d$ in~(\ref{eq:sl4seed}) when we look at more general tree invariants.  We must describe separately how to recursively handle each type of structure.

The first type of structure that can appear in place of a vector is a tensor product involving 5 points:
\begin{align}
\label{eq:type1}
\lbrack (ab)\cap(cde) \rbrack^{\mu}:=\left( \frac{1}{2} a^\nu b^\rho \epsilon_{\nu\rho\sigma\tau} \right) \Big(
c^\alpha d^{\beta} e^{\gamma} \epsilon_{\alpha\beta\gamma\delta} \Big)\, \epsilon^{\sigma\tau\delta\mu}
\end{align}
where all Greek subscripts and superscripts run from 1 to 4 and $\epsilon$ is the antisymmetric Levi-Civita symbol with $\epsilon_{1234}=\epsilon^{1234}=1$; recall~(\ref{eq:shorthand}).  For invariants involving one of these we have
\begin{align}
\begin{split}
X_L(\langle{a,b,c,(d,e)\cap(f,g,h)}\rangle) &=
X_L(\langle{a,b,c,e}\rangle) + X_L(\langle{d,e,g,h}\rangle)\\
&\qquad \qquad + X_L(\langle{f,g,h,c}\rangle) - X_L(\langle{c,e,g,h}\rangle)\,, \\
X_R(\langle{a,b,c,(d,e)\cap(f,g,h)}\rangle) &=
 X_R(\langle{a,b,d,e}\rangle) + X_R(\langle{d,f,g,h}\rangle)\\
&\qquad \qquad + X_R(\langle{f,a,b,c}\rangle) - X_R(\langle{a,b,d,f}\rangle)\,.
\end{split}
\label{eq:sl4example}
\end{align}
We could also have a structure like~(\ref{eq:type1}) but with a lower $\mu$ index---as would be the case for example if all of $a,\ldots,e$ were not individual vectors but triple-products like $a_\mu = \epsilon_{\mu\nu\rho\sigma} a_1^\nu a_2^\rho a_3^\sigma$ for some $a_i$.  (Note that $a_\mu$ represents the plane in $\mathbb{P}^3$ containing the three $a_i$.).  In this case we would have
\begin{align}
\begin{split}
X_L(\langle{a,b,c,(d,e)\cap(f,g,h)}\rangle) &=
X_L(\langle{a,b,d,e}\rangle) + X_L(\langle{d,f,g,h}\rangle)\\
&\qquad \qquad + X_L(\langle{f,a,b,c}\rangle) - X_L(\langle{a,b,d,f}\rangle)\,,\\
X_R(\langle{a,b,c,(d,e)\cap(f,g,h)}\rangle) &=
X_R(\langle{a,b,c,e}\rangle) + X_R(\langle{d,e,g,h}\rangle)\\
& \qquad \qquad + X_R(\langle{f,g,h,c}\rangle) - X_R(\langle{c,e,g,h}\rangle)\,.
\end{split}
\label{eq:sl4Xdefone}
\end{align}

The second type of structure that can appear in tree invariants is a tensor product involving 9 points:
\begin{multline}
\lbrack (a_1a_2a_3) (b_1b_2b_3) (c_1c_2c_3) \rbrack^\mu :=\\
(a_1^{\nu_1} a_2^{\nu_2} a_3^{\nu_3}\epsilon_{\nu_1\nu_2\nu_3\nu_4}) (b_1^{\rho_1} b_2^{\rho_2} b_3^{\rho_3} \epsilon_{\rho_1 \rho_2 \rho_3 \rho_4}) (c_1^{\tau_1} c_2^{\tau_2} c_3^{\tau_3} \epsilon_{\tau_1 \tau_2 \tau_3 \tau_4}) \epsilon^{\nu_4 \rho_4 \tau_4 \mu}\,,
\label{eq:9points}
\end{multline}
which can appear in combinations of the form (again recall~(\ref{eq:shorthand}))
\begin{align}
C_{abc} := \langle{(a_1 a_2 a_3), (b_1 b_2 b_3), (c_1 c_2 c_3), (d_1 d_2 d_3)}\rangle=\langle{a_1,a_2,a_3,(b_1 b_2 b_3)(c_1 c_2 c_3)(d_1 d_2 d_3)}\rangle\,.
\end{align}
For this kind of invariant we have
\begin{align}
\begin{split}
X_L(C_{abc}) &=
X_L (\langle{a_1,a_2,a_3,b_3}\rangle) + X_L(\langle{b_1,b_2,b_3,c_3}\rangle) + X_L(\langle{c_1,c_2,c_3,d_3}\rangle) \\
& + X_L(\langle{d_1,d_2,d_3,a_3}\rangle)
 + X_L(\langle{a_2,a_3,c_2,c_3}\rangle)
               + X_L(\langle{b_2,b_3,d_2,d_3}\rangle)\\
	       &- X_L(\langle{a_2,a_3,b_3,c_3}\rangle)
	       - X_L(\langle{b_2,b_3,c_3,d_3}\rangle)
 - X_L(\langle{c_2,c_3,d_3,a_3}\rangle)\\
                &- X_L(\langle{d_2,d_3,a_3,b_3}\rangle)
		+ X_L(\langle{a_3,b_3,c_3,d_3}\rangle)\,,
\\
X_R(C_{abc}) &=
 X_R(\langle{a_1,b_1,b_2,b_3}\rangle) + X_R(\langle{b_1,c_1,c_2,c_3}\rangle) + X_R(\langle{c_1,d_1,d_2,d_3}\rangle)\\
 &+ X_R(\langle{d_1,a_1,a_2,a_3}\rangle)
 + X_R(\langle{a_1,a_2,c_1,c_2}\rangle)
         + X_R(\langle{b_1,b_2,d_1,d_2}\rangle)\\
	 &- X_R(\langle{a_1,b_1,c_1,c_2}\rangle) - X_R(\langle{b_1,c_1,d_1,d_2}\rangle)
 -X_R(\langle{c_1,d_1,a_1,a_2}\rangle)\\
    &      - X_R(\langle{d_1,a_1,b_1,b_2}\rangle) +X_R(\langle{a_1,b_1,c_1,d_2}\rangle)\,.
\end{split}
\label{eq:sl4Xdeftwo}
\end{align}
Alternatively, in the contravariant case (that is, when each of the nine entries in~(\ref{eq:9points}) represents a plane) we have
\begin{align}
\begin{split}
X_L(C_{abc}) &=
X_L(\langle{a_1,b_1,b_2,b_3}\rangle) + X_L(\langle{b_1,c_1,c_2,c_3}\rangle) + X_L(\langle{c_1,d_1,d_2,d_3}\rangle)\\
&+ X_L(\langle{d_1,a_1,a_2,a_3}\rangle)
 + X_L(\langle{a_1,a_2,c_1,c_2}\rangle)
               + X_L(\langle{b_1,b_2,d_1,d_2}\rangle)\\
	       &- X_L(\langle{a_1,b_1,c_1,c_2}\rangle) - X_L(\langle{b_1,c_1,d_1,d_2}\rangle)
 -X_L(\langle{c_1,d_1,a_1,a_2}\rangle)\\
& - X_L(\langle{d_1,a_1,b_1,b_2}\rangle) +X_L(\langle{a_1,b_1,c_1,d_2}\rangle)\,,\\
X_R(C_{abc}) &=
X_R (\langle{a_1,a_2,a_3,b_3}\rangle) + X_R(\langle{b_1,b_2,b_3,c_3}\rangle) + X_R(\langle{c_1,c_2,c_3,d_3}\rangle)\\
&+ X_R(\langle{d_1,d_2,d_3,a_3}\rangle)
 + X_R(\langle{a_2,a_3,c_2,c_3}\rangle)
          + X_R(\langle{b_2,b_3,d_2,d_3}\rangle)\\
	  &- X_R(\langle{a_2,a_3,b_3,c_3}\rangle) - X_R(\langle{b_2,b_3,c_3,d_3}\rangle)
 - X_R(\langle{c_2,c_3,d_3,a_3}\rangle)\\
& - X_R(\langle{d_2,d_3,a_3,b_3}\rangle) + X_R(\langle{a_3,b_3,c_3,d_3}\rangle)\,.
\end{split}
\label{eq:sl4Xdefthree}
\end{align}

All of the above relations are valid only when each invariant is read off from a tree diagram drawn in canonical order, as in the previous subsection.

\subsection{\texorpdfstring{$\mathfrak{sl}_4$ Non-Arborizable Invariants}{sl4 Non-Arborizable Invariants}}

Here we consider only inner loops which have no double edges.  We have found this to be sufficient to analyze the $\mathcal{C}(4,8)$ and $\mathcal{C}^\dagger(4,9)$ polytopes (see Sec.~\ref{sec:results}); it would be interesting to formulate a recursive rule for more general diagrams.  We find that single-edged inner loops of $\mathfrak{sl}_4$ diagrams can be unrolled with the rule
\begin{align}
\begin{split}
     X_L\left(\!\!\!
     \raisebox{-1.2cm}{
     \begin{tikzpicture}[scale=0.75]
          \def \rad{1};
          \def \dis{0.2};
          \foreach \x in {1,2,3,4}{
               \coordinate(v\x) at ({-60+60*\x}:\rad);
               \coordinate(B\x) at ({15+30*\x}:{1.8*\rad});}
          \draw[white] ({15+30*1}:{2*\rad+\dis}) circle (0.1) node[black]{\tikz{\node at (0,0) {$d$}}};
          \draw[white] ({15+30*2}:{2*\rad+\dis}) circle (0.1) node[black]{\tikz{\node at (0,0) {$c$}}};
          \draw[white] ({15+30*3}:{2*\rad+\dis}) circle (0.1) node[black]{\tikz{\node at (0,0) {$b$}}};
          \draw[white] ({15+30*4}:{2*\rad+\dis}) circle (0.1) node[black]{\tikz{\node at (0,0) {$a$}}};
          \draw[thick] (v1) -- (v2) -- (B1) (v2) -- (B2);
          \draw[thick] (v4) -- (v3) -- (B4) (v3) -- (B3);
          \draw[very thick, red] (v2) -- (v3);
          \draw[very thick, dashed, blue] (v4) arc (180:360:\rad);
          \foreach \x in {v2,v4,B3,B4}
               \filldraw[fill=black] (\x) circle (2.25pt);
          \foreach \x in {v1,v3,B1,B2}
               \filldraw[fill=white] (\x) circle (2.25pt);
     \end{tikzpicture}}\!\!\!\right) = X_L\left(\!\!\!
     \raisebox{-1.2cm}{
     \begin{tikzpicture}[scale=0.75]
          \def \rad{1};
          \def \dis{0.3};
          \coordinate (v1) at (\rad,0);
          \coordinate (v4) at (-\rad,0);
          \coordinate (v2) at (\rad,{0.5*\rad});
          \coordinate (v3) at (-\rad,{0.5*\rad});
          \foreach \x in {1,2,3,4,5,6,7,8,9,10}{
               \coordinate(B\x) at ({-9+18*\x}:{1.8*\rad});}
          \coordinate (A1) at (50:{1.2*\rad});
          \coordinate (A2) at (130:{1.2*\rad});
          \draw[white] ({-9+18*1}:{1.8*\rad+\dis}) circle (0.1) node[black]{\tikz{\node at (0,0) {$d$}}};
          \draw[white] ({-9+18*2}:{1.8*\rad+\dis}) circle (0.1) node[black]{\tikz{\node at (0,0) {$c$}}};
          \draw[white] ({-9+18*3}:{1.8*\rad+\dis}) circle (0.1) node[black]{\tikz{\node at (0,0) {$b$}}};
          \draw[white] ({-9+18*4}:{1.8*\rad+\dis}) circle (0.1) node[black]{\tikz{\node at (0,0) {$a$}}};
          \draw[white] ({-9+18*5-2}:{1.8*\rad+\dis}) circle (0.1) node[black]{\tikz{\node at (0,0) {$m_1$}}};
          \draw[white] ({-9+18*6+2}:{1.8*\rad+\dis}) circle (0.1) node[black]{\tikz{\node at (0,0) {$m_2$}}};
          \draw[white] ({-9+18*7}:{1.8*\rad+\dis}) circle (0.1) node[black]{\tikz{\node at (0,0) {$d$}}};
          \draw[white] ({-9+18*8}:{1.8*\rad+\dis}) circle (0.1) node[black]{\tikz{\node at (0,0) {$c$}}};
          \draw[white] ({-9+18*9}:{1.8*\rad+\dis}) circle (0.1) node[black]{\tikz{\node at (0,0) {$b$}}};
          \draw[white] ({-9+18*10}:{1.8*\rad+\dis}) circle (0.1) node[black]{\tikz{\node at (0,0) {$a$}}};
          \draw[thick] (v1) -- (v2) -- (B1) (v2) -- (B2);
          \draw[thick] (v4) -- (v3) -- (B10) (v3) -- (B9);
          \draw[thick,red] (v2) -- (A1) -- (B3) (B4) -- (A1) -- (B5);
          \draw[thick,red] (v3) -- (A2) -- (B8) (B6) -- (A2) -- (B7);
          \draw[very thick, dashed, blue] (v4) arc (180:360:\rad);
          \foreach \x in {v1,B1,B2,v3}
               \filldraw[fill=white] (\x) circle (2.25pt);
          \foreach \x in {v2,B9,B10,v4}
               \filldraw (\x) circle (2.25pt);
          \foreach \x in {A1,B6,B7,B8}
               \filldraw[fill=white,draw=red] (\x) circle (2.25pt);
          \foreach \x in {A2,B3,B4,B5}
               \filldraw[fill=red,draw=red] (\x) circle (2.25pt);
     \end{tikzpicture}}\!\!\!\right) - X_L\Big(\Big\langle{m_1,a,b,(c,d)\cap m_2}\Big\rangle\Big)\,,
\\
     X_R\left(\!\!\!
     \raisebox{-1.2cm}{
     \begin{tikzpicture}[scale=0.75]
          \def \rad{1};
          \def \dis{0.2};
          \foreach \x in {1,2,3,4}{
               \coordinate(v\x) at ({-60+60*\x}:\rad);
               \coordinate(B\x) at ({15+30*\x}:{1.8*\rad});}
          \draw[white] ({15+30*1}:{2*\rad+\dis}) circle (0.1) node[black]{\tikz{\node at (0,0) {$d$}}};
          \draw[white] ({15+30*2}:{2*\rad+\dis}) circle (0.1) node[black]{\tikz{\node at (0,0) {$c$}}};
          \draw[white] ({15+30*3}:{2*\rad+\dis}) circle (0.1) node[black]{\tikz{\node at (0,0) {$b$}}};
          \draw[white] ({15+30*4}:{2*\rad+\dis}) circle (0.1) node[black]{\tikz{\node at (0,0) {$a$}}};
          \draw[thick] (v1) -- (v2) -- (B1) (v2) -- (B2);
          \draw[thick] (v4) -- (v3) -- (B4) (v3) -- (B3);
          \draw[very thick, red] (v2) -- (v3);
          \draw[very thick, dashed, blue] (v4) arc (180:360:\rad);
          \foreach \x in {v2,v4,B3,B4}
               \filldraw[fill=white] (\x) circle (2.25pt);
          \foreach \x in {v1,v3,B1,B2}
               \filldraw[fill=black] (\x) circle (2.25pt);
     \end{tikzpicture}}\!\!\!\right) = X_R\left(\!\!\!
     \raisebox{-1.2cm}{
     \begin{tikzpicture}[scale=0.75]
          \def \rad{1};
          \def \dis{0.3};
          \coordinate (v1) at (\rad,0);
          \coordinate (v4) at (-\rad,0);
          \coordinate (v2) at (\rad,{0.5*\rad});
          \coordinate (v3) at (-\rad,{0.5*\rad});
          \foreach \x in {1,2,3,4,5,6,7,8,9,10}{
               \coordinate(B\x) at ({-9+18*\x}:{1.8*\rad});}
          \coordinate (A1) at (50:{1.2*\rad});
          \coordinate (A2) at (130:{1.2*\rad});
          \draw[white] ({-9+18*1}:{1.8*\rad+\dis}) circle (0.1) node[black]{\tikz{\node at (0,0) {$d$}}};
          \draw[white] ({-9+18*2}:{1.8*\rad+\dis}) circle (0.1) node[black]{\tikz{\node at (0,0) {$c$}}};
          \draw[white] ({-9+18*3}:{1.8*\rad+\dis}) circle (0.1) node[black]{\tikz{\node at (0,0) {$b$}}};
          \draw[white] ({-9+18*4}:{1.8*\rad+\dis}) circle (0.1) node[black]{\tikz{\node at (0,0) {$a$}}};
          \draw[white] ({-9+18*5-2}:{1.8*\rad+\dis}) circle (0.1) node[black]{\tikz{\node at (0,0) {$m_2$}}};
          \draw[white] ({-9+18*6+2}:{1.8*\rad+\dis}) circle (0.1) node[black]{\tikz{\node at (0,0) {$m_1$}}};
          \draw[white] ({-9+18*7}:{1.8*\rad+\dis}) circle (0.1) node[black]{\tikz{\node at (0,0) {$d$}}};
          \draw[white] ({-9+18*8}:{1.8*\rad+\dis}) circle (0.1) node[black]{\tikz{\node at (0,0) {$c$}}};
          \draw[white] ({-9+18*9}:{1.8*\rad+\dis}) circle (0.1) node[black]{\tikz{\node at (0,0) {$b$}}};
          \draw[white] ({-9+18*10}:{1.8*\rad+\dis}) circle (0.1) node[black]{\tikz{\node at (0,0) {$a$}}};
          \draw[thick] (v1) -- (v2) -- (B1) (v2) -- (B2);
          \draw[thick] (v4) -- (v3) -- (B10) (v3) -- (B9);
          \draw[thick,red] (v2) -- (A1) -- (B3) (B4) -- (A1) -- (B5);
          \draw[thick,red] (v3) -- (A2) -- (B8) (B6) -- (A2) -- (B7);
          \draw[very thick, dashed, blue] (v4) arc (180:360:\rad);
          \foreach \x in {v1,B1,B2,v3}
               \filldraw (\x) circle (2.25pt);
          \foreach \x in {v2,B9,B10,v4}
               \filldraw[fill=white] (\x) circle (2.25pt);
          \foreach \x in {A1,B6,B7,B8}
               \filldraw[fill=red,draw=red] (\x) circle (2.25pt);
          \foreach \x in {A2,B3,B4,B5}
               \filldraw[draw=red,fill=white] (\x) circle (2.25pt);
     \end{tikzpicture}}\!\!\!\right) - X_R\Big(\Big\langle{m_2\cap(a,b),c,d,m_1}\Big\rangle\Big)\,.
\end{split}
\label{eq:sl4unroll}
\end{align}
Here we need two reference points $m_1, m_2$ for each unrolled loop.  Actually, since $m_2$ is associated to a white vertex, it is better thought of as a reference plane, i.e. a triple of reference points.  All dependence on these reference points drops out of any invariant.  Of course, as always, the recursion rule can only be applied to canonically ordered diagrams.

By way of example let us apply $X_L$ to the non-arborizable $\Gr(4,8)$ web shown in Fig.~\ref{fig:nonarb}(b):
\begin{equation}
     \begin{aligned}
          & X_L\left(\!\!\!
          \raisebox{0.01cm}{
          \begin{tikzpicture}[baseline={([yshift=-.5ex]current bounding box.center)},scale=.6,xscale=-1]
               \def \cen{(0,0)};
               \def \rad{1.5};
               \draw {\cen} circle (\rad);

               \foreach \x in {1,2,3,4,5,6,7,8}{
                    \draw[white] \cen+({45* (\x)}:{\rad+0.4}) circle (0.1) node[black]{\tikz{\node at (0,0) {$\x$}}};
                    \coordinate(v\x) at (45*\x:1.5);}

               \foreach \x in {1,2,3,4}
                    \coordinate(A\x) at ({135/2-90+90*\x}:{\rad*0.7});

               \foreach \x in {1,2}{
                    \coordinate(B\x) at ({135/2-90+180*\x}:{\rad*0.3});
               }

               \draw[thick] (B1) -- (A3) -- (B2) -- (A1);
               \draw[very thick,red] (A1) -- (B1);
               \draw[thick] (v1) -- (A1) -- (v2) (v3) -- (A2) -- (v4) (v5) -- (A3) -- (v6) (v7) -- (A4) -- (v8);
               \foreach \x in {135/2+90,135/2-90}{
                    \draw[thick] ({\x+6}:{\rad*0.3}) -- ({\x+2}:{\rad*0.7});
                    \draw[thick] ({\x-6}:{\rad*0.3}) -- ({\x-2}:{\rad*0.7});}

               \foreach \x in {1,2,3,4}
                    \filldraw[fill=white] (A\x) circle (2pt);

               \foreach \x in {1,2,3,4,5,6,7,8}{
                    \filldraw[fill=black,draw=black] \cen+(45*\x:\rad) circle (2pt);}
               
               \foreach \x in {1,2}{
                    \filldraw (B\x) circle (2pt);
               }
          \end{tikzpicture}
          }\!\!\!\right)\\
&\qquad = X_L\Big( \Big\langle {(m_1 1 2)\cap(34),5,6,(78)\cap\big(1,2,(34)\cap(m_2)\big)} \Big\rangle \Big) - X_L(\langle{m_1,1,2,(34)\cap(m_2)}\rangle)\\
          &\qquad = X_L\big(\big\langle{(m_1 1 2)\cap(34)}\big\rangle,\,5,\,6,\,8\big) + X_L\big(\big\langle{7,\,8,\,2,\,(34)\cap(m_2)}\big\rangle\big)\\
	  & \qquad\qquad + X_L\big(\big\langle{6,1,2,(34)\cap(m_2)}\big\rangle\big)
          - X_L\big(\big\langle{6,8,2,(34)\cap(m_2)}\big\rangle\big)\\
	  & \qquad\qquad - X_L\big(\big\langle{m_1,1,2,(34)\cap(m_2)}\big\rangle\big)\\
          &\qquad = X_L(\langle{1246}\rangle) + X_L(\langle{3468}\rangle) + X_L(\langle{5682}\rangle) + X_L(\langle{7824}\rangle) - 2 X_L(\langle{2468}\rangle)\,,
     \end{aligned}
\end{equation}
which shows the cancellation of the reference points.

Like we discussed for the $\mathfrak{sl}_3$ case, it is of course important that the recursive relations we have given for the $k=4$ $X$-map respect the skein equivalences shown in Fig.~\ref{figskein34}, in particular applied to the arborization of an inner loop.  The proof is similar to the $k=3$ case but there are more cases of the equivalence to check.  We omit the details here and instead consider the illustrative example
\begin{equation}
     \begin{tikzpicture}[baseline={([yshift=-.5ex]current bounding box.center)},scale=.8,xscale=-1]
          \def \cen{(0,0)};
          \def \rad{1.5};
          \draw {\cen} circle (\rad);

          \foreach \x in {1,2,3,4,5,6,7,8}{
               \draw[white] \cen+({45* (\x)}:{\rad+0.4}) circle (0.1) node[black]{\tikz{\node at (0,0) {$\x$}}};
               \coordinate(v\x) at (45*\x:1.5);}

          \foreach \x in {1,2,3,4}
               \coordinate(A\x) at ({135/2-90+90*\x}:{\rad*0.7});

          \coordinate(A1) at (80:{\rad*0.55});
          \coordinate(A2) at (160:{\rad*0.6});

          \coordinate (B2) at ({135/2-90+180*2}:{\rad*0.3});
          \coordinate (B1) at (160:{\rad*0.3});

          \draw[thick] (A3) -- (B2) -- (A1);
          \draw[very thick,red] (A1) -- (B1);
          \draw[thick] (v1) -- (A1) -- (v3) (v5) -- (A3) -- (v6) (v7) -- (A4) -- (v8);
          \draw[very thick,blue] (v3) -- (A2) -- (v4) (B1) -- (A3);
          \foreach \x in {160}{
               \draw[thick,blue] ({\x+6}:{\rad*0.3}) -- ({\x+2}:{\rad*0.6});
               \draw[thick,blue] ({\x-6}:{\rad*0.3}) -- ({\x-2}:{\rad*0.6});}

          \foreach \x in {135/2-90}{
               \draw[thick] ({\x+6}:{\rad*0.3}) -- ({\x+2}:{\rad*0.7});
               \draw[thick] ({\x-6}:{\rad*0.3}) -- ({\x-2}:{\rad*0.7});}

          \foreach \x in {1,2,3,4}
               \filldraw[fill=white] (A\x) circle (2pt);

          \foreach \x in {1,2}
               \filldraw (B\x) circle (2pt);

          \foreach \x in {1,2,3,4,5,6,7,8}{
               \filldraw[fill=black,draw=black] \cen+(45*\x:\rad) circle (2pt);}

          \node at (-3,0){$=$};
     \end{tikzpicture}
     \begin{tikzpicture}[baseline={([yshift=-.5ex]current bounding box.center)},scale=.8,xscale=-1]
          \def \cen{(0,0)};
          \def \rad{1.5};
          \draw {\cen} circle (\rad);

          \foreach \x in {1,2,3,4,5,6,7,8}{
               \draw[white] \cen+({45* (\x)}:{\rad+0.4}) circle (0.1) node[black]{\tikz{\node at (0,0) {$\x$}}};
               \coordinate(v\x) at (45*\x:1.5);}

          \foreach \x in {1,2,3,4}
               \coordinate(A\x) at ({135/2-90+90*\x}:{\rad*0.7});

          \coordinate(A1) at (80:{\rad*0.55});
          \coordinate(A2) at (160:{\rad*0.6});

          \coordinate (B2) at ({135/2-90+180*2}:{\rad*0.3});
          \coordinate (B1) at (160:{\rad*0.3});

          \draw[thick] (A3) -- (B2) -- (A1);
          \draw[thick] (v1) -- (A1) -- (v3) (v5) -- (A3) -- (v6) (v7) -- (A4) -- (v8);
          \draw[very thick,blue] (v4) -- (A1) (A3) -- (v3);

          \foreach \x in {135/2-90}{
               \draw[thick] ({\x+6}:{\rad*0.3}) -- ({\x+2}:{\rad*0.7});
               \draw[thick] ({\x-6}:{\rad*0.3}) -- ({\x-2}:{\rad*0.7});}

          \foreach \x in {1,3,4}
               \filldraw[fill=white] (A\x) circle (2pt);

          \foreach \x in {2}
               \filldraw (B\x) circle (2pt);

          \foreach \x in {1,2,3,4,5,6,7,8}{
               \filldraw[fill=black,draw=black] \cen+(45*\x:\rad) circle (2pt);}
     \end{tikzpicture}.
\end{equation}
According to~(\ref{eq:sl4unroll}), the $X_L$ image of the web on the left is
\begin{multline}
X_L(\langle{1346}\rangle) + X_L(\langle{3468}\rangle) + X_L(\langle{5683}\rangle) + X_L(\langle{7834}\rangle) - 2 X_L(\langle{3468}\rangle)\\
 = X_L(\langle{1346}\rangle) + X_L(\langle{5683}\rangle) + X_L(\langle{7834}\rangle) - X_L(\langle{3468}\rangle)\,,
\end{multline}
which agrees as required with the $X_L$ image of the web on the right computed from~(\ref{eq:sl4example}).

\subsection{\texorpdfstring{General Relations Between $X_L$ and $X_R$}{A General Relation Between XL and XR}}
\label{sec:generalrelation}

Here we point out a few general relations between $X_L$ and $X_R$ that can be derived from the above definitions.  First, for $k=3$, it follows from~(\ref{eq:sl3vector}) and~(\ref{eq:sl3covector}) that
\begin{equation}\label{eq:relxlxr3}
X_L(W) = X_R(W^{\sharp_3})\,,
\end{equation}
where $W$ is any web invariant, either arborizable or non-arborizable, and $W^{\sharp_3}$ is obtained from $W$ by the replacement
\begin{equation}\label{eq:pc3ltor}
a\to(a{+}1)\times (a{+}2)
\end{equation}
(which exchanges vectors and covectors).  Note that if $W$ involves something like $b \times (b{+}1)$ then according to~(\ref{eq:pc3ltor}) we have
\begin{equation}
b\times(b{+}1) \to \lbrack (b{+}1)\times (b{+}2)\rbrack\times \lbrack(b{+}2)\times(b{+}3)\rbrack=\langle{b{+}1,b{+}2,b{+}3}\rangle (b{+}2)\,.
\end{equation}
Apart from a trivial overall factor of the frozen variable $\langle{b{+}1,b{+}2,b{+}3}\rangle$ we can effectively regard the above transformation as
\begin{equation}
b\times (b{+}1)\to b{+}2\,.
\end{equation}
One can check that this makes sense because for both $X_L$ and $X_R$ it follows from~(\ref{eq:sl3vector}) and~(\ref{eq:sl3covector}) that
\begin{align}
X_L(\langle{...,\,\lbrack (b{+}1)\times (b{+}2)\rbrack\times \lbrack(b{+}2)\times(b{+}3)\rbrack,\, ...}\rangle) &= X_L(\langle{...,\,b{+}2,\, ...}\rangle)\,,\\
X_R(\langle{...,\,\lbrack (b{+}1)\times (b{+}2)\rbrack\times \lbrack(b{+}2)\times(b{+}3)\rbrack,\, ...}\rangle) &= X_R(\langle{...,\,b{+}2,\, ...}\rangle)\,.
\end{align}
Therefore we can phrase the inverse of~(\ref{eq:relxlxr3}) and~(\ref{eq:pc3ltor}) as
\begin{equation}\label{eq:relxrxl3}
X_R(W) = X_L(W^{\flat_3})\,,
\end{equation}
where $W^{\flat_3}$ is obtained from $W$ by taking
\begin{equation}\label{eq:pc3rtol}
a \to (a{-}2)\times(a{-}1)\,.
\end{equation}

Another relation between $X_L$ and $X_R$ involves reflection.  For an arbitrary web invariant $W$, either arborizable or non-arborizable, we have
\begin{equation}\label{eq:refl3}
\textbf{Ref}_s(X_L(W)) = X_R(\textbf{Ref}(W))\,,
\end{equation}
where \textbf{Ref} relabels the external vertices of $W$ according to $a\to n\!+\!1\!-\!a$, and $\textbf{Ref}_s$ acts on kinematic space by $s_{a,b,c}\to s_{n\!+\!1\!-c,n\!+\!1\!-\!b,n\!+\!1\!-\!a}$.  By combining~(\ref{eq:relxlxr3}) and~(\ref{eq:refl3}) we can also say that
\begin{equation}\label{eq:refpc3}
\textbf{Ref}_s(X_L(W)) = X_R(\textbf{Ref}(W)) = X_L(\textbf{Ref}(W)^{\flat_3}) = X_L(\textbf{Ref}(W^{\sharp_3}))\,.
\end{equation}

Analogous relations also exist for $k=4$.  First we have
\begin{align}
\begin{split}
X_L(W) &= X_R(W^{\sharp_4})\,,\\
X_R(W) &= X_L(W^{\flat_4})\,,
\end{split}
\label{eq:relxlxr4}
\end{align}
where $W^{\sharp_4}$ and $W^{\flat_4}$ are defined respectively by
\begin{equation}
a \to (a{+}1,\,a{+}2,\,a{+}3)
\end{equation}
and
\begin{equation}
a \to (a{-}3,\,a{-}2,\,a{-}1)\,.
\end{equation}
The composition of $\sharp_4$ and $\flat_4$ is again equivalent to the identity transformation, up to overall factors of frozen variables.  For $k=4$ we again have
\begin{equation}\label{eq:refl4}
\textbf{Ref}_s(X_L(W)) = X_R(\textbf{Ref}(W))
\end{equation}
(where $\textbf{Ref}_s$ acts in the obvious way on indices of Mandelstam variables) and hence
\begin{equation}\label{eq:refpc4}
\textbf{Ref}_s(X_L(W)) = X_R(\textbf{Ref}(W)) = X_L(\textbf{Ref}(W)^{\flat_4}) = X_L(\textbf{Ref}(W^{\sharp_4}))\,.
\end{equation}

\subsection{Kinematic Length}
\label{sec:kinlength}

It is intuitively clear that ``more complicated'' webs have ``more complicated'' invariants, and are assigned by the $X$-map to ''more complicated'' kinematic functions.  In this section we formalize this notion of complexity in a way that will play an important role in Sec.~\ref{sec:results}.  To that end we first define two bases of the generalized kinematic space $\mathcal{K}_{k,n}$ (different from the ABHY basis constructed in Sec.~\ref{sec:kinematicspace}).  The \emph{left kinematic basis} is the set $\{X_L(p) : p \text{ is a non-frozen Pl\"ucker coordinate}\}$, and the \emph{right kinematic basis} is defined analogously using $X_R$.  Note that each set contains $\binom{n}{k}-k$ elements, the same as the dimension of $\mathcal{K}_{k,n}$, and one can check that they are linearly independent in $\mathcal{K}_{k,n}$, so each is indeed a basis.

We define the \emph{left (right) kinematic length} of a kinematic function $F$ as the sum of the coefficients of $F$ when expressed in the left (right) kinematic basis.  Next, we define the \emph{cluster length} of a web invariant $[W]$ to be equal to $1/k$ times the number of external legs of the associated web; this is the same as the degree of $[W]$ when expressed as a polynomial in $\Gr(k,n)$ Pl\"ucker coordinates.

Now the notion that more complicated invariants are associated to more complicated kinematic functions is formalized in the statement that the left (or right) kinematic length of $X_L([W])$ (or $X_R([W])$) is equal to the cluster length of $[W]$.  It is easy to verify that this statement is true by recursion.  If $p$ is a non-frozen Pl\"ucker coordinate then by definition its cluster length is 1 and the left (or right) kinematic length of $X_L(p)$ (or $X_R(p)$) is 1.  For more complicated invariants, we note that in all of the recursive definitions~(\ref{eq:sl3vector}), (\ref{eq:sl3covector}), (\ref{eq:sl4Xdefone}), (\ref{eq:sl4Xdeftwo}) and~(\ref{eq:sl4Xdefthree}), each side is linear in both kinematic and cluster length; this establishes the equality.

\section{Web Series}
\label{sec:series}

A \emph{web series} $\mathcal{W}$ is a formal power series of webs $W_1,W_2,\ldots$
\begin{align}
\mathcal{W} = 1 + \sum_{m=1}^\infty t^m W_m\,,
\end{align}
to which we associate the invariant
\begin{align}
[\mathcal{W}] = 1 + \sum_{m=1}^\infty t^m [W_m]\,.
\end{align}
We are interested in web series whose invariants are cluster series of the type reviewed in Sec.~\ref{sec:clusteralgebraicseries}.  Once a cluster algebra basis is specified (for example, the one provided by the character formula of~\cite{chang2020quantum}), then each such series is (in principle) completely determined by its first nontrivial term $W_1$.  Therefore, we are interested to study natural ways to associate an entire web series $\mathcal{W}(W)$ to a single web $W$, with $W_1 = W$ and with the higher-order terms $W_2,W_3,\ldots$ being determined from $W$ in some manner.

One simple way to do this is via the ``web thickening'' procedure of~\cite[Definition 10.8]{fomin2016tensor}.  If $W$ is any web and we take $W_m$ to be the combination of $m$ copies of $W$, then $[W_m] = [W]^m$ and the web series invariant is geometric:
\begin{align}
1 + \sum_{m=1}^\infty t^m [W_m] = \frac{1}{1 - t [W]}\,,
\end{align}
just like the series~(\ref{eq:geometric}) for cluster variables.  According to the FP conjectures, $[W]$ is a cluster variable precisely when $W$ is an indecomposable arborizable web, so for such $W$ we define the web series $\mathcal{W}(W)$ by the aforementioned thickening procedure.

However if $W$ is a non-arborizable web we seek a different definition of the web series $\mathcal{W}(W)$ because we want its invariant to not be geometric, but rather to evaluate to more complicated rational functions such as~(\ref{eq:firstseries}) or~(\ref{eq:newseries}).  In Sec.~\ref{sec:slkwebseries} we provide such a definition for a class of indecomposable webs that we call \emph{almost arborizable}---these are non-arborizable webs that can be converted, via skein relations, to tensor diagrams (possibly non-planar) with a single closed inner loop.  (For $k>3$ we further require every edge of that closed inner loop to be a single line.)  We leave for future work the study of web series associated to more complicated non-arborizable webs.  In order to connect to the notation used in Sec.~\ref{sec:clusteralgebraicseries} let us define $A(W) = [W]$, the usual web invariant, and now take a short diversion to define a new type of invariant $B(W)$ that we can associate to certain almost arborizable webs.

\subsection{\texorpdfstring{$\mathfrak{sl}_3$ Almost Arborizable Webs}{sl3 Almost Arborizable Webs}}

Let $W$ be an almost arborizable $\mathfrak{sl}_3$ web and let $D$ be the equivalent tensor diagram with exactly one inner loop.  We first define $B_1(W)$ and $B_2(W)$ as follows:
\begin{itemize}
     \item (a) Starting with $D$, delete all the edges on the loop that go in clockwise order from a white vertex to a black vertex.
     \item (b) Now all vertices originally on the loop are divalent. Delete those vertices, fusing the two edges at each vertex into a single edge.
     \item (c) Now we have a new tensor diagram where all inner vertices are trivalent, and all edges connect a black vertex with a white vertex. Define $B_1(W)$ to be the invariant of this diagram.
     \item (d) Repeat steps (a)--(c) but delete edges that go from white vertices to black vertices in counter-clockwise order.  Define $B_2(W)$ to be the invariant of this diagram.
\end{itemize}
Then define $B(W) = B_1(W) B_2(W)$.  As an example let $W^{(1)}$ be our old friend, Fig.~\ref{fig:nonarb}(a), whose invariant is
\begin{align}
A(W^{(1)}) = \langle 145 \rangle\langle 278 \rangle\langle 369 \rangle - \langle 245 \rangle\langle 178 \rangle\langle 369 \rangle - \langle 123 \rangle\langle 456 \rangle\langle 789\rangle - \langle 129 \rangle\langle 345 \rangle\langle 678 \rangle\,.
\end{align}
To compute $B_1(W^{(1)})$ we delete the edges shown here as dotted red lines, and then remove the divalent vertices, which gives:
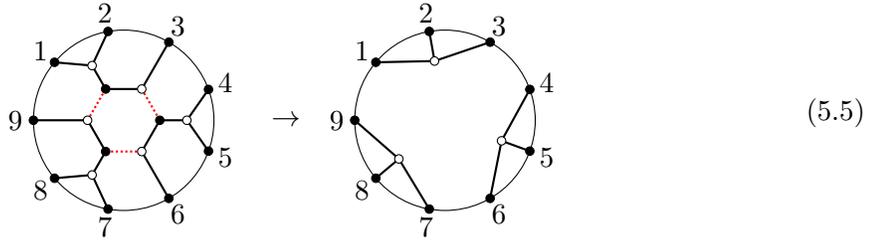
\begin{equation}
     \begin{tikzpicture}[baseline={(current bounding box.center)},scale=0.8,xscale=-1]
          \def \cen{(0,0)};
          \def \rad{1.5};
          \draw {\cen} circle (\rad);
          
          \foreach \x in {1,2,3,4,5,6,7,8,9}{
               \filldraw[fill=black,draw=black] \cen+(40*\x:\rad) circle (2pt);
               \draw[white] \cen+({40* (\x)}:{\rad+0.3}) circle (0.1) node[black]{\tikz{\node at (0,0) {$\x$}}};
               \coordinate(v\x) at (40*\x:1.5);}
          
          \foreach \x in {1,2,3,4,5,6}{
               \coordinate(A\x) at ({60*\x}:{.4*\rad});}
          
          \coordinate(B1) at (60:{.7*\rad});
          \coordinate(B2) at (180:{.7*\rad});
          \coordinate(B3) at (300:{.7*\rad});
          
          \draw[thick] (B1) -- (A1) (B2) -- (A3) (B3) -- (A5);
          \draw[thick] (A2) -- (v3) (A4) -- (v6) (A6) -- (v9);
          
          \draw[thick] (v1) -- (B1) -- (v2) (v4) -- (B2) -- (v5)  (v7) -- (B3) -- (v8);

          \draw[thick] (A1) -- (A2) (A3) -- (A4) (A5) -- (A6);
          \draw[thick,red,densely dotted] (A2) -- (A3) (A4) -- (A5) (A6) -- (A1);

          \foreach \x in {1,3,5}{
               \filldraw[fill=black,draw=black] (A\x) circle (2pt);
               \filldraw[fill=white,draw=black] ({60+60*\x}:{.4*\rad}) circle (2pt);}
          \foreach \x in {1,2,3}{
               \filldraw[fill=white,draw=black] (B\x) circle (2pt);}

          \node at (-2.7,0) {$\rightarrow$};
     \end{tikzpicture}
     \begin{tikzpicture}[baseline={(current bounding box.center)},scale=0.8,xscale=-1]
          \def \cen{(0,0)};
          \def \rad{1.5};
          \draw {\cen} circle (\rad);
          
          \foreach \x in {1,2,3,4,5,6,7,8,9}{
               \filldraw[fill=black,draw=black] \cen+(40*\x:\rad) circle (2pt);
               \draw[white] \cen+({40* (\x)}:{\rad+0.3}) circle (0.1) node[black]{\tikz{\node at (0,0) {$\x$}}};
               \coordinate(v\x) at (40*\x:1.5);}

          \foreach \x in {1,2,3}{
               \coordinate(A\x) at ({40*3*\x-40}:1);
          }

     \draw[thick] (v1) -- (A1) -- (v2) (A1) -- (v3);
     \draw[thick] (v4) -- (A2) -- (v5) (A2) -- (v6);
     \draw[thick] (v7) -- (A3) -- (v8) (A3) -- (v9);

     \foreach \x in {1,2,3}{
          \filldraw[fill=white] (A\x) circle (2pt);
     }

     \end{tikzpicture}
\end{equation}
Similarly for $B_2(W^{(1)})$ we look at
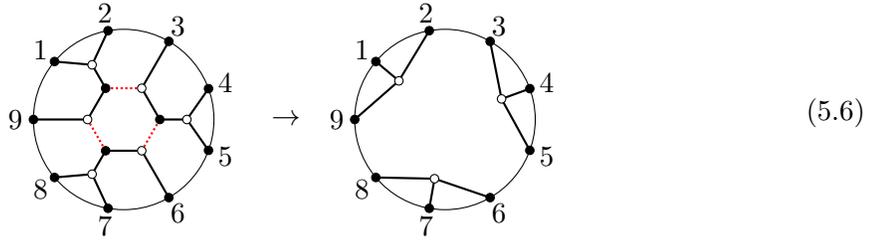
\begin{equation}
     \begin{tikzpicture}[baseline={(current bounding box.center)},scale=0.8,xscale=-1]
          \def \cen{(0,0)};
          \def \rad{1.5};
          \draw {\cen} circle (\rad);
          
          \foreach \x in {1,2,3,4,5,6,7,8,9}{
               \filldraw[fill=black,draw=black] \cen+(40*\x:\rad) circle (2pt);
               \draw[white] \cen+({40* (\x)}:{\rad+0.3}) circle (0.1) node[black]{\tikz{\node at (0,0) {$\x$}}};
               \coordinate(v\x) at (40*\x:1.5);}
          
          \foreach \x in {1,2,3,4,5,6}{
               \coordinate(A\x) at ({60*\x}:{.4*\rad});}
          
          \coordinate(B1) at (60:{.7*\rad});
          \coordinate(B2) at (180:{.7*\rad});
          \coordinate(B3) at (300:{.7*\rad});
          
          \draw[thick] (B1) -- (A1) (B2) -- (A3) (B3) -- (A5);
          \draw[thick] (A2) -- (v3) (A4) -- (v6) (A6) -- (v9);
          
          \draw[thick] (v1) -- (B1) -- (v2) (v4) -- (B2) -- (v5)  (v7) -- (B3) -- (v8);

          \draw[thick,red,densely dotted] (A2) -- (A1) (A4) -- (A3) (A6) -- (A5);
          \draw[thick] (A2) -- (A3) (A4) -- (A5) (A6) -- (A1);

          \foreach \x in {1,3,5}{
               \filldraw[fill=black,draw=black] (A\x) circle (2pt);
               \filldraw[fill=white,draw=black] ({60+60*\x}:{.4*\rad}) circle (2pt);}
          \foreach \x in {1,2,3}{
               \filldraw[fill=white,draw=black] (B\x) circle (2pt);}

          \node at (-2.7,0) {$\rightarrow$};
     \end{tikzpicture}
     \begin{tikzpicture}[baseline={(current bounding box.center)},scale=0.8,xscale=-1]
          \def \cen{(0,0)};
          \def \rad{1.5};
          \draw {\cen} circle (\rad); 
          
          \foreach \x in {1,2,3,4,5,6,7,8,9}{
               \filldraw[fill=black,draw=black] \cen+(40*\x:\rad) circle (2pt);
               \draw[white] \cen+({40* (\x)}:{\rad+0.3}) circle (0.1) node[black]{\tikz{\node at (0,0) {$\x$}}};
               \coordinate(v\x) at (40*\x:1.5);}

          \foreach \x in {1,2,3}{
               \coordinate(A\x) at ({40*3*\x+40}:1);
          }

     \draw[thick] (v4) -- (A1) -- (v5) (A1) -- (v3);
     \draw[thick] (v7) -- (A2) -- (v8) (A2) -- (v6);
     \draw[thick] (v1) -- (A3) -- (v2) (A3) -- (v9);

     \foreach \x in {1,2,3}{
          \filldraw[fill=white] (A\x) circle (2pt);
     }
     
     \end{tikzpicture}
\end{equation}
and so we can read off
\begin{align}
B(W^{(1)}) = \langle 123 \rangle\langle 456 \rangle \langle 789 \rangle \langle 345 \rangle \langle 678 \rangle \langle 129 \rangle \,.
\end{align}

To see a less trivial example, we move on to $n=10$ and consider the web $W^{(2)}$ given by
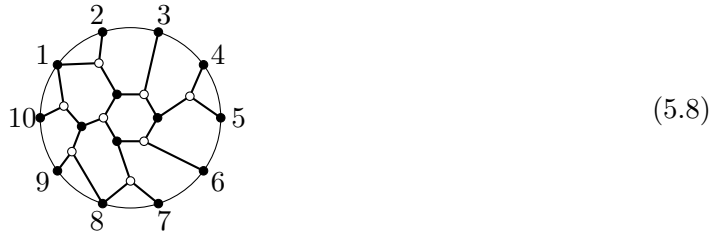
\begin{equation}
     \begin{tikzpicture}[baseline={(current bounding box.center)},scale=0.8,xscale=-1]
          \def \cen{(0,0)};
          \def \mov{{-.2,0}};
          \def \rad{1.5};
          \draw {\cen} circle (\rad);
          
          \foreach \x in {1,2,3,4,5,6,7,8,9,10}{
               \filldraw[fill=black,draw=black] \cen+(36*\x:\rad) circle (2pt);
               \draw[white] \cen+({36* (\x)}:{\rad+0.3}) circle (0.1) node[black]{\tikz{\node at (0,0) {$\x$}}};
               \coordinate(v\x) at (36*\x:1.5);}
          
          \foreach \x in {1,2,3,4,5,6}{
               \coordinate(A\x) at  ({60*\x}:{.3*\rad});
               \draw[thick] ({60*\x}:{.3*\rad}) -- ({60+60*\x}:{.3*\rad});}
          
          \coordinate(B1) at (60:{.7*\rad});
          \coordinate(B2) at (160:{.7*\rad});
          \coordinate(B3) at (270:{.7*\rad});
          \coordinate(B4) at (330:{.75*\rad});
          \coordinate(B5) at (10:{.75*\rad});
          \coordinate(C1) at (350:{.55*\rad});
          
          \draw[thick] (B1) -- (A1) (B2) -- (A3) (B3) -- (A5);
          \draw[thick] (A2) -- (v3) (A4) -- (v6);
          \draw[thick] (v8) -- (B4) -- (v9) (v10) -- (B5) -- (v1) (B4) -- (C1) -- (B5) (C1) -- (A6);
          
          \draw[thick] (v1) -- (B1) -- (v2) (v4) -- (B2) -- (v5)  (v7) -- (B3) -- (v8);

          \foreach \x in {1,3,5}{
               \filldraw[fill=black,draw=black] (A\x) circle (2pt);
               \filldraw[fill=white,draw=black] ({60+60*\x}:{.3*\rad}) circle (2pt);}
          \foreach \x in {1,2,3,4,5}
               \filldraw[fill=white,draw=black] (B\x) circle (2pt);
          \filldraw[fill=black,draw=black] (C1) circle (2pt);
     \end{tikzpicture}
\end{equation}
whose invariant is
\begin{multline}
A(W^{(2)}) = \langle 3\times6, 8\times9, 10\times 1\rangle \langle 1\times2, 4\times5, 7\times8 \rangle - \langle 1,2,3 \rangle\langle 4,5,6 \rangle\langle 7,8,9 \rangle\langle 1,8,10 \rangle\\
     - \langle 3,4,5 \rangle\langle 6,7,8 \rangle\langle 1,8,9 \rangle \langle 1,2,10 \rangle\,.
     \end{multline}
To compute $B_1(W^{(2)})$ we look at
\begin{equation}
     \begin{tikzpicture}[baseline={(current bounding box.center)},scale=0.8,xscale=-1]
          \def \cen{(0,0)};
          \def \mov{{-.2,0}};
          \def \rad{1.5};
          \draw {\cen} circle (\rad); 
          
          \foreach \x in {1,2,3,4,5,6,7,8,9,10}{
               \filldraw[fill=black,draw=black] \cen+(36*\x:\rad) circle (2pt);
               \draw[white] \cen+({36* (\x)}:{\rad+0.3}) circle (0.1) node[black]{\tikz{\node at (0,0) {$\x$}}};
               \coordinate(v\x) at (36*\x:1.5);}
          
          \foreach \x in {1,2,3,4,5,6}{
               \coordinate(A\x) at  ({60*\x}:{.3*\rad});}
          
          \coordinate(B1) at (60:{.7*\rad});
          \coordinate(B2) at (160:{.7*\rad});
          \coordinate(B3) at (270:{.7*\rad});
          \coordinate(B4) at (330:{.75*\rad});
          \coordinate(B5) at (10:{.75*\rad});
          \coordinate(C1) at (350:{.55*\rad});
          
          \draw[thick] (B1) -- (A1) (B2) -- (A3) (B3) -- (A5);
          \draw[thick] (A2) -- (v3) (A4) -- (v6);
          \draw[thick] (v8) -- (B4) -- (v9) (v10) -- (B5) -- (v1) (B4) -- (C1) -- (B5) (C1) -- (A6);
          
          \draw[thick] (v1) -- (B1) -- (v2) (v4) -- (B2) -- (v5)  (v7) -- (B3) -- (v8);

          \draw[thick,red,densely dotted] (A2) -- (A3) (A4) -- (A5) (A6) -- (A1);
          \draw[thick] (A1) -- (A2) (A3) -- (A4) (A5) -- (A6);

          \foreach \x in {1,3,5}{
               \filldraw[fill=black,draw=black] (A\x) circle (2pt);
               \filldraw[fill=white,draw=black] ({60+60*\x}:{.3*\rad}) circle (2pt);}
          \foreach \x in {1,2,3,4,5}
               \filldraw[fill=white,draw=black] (B\x) circle (2pt);
          \filldraw[fill=black,draw=black] (C1) circle (2pt);

          \node at (-2.7,0) {$\rightarrow$};
          
     \end{tikzpicture}
     \begin{tikzpicture}[baseline={(current bounding box.center)},scale=0.8,xscale=-1]
          \def \cen{(0,0)};
          \def \mov{{-.2,0}};
          \def \rad{1.5};
          \draw {\cen} circle (\rad); 
          
          \foreach \x in {1,2,3,4,5,6,7,8,9,10}{
               \filldraw[fill=black,draw=black] \cen+(36*\x:\rad) circle (2pt);
               \draw[white] \cen+({36* (\x)}:{\rad+0.3}) circle (0.1) node[black]{\tikz{\node at (0,0) {$\x$}}};
               \coordinate(v\x) at (36*\x:1.5);}

          \coordinate(B1) at (72:{.6*\rad});
          \coordinate(B2) at (180:{.6*\rad});
          \coordinate(B3) at (270:{.6*\rad});
          \coordinate(B4) at (306:{.6*\rad});
          \coordinate(B5) at (370:{.5*\rad});
          \coordinate(C1) at (320:{.25*\rad});

          \draw[thick] (v8) -- (B4) -- (v9) (v10) -- (B5) -- (v1) (B4) -- (C1) -- (B5);
          
          \draw[thick] (v1) -- (B1) -- (v2) (v4) -- (B2) -- (v5)  (v7) -- (B3) -- (v8);

          \draw[thick] (B1) -- (v3) (B2) -- (v6) (B3) -- (C1);

          \foreach \x in {1,2,3,4,5}
               \filldraw[fill=white,draw=black] (B\x) circle (2pt);
          \filldraw[fill=black,draw=black] (C1) circle (2pt);

          \node at (-3.5,0) {$\xlongequal[]{\text{skein relation}}$};
     \end{tikzpicture}
     \begin{tikzpicture}[baseline={(current bounding box.center)},scale=0.8,xscale=-1]
          \def \cen{(0,0)};
          \def \mov{{-.2,0}};
          \def \rad{1.5};
          \draw {\cen} circle (\rad); 
          
          \foreach \x in {1,2,3,4,5,6,7,8,9,10}{
               \filldraw[fill=black,draw=black] \cen+(36*\x:\rad) circle (2pt);
               \draw[white] \cen+({36* (\x)}:{\rad+0.3}) circle (0.1) node[black]{\tikz{\node at (0,0) {$\x$}}};
               \coordinate(v\x) at (36*\x:1.5);}

          \coordinate(B1) at (72:{.6*\rad});
          \coordinate(B2) at (180:{.6*\rad});
          \coordinate(B3) at (270:{.6*\rad});
          \coordinate(B4) at (306:{.6*\rad});
          \coordinate(B5) at (370:{.5*\rad});
          \coordinate(C1) at (300:{.4*\rad});

          \draw[thick] (v10) -- (B5) -- (v1) (B5) -- (v8);

          \draw[thick] (v1) -- (B1) -- (v2) (v4) -- (B2) -- (v5);

          \draw[thick] (v7) -- (C1) -- (v8) (C1) -- (v9);

          \draw[thick] (B1) -- (v3) (B2) -- (v6);

          \foreach \x in {1,2,5}
               \filldraw[fill=white,draw=black] (B\x) circle (2pt);
          \filldraw[fill=white,draw=black] (C1) circle (2pt);

     \end{tikzpicture}
\end{equation}
and to compute $B_2(W^{(2)})$ we look at
\begin{equation}
     \begin{tikzpicture}[baseline={(current bounding box.center)},scale=0.8,xscale=-1]
          \def \cen{(0,0)};
          \def \mov{{-.2,0}};
          \def \rad{1.5};
          \draw {\cen} circle (\rad);
          
          \foreach \x in {1,2,3,4,5,6,7,8,9,10}{
               \filldraw[fill=black,draw=black] \cen+(36*\x:\rad) circle (2pt);
               \draw[white] \cen+({36* (\x)}:{\rad+0.3}) circle (0.1) node[black]{\tikz{\node at (0,0) {$\x$}}};
               \coordinate(v\x) at (36*\x:1.5);}
          
          \foreach \x in {1,2,3,4,5,6}{
               \coordinate(A\x) at  ({60*\x}:{.3*\rad});}
          
          \coordinate(B1) at (60:{.7*\rad});
          \coordinate(B2) at (160:{.7*\rad});
          \coordinate(B3) at (270:{.7*\rad});
          \coordinate(B4) at (330:{.75*\rad});
          \coordinate(B5) at (10:{.75*\rad});
          \coordinate(C1) at (350:{.55*\rad});
          
          \draw[thick] (B1) -- (A1) (B2) -- (A3) (B3) -- (A5);
          \draw[thick] (A2) -- (v3) (A4) -- (v6);
          \draw[thick] (v8) -- (B4) -- (v9) (v10) -- (B5) -- (v1) (B4) -- (C1) -- (B5) (C1) -- (A6);
          
          \draw[thick] (v1) -- (B1) -- (v2) (v4) -- (B2) -- (v5)  (v7) -- (B3) -- (v8);

          \draw[thick] (A2) -- (A3) (A4) -- (A5) (A6) -- (A1);
          \draw[thick,red,densely dotted] (A1) -- (A2) (A3) -- (A4) (A5) -- (A6);
          
          \foreach \x in {1,3,5}{
               \filldraw[fill=black,draw=black] (A\x) circle (2pt);
               \filldraw[fill=white,draw=black] ({60+60*\x}:{.3*\rad}) circle (2pt);}
          \foreach \x in {1,2,3,4,5}
               \filldraw[fill=white,draw=black] (B\x) circle (2pt);
          \filldraw[fill=black,draw=black] (C1) circle (2pt);

          \node at (-2.7,0) {$\rightarrow$};
     \end{tikzpicture}
     \begin{tikzpicture}[baseline={(current bounding box.center)},scale=0.8,xscale=-1]
          \def \cen{(0,0)};
          \def \mov{{-.2,0}};
          \def \rad{1.5};
          \draw {\cen} circle (\rad); 
          
          \foreach \x in {1,2,3,4,5,6,7,8,9,10}{
               \filldraw[fill=black,draw=black] \cen+(36*\x:\rad) circle (2pt);
               \draw[white] \cen+({36* (\x)}:{\rad+0.3}) circle (0.1) node[black]{\tikz{\node at (0,0) {$\x$}}};
               \coordinate(v\x) at (36*\x:1.5);}

          \coordinate(B1) at (54:{.5*\rad});
          \coordinate(B2) at (144:{.6*\rad});
          \coordinate(B3) at ({36*7}:{.6*\rad});
          \coordinate(B4) at (306:{.6*\rad});
          \coordinate(B5) at (370:{.5*\rad});
          \coordinate(C1) at (320:{.25*\rad});

          \draw[thick] (v8) -- (B4) -- (v9) (v10) -- (B5) -- (v1) (B4) -- (C1) -- (B5);
          
          \draw[thick] (v1) -- (B1) -- (v2) (v4) -- (B2) -- (v5)  (v7) -- (B3) -- (v8);

          \draw[thick] (B1) -- (C1) (B2) -- (v3) (B3) -- (v6);

          \foreach \x in {1,2,3,4,5}
               \filldraw[fill=white,draw=black] (B\x) circle (2pt);
          \filldraw[fill=black,draw=black] (C1) circle (2pt);

          \node at (-3.5,0) {$\xlongequal[]{\text{skein relation}}$};
     \end{tikzpicture}
     \begin{tikzpicture}[baseline={(current bounding box.center)},scale=0.8,xscale=-1]
          \def \cen{(0,0)};
          \def \mov{{-.2,0}};
          \def \rad{1.5};
          \draw {\cen} circle (\rad); 
          
          \foreach \x in {1,2,3,4,5,6,7,8,9,10}{
               \filldraw[fill=black,draw=black] \cen+(36*\x:\rad) circle (2pt);
               \draw[white] \cen+({36* (\x)}:{\rad+0.3}) circle (0.1) node[black]{\tikz{\node at (0,0) {$\x$}}};
               \coordinate(v\x) at (36*\x:1.5);}
          \coordinate(B1) at (54:{.5*\rad});
          \coordinate(B2) at (144:{.6*\rad});
          \coordinate(B3) at ({36*7}:{.6*\rad});
          \coordinate(B4) at (310:{.5*\rad});
          \coordinate(B5) at (370:{.5*\rad});
          \coordinate(C1) at (36:{.4*\rad});

          \draw[thick] (v8) -- (B4) -- (v9) (v10) -- (C1) -- (v1) (C1) -- (v2) (B4) -- (v1);
          
          \draw[thick] (v4) -- (B2) -- (v5)  (v7) -- (B3) -- (v8);

          \draw[thick] (B2) -- (v3) (B3) -- (v6);

          \foreach \x in {2,3,4}
               \filldraw[fill=white,draw=black] (B\x) circle (2pt);
          \filldraw[fill=white,draw=black] (C1) circle (2pt);

     \end{tikzpicture}
\end{equation}
from which we read off
\begin{align}
B(W^{(2)}) = \langle 1,2,3\rangle \langle 4,5,6 \rangle \langle 7,8,9\rangle \langle 1,8,10\rangle \langle 1,2,10\rangle \langle 3,4,5\rangle \langle 6,7,8\rangle \langle 1,8,9\rangle\,.
\end{align}

\subsection{\texorpdfstring{$\mathfrak{sl}_4$ Almost Arborizable Webs}{sl4 Almost Arborizable Webs}}
\label{sec:sl4aawebs}

Let $W$ be an almost arborizable $\mathfrak{sl}_4$ web and let $D$ be the equivalent tensor diagram with exactly one inner loop.  As mentioned above, require that all edges on the inner loop of $D$ are single lines; this is sufficient to cover all webs we encounter in Sec.~\ref{sec:results}.  In such cases we define $B_1(W)$ and $B_2(W)$ as follows:
\begin{itemize}
\item (a) Starting with $D$, delete all the edges on the loop that go in clockwise order from a white vertex to a black vertex.
\item (b) If there were initially $e$ edges on the inner loop, then there are now $e/2$ pairs of connected trivalent vertices where the loop used to be.  For each pair of vertices we add another edge, giving altogether $e/2$ 2-cycles, and then multiply each new 2-cycle by a factor of $1/2$.
\item (c) Then use the skein relations to cancel each 2-cycle, which removes the factors of 1/2 introduced in the previous step.
\item (d) At this stage we have a valid $\mathfrak{sl}_4$ tensor diagram (all inner vertices are quadrivalent, and all edges connect a black vertex to a white vertex).  Define $B_1$ to be the invariant of this diagram.
\item (e) Repeat steps (a)--(d) but delete edges that go from white vertices to black vertices in counter-clockwise order.  Define $B_2(W)$ to be the invariant obtained in this way.
\end{itemize}
Then define $B(W) = B_1(W) B_2(W)$.  To illustrate these definitions let $W^{(3)}$ be our other friend, Fig.~\ref{fig:nonarb}(b), whose invariant is
\begin{align}
A(W^{(3)}) = \langle 1\,2\,5\,6 \rangle \langle 3\,4\,7\,8\rangle - \langle 1\,2\,7\,8 \rangle \langle 3\,4\,5\,6 \rangle - \langle 1\,2\,3\,4 \rangle \langle 5\,6\,7\,8 \rangle\,.
\label{eq:A}
\end{align}
Then to compute $B_1$ we look at
\begin{equation}
     \begin{tikzpicture}[baseline={(current bounding box.center)},xscale=-1,scale=.8]
          \def \cen{(0,0)};
          \def \rad{1.5};
          \draw {\cen} circle (\rad);

          \foreach \x in {1,2,3,4,5,6,7,8}{
               \draw[white] \cen+({45* (\x)}:{\rad+0.4}) circle (0.1) node[black]{\tikz{\node at (0,0) {$\x$}}};
               \coordinate(v\x) at (45*\x:1.5);}

          \foreach \x in {1,2,3,4}
               \coordinate(A\x) at ({135/2-90+90*\x}:{\rad*0.7});

          \foreach \x in {1,2}{
               \coordinate(B\x) at ({135/2-90+180*\x}:{\rad*0.3});
          }

          \draw[thick,red,densely dotted] (A1) -- (B1) (A3) -- (B2);
          \draw[thick] (A3) -- (B1) (A1) -- (B2);
          \draw[thick] (v1) -- (A1) -- (v2) (v3) -- (A2) -- (v4) (v5) -- (A3) -- (v6) (v7) -- (A4) -- (v8);
          \foreach \x in {135/2+90,135/2-90}{
               \draw[thick] ({\x+6}:{\rad*0.3}) -- ({\x+2}:{\rad*0.7});
               \draw[thick] ({\x-6}:{\rad*0.3}) -- ({\x-2}:{\rad*0.7});}

          \foreach \x in {1,2,3,4}
               \filldraw[fill=white] (A\x) circle (2pt);

          \foreach \x in {1,2,3,4,5,6,7,8}{
               \filldraw[fill=black,draw=black] \cen+(45*\x:\rad) circle (2pt);}
          
          \foreach \x in {1,2}{
               \filldraw (B\x) circle (2pt);
          }

          \node at (-2.7,0) {$\rightarrow$};
     \end{tikzpicture}
     \begin{tikzpicture}[baseline={(current bounding box.center)},xscale=-1,scale=.8]
          \def \cen{(0,0)};
          \def \rad{1.5};
          \draw {\cen} circle (\rad); 

          \foreach \x in {1,2,3,4,5,6,7,8}{
               \draw[white] \cen+({45* (\x)}:{\rad+0.4}) circle (0.1) node[black]{\tikz{\node at (0,0) {$\x$}}};
               \coordinate(v\x) at (45*\x:1.5);}

          \foreach \x in {1,2,3,4}
               \coordinate(A\x) at ({135/2-90+90*\x}:{\rad*0.7});

          \foreach \x in {1,2}{
               \coordinate(B\x) at ({135/2-90+180*\x}:{\rad*0.3});
          }
          \coordinate(C1) at ({135/2+135+10}:{\rad*0.5});
          \coordinate(C2) at ({135/2+135+180+10}:{\rad*0.5});
          \coordinate(D1) at ({135/2+135+30+5}:{\rad*0.25});
          \coordinate(D2) at ({135/2+135+180+30+5}:{\rad*0.25});

          \draw[thick,red] (A3) .. controls (C1) .. (B1);
          \draw[thick,red] (A1) .. controls (C2) .. (B2);
          \draw[thick,red] (A3) .. controls (D1) .. (B1);
          \draw[thick,red] (A1) .. controls (D2) .. (B2);
          \draw[thick] (v1) -- (A1) -- (v2) (v3) -- (A2) -- (v4) (v5) -- (A3) -- (v6) (v7) -- (A4) -- (v8);
          \foreach \x in {135/2+90,135/2-90}{
               \draw[thick] ({\x+6}:{\rad*0.3}) -- ({\x+2}:{\rad*0.7});
               \draw[thick] ({\x-6}:{\rad*0.3}) -- ({\x-2}:{\rad*0.7});}

          \foreach \x in {1,2,3,4}
               \filldraw[fill=white] (A\x) circle (2pt); 

          \foreach \x in {1,2,3,4,5,6,7,8}{
               \filldraw[fill=black,draw=black] \cen+(45*\x:\rad) circle (2pt);}
          
          \foreach \x in {1,2}{
               \filldraw (B\x) circle (2pt);
          }

          \draw[thick] (-2,2) -- (-2.2,2) -- (-2.2,-2) -- (-2,-2);
          \draw[thick] (2,2) -- (2.2,2) -- (2.2,-2) -- (2,-2);

          \node at (-3.7,0) {$\xlongequal[]{\text{skein relation}}$};
          \node at (2.7,0) {$\dfrac{1}{4} *$}; 
     \end{tikzpicture}
     \begin{tikzpicture}[baseline={(current bounding box.center)},xscale=-1,scale=.8]
          \def \cen{(0,0)};
          \def \rad{1.5};
          \draw {\cen} circle (\rad); 

          \foreach \x in {1,2,3,4,5,6,7,8}{
               \draw[white] \cen+({45* (\x)}:{\rad+0.4}) circle (0.1) node[black]{\tikz{\node at (0,0) {$\x$}}};
               \coordinate(v\x) at (45*\x:1.5);}

          \coordinate(A1) at ({45* (0.5)}:{0.7});
          \coordinate(A2) at ({45*(0.5)+180}:{0.7});

          \draw[thick] (v1) -- (A1) -- (v2) (v7) -- (A1) -- (v8);
          \draw[thick] (v5) -- (A2) -- (v6) (v3) -- (A2) -- (v4);

          \foreach \x in {1,2}
               \filldraw[fill=white] (A\x) circle (2pt); 

          \foreach \x in {1,2,3,4,5,6,7,8}{
               \filldraw[fill=black,draw=black] \cen+(45*\x:\rad) circle (2pt);}
          
     \end{tikzpicture}
\end{equation}
and to compute $B_2$ we look at
\begin{equation}
     \begin{tikzpicture}[baseline={(current bounding box.center)},xscale=-1,scale=.8]
          \def \cen{(0,0)};
          \def \rad{1.5};
          \draw {\cen} circle (\rad); 

          \foreach \x in {1,2,3,4,5,6,7,8}{
               \draw[white] \cen+({45* (\x)}:{\rad+0.4}) circle (0.1) node[black]{\tikz{\node at (0,0) {$\x$}}};
               \coordinate(v\x) at (45*\x:1.5);}

          \foreach \x in {1,2,3,4}
               \coordinate(A\x) at ({135/2-90+90*\x}:{\rad*0.7});

          \foreach \x in {1,2}{
               \coordinate(B\x) at ({135/2-90+180*\x}:{\rad*0.3});
          }

          \draw[thick,red,densely dotted] (A1) -- (B2) (A3) -- (B1);
          \draw[thick] (A3) -- (B2) (A1) -- (B1);
          \draw[thick] (v1) -- (A1) -- (v2) (v3) -- (A2) -- (v4) (v5) -- (A3) -- (v6) (v7) -- (A4) -- (v8);
          \foreach \x in {135/2+90,135/2-90}{
               \draw[thick] ({\x+6}:{\rad*0.3}) -- ({\x+2}:{\rad*0.7});
               \draw[thick] ({\x-6}:{\rad*0.3}) -- ({\x-2}:{\rad*0.7});}

          \foreach \x in {1,2,3,4}
               \filldraw[fill=white] (A\x) circle (2pt);

          \foreach \x in {1,2,3,4,5,6,7,8}{
               \filldraw[fill=black,draw=black] \cen+(45*\x:\rad) circle (2pt);}
          
          \foreach \x in {1,2}{
               \filldraw (B\x) circle (2pt);
          }

          \node at (-2.7,0) {$\rightarrow$};
     \end{tikzpicture}
     \begin{tikzpicture}[baseline={(current bounding box.center)},xscale=-1,scale=.8]
          \def \cen{(0,0)};
          \def \rad{1.5};
          \draw {\cen} circle (\rad);

          \foreach \x in {1,2,3,4,5,6,7,8}{
               \draw[white] \cen+({45* (\x)}:{\rad+0.4}) circle (0.1) node[black]{\tikz{\node at (0,0) {$\x$}}};
               \coordinate(v\x) at (45*\x:1.5);}

          \foreach \x in {1,2,3,4}
               \coordinate(A\x) at ({135/2-90+90*\x}:{\rad*0.7});

          \foreach \x in {1,2}{
               \coordinate(B\x) at ({135/2-90+180*\x}:{\rad*0.3});
          }
          \coordinate(C1) at ({180-135/2-135-10-45}:{\rad*0.5});
          \coordinate(C2) at ({180-135/2-135-180-10-45}:{\rad*0.5});
          \coordinate(D1) at ({180-135/2-135-30-5-45}:{\rad*0.25});
          \coordinate(D2) at ({180-135/2-135-180-30-5-45}:{\rad*0.25});

          \draw[thick,red] (A1) .. controls (C2) .. (B1);
          \draw[thick,red] (A3) .. controls (C1) .. (B2);
          \draw[thick,red] (A1) .. controls (D2) .. (B1);
          \draw[thick,red] (A3) .. controls (D1) .. (B2);
          \draw[thick] (v1) -- (A1) -- (v2) (v3) -- (A2) -- (v4) (v5) -- (A3) -- (v6) (v7) -- (A4) -- (v8);
          \foreach \x in {135/2+90,135/2-90}{
               \draw[thick] ({\x+6}:{\rad*0.3}) -- ({\x+2}:{\rad*0.7});
               \draw[thick] ({\x-6}:{\rad*0.3}) -- ({\x-2}:{\rad*0.7});}

          \foreach \x in {1,2,3,4}
               \filldraw[fill=white] (A\x) circle (2pt);

          \foreach \x in {1,2,3,4,5,6,7,8}{
               \filldraw[fill=black,draw=black] \cen+(45*\x:\rad) circle (2pt);}
          
          \foreach \x in {1,2}{
               \filldraw (B\x) circle (2pt);
          }

          \draw[thick] (-2,2) -- (-2.2,2) -- (-2.2,-2) -- (-2,-2);
          \draw[thick] (2,2) -- (2.2,2) -- (2.2,-2) -- (2,-2);

          \node at (-3.7,0) {$\xlongequal[]{\text{skein relation}}$};
          \node at (2.7,0) {$\dfrac{1}{4} *$}; 
     \end{tikzpicture}
     \begin{tikzpicture}[baseline={(current bounding box.center)},xscale=-1,scale=.8]
          \def \cen{(0,0)};
          \def \rad{1.5};
          \draw {\cen} circle (\rad);

          \foreach \x in {1,2,3,4,5,6,7,8}{
               \draw[white] \cen+({45* (\x)}:{\rad+0.4}) circle (0.1) node[black]{\tikz{\node at (0,0) {$\x$}}};
               \coordinate(v\x) at (45*\x:1.5);}

          \coordinate(A1) at ({45* (2.5)}:{0.7});
          \coordinate(A2) at ({45*(2.5)+180}:{0.7});

          \draw[thick] (v1) -- (A1) -- (v2) (v3) -- (A1) -- (v4);
          \draw[thick] (v5) -- (A2) -- (v6) (v7) -- (A2) -- (v8);

          \foreach \x in {1,2}
               \filldraw[fill=white] (A\x) circle (2pt);

          \foreach \x in {1,2,3,4,5,6,7,8}{
               \filldraw[fill=black,draw=black] \cen+(45*\x:\rad) circle (2pt);}

     \end{tikzpicture}
\end{equation}
from which we see that
\begin{align}
B(W^{(3)}) &= \langle 1\,2\,3\,4\rangle \langle 5\,6\,7\,8 \rangle \langle 1\,2\,7\,8\rangle \langle 3\,4\,5\,6 \rangle\,.
\label{eq:B}
\end{align}
Note that~(\ref{eq:A}) and~(\ref{eq:B}) agree with~(\ref{eq:ABforfirstray}).

\subsection{\texorpdfstring{A Web Series for Almost Arborizable Webs}{A Web Series for Almost Arborizable Webs}}
\label{sec:slkwebseries}

We now define a web series $\mathcal{W}(W)$ associated to every almost-arborizable web $W$ via a slight modification of the thickening procedure of~\cite{fomin2016tensor}.  Let $W$ be a given almost-arborizable web and let $D$ be a tensor diagram with a single inner loop such that $[D]=[W]$.  To define the ${\cal{O}}(t^2)$ term $W_{(2)}$ in the web series we first draw a combination of two copies of $D$ and then connect the two inner loops by twisting any pair of edges.  For example, if the inner loop is a hexagon then we take
\begin{equation}\label{eq:nontrivialcopy}
     \begin{tikzpicture}[scale=0.7,baseline={(current bounding box.center)}]
          \def \rads{0.85};
          \def \radb{1.5};

          \foreach \x in {1,2,3,4,5,6}{
               \coordinate (v\x) at ({60*\x}:\rads);
               \coordinate (u\x) at ({60*\x}:\radb);

               \draw[thick] ({60*\x}:\rads) -- ({60*(\x+1)}:\rads);
               \draw[thick] ({60*\x}:\radb) -- ({60*(\x+1)}:\radb);
          }

          \draw[very thick,blue] (v1) -- (v6) (u1) -- (u6);
          \foreach \x in {1,3,5}{
               \filldraw[fill=black] (v\x) circle (2pt);
               \filldraw[fill=black] (u\x) circle (2pt);
          }
          \foreach \x in {2,4,6}{
               \filldraw[fill=white] (v\x) circle (2pt);
               \filldraw[fill=white] (u\x) circle (2pt);
          }

          \node at ({\rads+\radb},0) {$\Longrightarrow\quad$};
     \end{tikzpicture}
     \begin{tikzpicture}[scale=0.7,baseline={(current bounding box.center)}]
          \def \rads{0.85};
          \def \radb{1.5};

          \foreach \x in {1,2,3,4,5,6}{
               \coordinate (v\x) at ({60*\x}:\rads);
               \coordinate (u\x) at ({60*\x}:\radb);
          }

          \draw[thick] (v1) -- (v2) -- (v3) -- (v4) -- (v5) -- (v6) (u1) -- (u2) -- (u3) -- (u4) -- (u5) -- (u6);
          \draw[very thick,blue] (v1) -- (u6) (u1) -- (v6);
          \foreach \x in {1,3,5}{
               \filldraw[fill=black] (v\x) circle (2pt);
               \filldraw[fill=black] (u\x) circle (2pt);
          }
          \foreach \x in {2,4,6}{
               \filldraw[fill=white] (v\x) circle (2pt);
               \filldraw[fill=white] (u\x) circle (2pt);
          }
     \end{tikzpicture}
\end{equation}
where we suppress the rest of the diagram, showing only the internal loop.  The generalization is clear:  $W_{(m)}$ is defined by combining $m$ copies of $W$, cutting one identical edge on each of the $m$ inner loops, and gluing them back together after a (cyclic) ``shift-by-one'' permutation.  This was called the bracelet operation in~\cite{lamberti2020tensor}, where web series and invariants constructed in this way were studied for $\mathfrak{sl}_3$.

Using skein relations and the definitions given in the previous two subsections, it is easy to see that
\vspace*{-0.021cm}
\begin{align}
\begin{split}
[W_{(2)}] &= A(W)^2 - 2 \, B(W)\,,\\
[W_{(m)}] &= A(W) \, [W_{(m-1)}] - B(W) \, [W_{(m-2)}] \qquad  m > 2\,,
\end{split}
\label{eq:seriesrecursion}
\end{align}
where $A(W) = [W]$ is the web invariant we start with and
$B(W) = B_1(W) B_2(W)$.  Thanks to~(\ref{eq:seriesrecursion}), the invariant of the web series $\mathcal{W}(W)$ can be written in the form of~(\ref{eq:newseries}):
\begin{align}
[\mathcal{W}(W)] = \frac{1 - B(W)\, t^2}{1 - A(W)\,t + B(W)\,t^2}\,.
\label{eq:webseries}
\end{align}

To summarize: we have shown that there is a natural web series $\mathcal{W}(W)$ one can associate to any almost arborizable web $W$, and that the invariant of this series evaluates to~(\ref{eq:webseries}) in terms of the usual web invariant $A(W) = [W]$ and a second quantity $B(W)$ that admits a simple diagrammatic definition.  (The definition of the web series can be considered to include arborizable webs as a special case for which $B(W) = 0$.)

We conjecture that for any almost arborizable $W$, $A(W)$ and $B(W)$ agree with the quantities $A$ and $B$ appearing in the series~(\ref{eq:firstseries}) associated (via the character formula of~\cite{chang2020quantum}) to the ray $\mathbb{R}^+ {\bf y}$, where ${\bf y}$ is the ${\bf g}$-vector of $[W]$.  It is furthermore natural to speculate that more complicated webs (that are not almost arborizable) are associated to series with higher-order polynomials in their denominators.  We leave such questions to future work.

\section{Results and Discussion}
\label{sec:results}

Now that all the pieces are finally in place we detail the application of our algorithm to the polytopes $\mathcal{C}(3, n \le 10)$, $\mathcal{C}^\dagger(3, n \le 10)$, $\mathcal{C}(4, n \le 8)$ and $\mathcal{C}^\dagger(4, n \le 9)$, the definitions of which are reviewed in Sec.~\ref{sec:polytopes}.  All of these have been constructed or analyzed in the literature, to varying degrees, and by various methods, although only the $k=4$ polytopes are of direct relevance to SYM theory.  In particular, the cluster variables associated to $\mathcal{C}(4,8)$ and $\mathcal{C}^\dagger(4,8)$ were determined in~\cite{Drummond:2019cxm,Arkani-Hamed:2019rds,Henke:2019hve} and the cluster series associated to them were determined in~\cite{Arkani-Hamed:2019rds}.  These prior results serve as checks on the correctness of our methods; however our results for $\mathcal{C}(3,10)$, $\mathcal{C}^\dagger(3,10)$ and $\mathcal{C}^\dagger(4,9)$ are genuinely new\footnote{We thank N.~Henke for independently corroborating our results for $\mathcal{C}^\dagger(4,9)$; see~\cite{HenkeToAppear}.}.

For each polytope $\mathcal{P}$ on the above list, our algorithm (summarized in Fig.~\ref{fig:outline}) proceeds as follows.  Let $i=1,2,\ldots$ index the facets of $\mathcal{P}$, with ${\bf y}_i$ being the generator of the ray normal to facet $i$\footnote{This data, which is the ``raw input'' to our computation, was first computed in~\cite{speyer2005tropical} for $(3,6)$ and $(3,7)$ (see also~\cite{herrmann2008draw}), \cite{Arkani-Hamed:2019rds,Cachazo:2019xjx} for $(3,8)$, \cite{Cachazo:2019xjx} for $(3,9)$, \cite{Drummond:2019qjk,Drummond:2019cxm,Arkani-Hamed:2019rds,Henke:2019hve,Cachazo:2019xjx} for $(4,8)$, \cite{He:2020ray} for $(3,10)$ and \cite{Cachazo:2019xjx} for (4,9).} and $F_{{\bf y}_i}$ being the associated kinematic function.  Our first goal is to assign a web $W_i$ to each facet such that $X_R(W_i) = F_{{\bf y}_i}$\footnote{We choose $X_R$ to match the conventions of~\cite{chang2020quantum}, following~\cite{Arkani-Hamed:2019rds}.}.  A priori it is not guaranteed that it always possible to find such a $W_i$.  In practice, searching for $W_i$ is feasible since for any given $F_{{\bf y}_i}$, we only need to scan over a manifestly finite set of sufficiently simple candidate webs---specifically, those whose length (defined in Sec.~\ref{sec:kinlength}) is at most that of $F_{{\bf y}_i}$.  Actually we can exploit the general relations derived in Sec.~\ref{sec:generalrelation} for considerable simplification: for each facet $i$ we only need to scan up to the length set by the \emph{shortest} image of $i$ under the $D_n$ dihedral group.  In this manner we have found webs associated to all facets of $\mathcal{C}(3,10)$, $\mathcal{C}^\dagger(3,10)$ and $\mathcal{C}^\dagger(4,9)$ by scanning webs of length up to 7, 5, and 6, respectively.

The webs $W_i$ we encounter fall into three types:
\begin{enumerate}
\item{If $W_i$ is arborizable, then (according to the FP conjectures) ${\bf y}_i$ is a ${\bf g}$-vector of the $\Gr(k,n)$ cluster algebra and $[W_i]$ is a cluster variable that we associate to facet $i$.  (Equivalently, we associate to facet $i$ the cluster series $1/(1 - t [W_i])$).}
\item{If $W_i$ is almost arborizable, then we can compute $A(W_i)$ and $B(W_i)$ as described in Sec.~\ref{sec:series} and the cluster series associated to facet $i$ is $(1 - B(W_i)\, t^2)/(1 - A(W_i)\,t + B(W_i)\, t^2)$.}
\item{In other cases we don't yet know how to associate a web series to $W_i$, although we conjecture that there exists a natural way to do so; the cluster series associated to these facets may have polynomials of degree higher than 2 in their denominators.}
\end{enumerate}
We summarize the number of facets of each type for various polytopes in Tab.~\ref{tab:summary}.  We also include ancillary files that list, for each of these polytopes, the kinematic function $F_{{\bf y}_i}$ and web invariant $[W_i]$ associated to each facet.  For each almost arborizable web we also include the $A$ and $B$ invariants appearing in the associated series~(\ref{eq:webseries}).

\begin{table}
\begin{center}
\begin{tabular}{cccc}
& (1) & (2) & (3) \\
web type: & arborizable  & almost arborizable & neither \\
cluster series type: & (\ref{eq:geometric}) (i.e., cluster variable) & (\ref{eq:webseries}) & unknown \\
\hline
$\mathcal{C}(3,5) = \mathcal{C}^\dagger(3,5)$ & 5 & 0 & 0 \\
$\mathcal{C}(3,6)$, $\mathcal{C}^\dagger(3,6)$ & 16 & 0 & 0 \\
$\mathcal{C}(3,7)$, $\mathcal{C}^\dagger(3,7)$ & 42 & 0 & 0 \\
$\mathcal{C}^\dagger(3,8)$ & 112 & 0 & 0 \\
$\mathcal{C}(3,8)$ & 120 & 0 & 0 \\
$\mathcal{C}^\dagger(3,9)$ & 327 & 0 & 0 \\
$\mathcal{C}(3,9)$ & 468 & 3 & 0 \\
$\mathcal{C}^\dagger(3,10)$ & 1060 & 0 & 0 \\
$\mathcal{C}(3,10)$ & 2860 & 280 & 0 \\
$\mathcal{C}^\dagger(4,6)$, $\mathcal{C}(4,6)$ & 9 & 0 & 0 \\
$\mathcal{C}^\dagger(4,7)$, $\mathcal{C}(4,7)$ & 42 & 0 & 0 \\
$\mathcal{C}^\dagger(4,8)$ & 272 & 2 & 0 \\
$\mathcal{C}(4,8)$ & 356 & 4 & 0 \\
$\mathcal{C}^\dagger(4,9)$ & 3078 & 324 & 27
\end{tabular}
\end{center}
\caption{The number of facets of types (1), (2) and (3) (defined in the text) for various polytopes.  Note that the facets of $\mathcal{C}^\dagger(k,n)$ are always (by construction) a subset of those of $\mathcal{C}(k,n)$.  The set of facets associated to each polytope is closed under the action of the $\mathbb{Z}_n$ cyclic group, and for the $\mathcal{C}^\dagger$ polytopes they are closed under the full $D_n$ dihedral group as well as under parity (see Appendix~A of~\cite{Golden:2013xva} for a discussion of parity symmetry.)}
\label{tab:summary}
\end{table}

A few important comments about our algorithm are in order.  First of all, we cannot exclude in generality the possibility that there might exist two webs $W_1$, $W_2$ with $[W_1] \ne [W_2]$ that have the same image $X_R(W_1) = X_R(W_2) = F$.  If this were to happen for a kinematic function $F$ associated to some facet of a polytope of interest, then we would not know which web to assign to that facet.  However, we have not encountered such a situation as far as we have computed:  for given $F$, we have always found there is (up to skein relations, of course) precisely one web $W$ such that $X_R(W) = F$ (among all possible webs below the maximum lengths we have checked).

Second, we must of course mention the possibility that the FP conjectures could be wrong for $(k,n) = (3,10)$ or $(4,9)$.  Then we would have to worry that there could be some web $W$ and some non-web $D$ such that (1) $X_R(W) = X_R(D) = F$ and (2) $[D]$ is a cluster variable but $[W]$ is not.  In such a case our algorithm would suggest associating $W$ to the facet $F$, when it might be more appropriate to associate $D$ instead.  The fact that we have not encountered any apparent inconsistency in our calculations for $\mathcal{C}^\dagger(4,9)$, which furthermore are corroborated by the independent work of~\cite{HenkeToAppear}, suggests that such worries may be postponed to higher $(k,n)$, if not indefinitely.

Next let us comment on a few interesting features of our results.  First of all we note that while $\mathcal{C}(3,n)$ has facets associated to non-arborizable webs for $n=9,10$ (and, presumably, for all $n \ge 9$), these are absent from the $\mathcal{C}^\dagger(3,n)$ polytopes that we have studied: all facets of $\mathcal{C}^\dagger(3,n\le 10)$ are associated to cluster variables.  It would be interesting to see if this continues to hold for higher $n$.

The 3 non-arborizable webs associated to $\mathcal{C}(3,9)$ are the three cyclic images of Fig.~\ref{fig:nonarb}(a) and the 4 non-arborizable webs associated to $\mathcal{C}(4,8)$ are the four cyclic images of Fig.~\ref{fig:nonarb}(b).  Out of the 324 almost arborizable webs associated to $\mathcal{C}^\dagger(4,9)$, 315 have an inner quadrilateral loop and 9 have an inner hexagon.  The latter are the cyclic images of
\begin{equation}
    \begin{tikzpicture}[scale=0.8,xscale=-1,baseline={(current bounding box.center)}]
        \def \cen{(0,0)};
        \def \mov{{-.2,0}};
        \def \rad{1.5};
        \draw {\cen} circle (\rad);
        
        \foreach \x in {1,2,3,4,5,6,7,8,9}{
            \filldraw[fill=black,draw=black] \cen+(40*\x:\rad) circle (2pt);
            \draw[white] \cen+({40* (\x)}:{\rad+0.3}) circle (0.1) node[black]{\tikz{\node at (0,0) {$\x$}}};
            \coordinate(v\x) at (40*\x:1.5);}

        \foreach \x in {2,3,4,5,6}{
            \coordinate(A\x) at ({60*\x+15}:0.75);
        }
        \coordinate(A1) at ({60+10}:0.75);

        \coordinate (B1) at ({60+5}:1.1);
        \coordinate (B2) at ({60*4+15}:1.1);
        \coordinate (B3) at ({60*5.5+12}:.8);
        \coordinate (B4) at ({60*5.5+12}:1.2);

        \coordinate(C) at (0,0);

        \draw[thick] (A1) -- (A2) -- (C) -- (A3) -- (A4) -- (A5) -- (C) -- (A6) -- (A1);
        \draw[thick] (v8) -- (A5) -- (B3) -- (A6) -- (v9);
        \draw[thick,double] (B1) -- (A1) (B2) -- (A4) (B3) -- (B4);
        \draw[thick] (v1) -- (B1) -- (v2) (v3) -- (A2) -- (v4) (v4) -- (A3) -- (v5) (v6) -- (B2) -- (v7) (v8) -- (B4) -- (v9);

        \foreach \x in {A1,A4,C,B3}{
            \filldraw (\x) circle (2pt);}
        \foreach \x in {A2,A3,A5,A6,B1,B2,B4}{
            \filldraw[fill=white] (\x) circle (2pt);
        }
        
        \node at (-2.35,0) {$=$};
        
    \end{tikzpicture}
    \begin{tikzpicture}[scale=0.8,baseline={(current bounding box.center)},xscale=-1]
        \def \cen{(0,0)};
        \def \mov{{-.2,0}};
        \def \rad{1.5};
        \draw {\cen} circle (\rad); 
        
        \foreach \x in {1,2,3,4,5,6,7,8,9}{
            \filldraw[fill=black,draw=black] \cen+(40*\x:\rad) circle (2pt);
            \draw[white] \cen+({40* (\x)}:{\rad+0.3}) circle (0.1) node[black]{\tikz{\node at (0,0) {$\x$}}};
            \coordinate(v\x) at (40*\x:1.5);}

        \foreach \x in {2,3,4,6}{
            \coordinate(A\x) at ({60*\x+15}:0.75);
        }
        \coordinate(A1) at ({60+3}:0.75);
        \coordinate(A5) at ({60*5+25}:0.75);

        \coordinate (B1) at ({60+3}:1.1);
        \coordinate (B2) at ({60*4+18}:1.1);

        \coordinate(C) at (0,0);

        \draw[thick,red] (A1) -- (A2) -- (C) -- (A3) -- (A4) -- (A6) -- cycle;
        \draw[thick,double] (B1) -- (A1) (C) -- (A5) (B2) -- (A4);
        \draw[thick] (v1) -- (B1) -- (v2) (v3) -- (A2) -- (v4) (v4) -- (A3) -- (v5) (v6) -- (B2) -- (v7) (v8) -- (A5) -- (v9) (v8) -- (A6) -- (v9);

        \foreach \x in {A1,A4,C}{
            \filldraw (\x) circle (2pt);}
        \foreach \x in {A2,A3,A5,A6,B1,B2}{
            \filldraw[fill=white] (\x) circle (2pt);
        }

    \end{tikzpicture},
\end{equation}
which are skein equivalent.  The figure on the left is a web with many inner loops which is almost arborizable, as apparent by the skein-equivalent figure on the right where the inner hexagon is highlighted in red.

The 27 webs of type (3) listed for $\mathcal{C}^\dagger(4,9)$ fall into three cyclic classes.  Instead of drawing the (very complicated) webs, we display here a 2-loop non-planar tensor diagram for one representative of each cyclic class:
\begin{equation}
    \begin{tikzpicture}[scale=0.8,baseline={(current bounding box.center)},xscale=-1]
        \def \cen{(0,0)};
        \def \mov{{-.2,0}};
        \def \rad{1.5};
        \draw {\cen} circle (\rad);
        
        \foreach \x in {1,2,3,4,5,6,7,8,9}{
            \filldraw[fill=black,draw=black] \cen+(40*\x:\rad) circle (2pt);
            \draw[white] \cen+({40* (\x)}:{\rad+0.3}) circle (0.1) node[black]{\tikz{\node at (0,0) {$\x$}}};
            \coordinate(v\x) at (40*\x:1.5);}

        \foreach \x in {1,3,4}{
            \coordinate(A\x) at ({72*\x-15}:1);
        }
        \coordinate (A2) at ({72*2-35}:1);
        \coordinate (A5) at ({72*5-12}:1.15);
  
        \foreach \x in {1,2}{
            \coordinate(B\x) at ({40*6-15+180*\x}:0.3);
            \coordinate(C\x) at ({40*8-5+180*\x}:.65);
        }
        \coordinate (B1) at ({40*2}:.6);

        \coordinate(D) at ({72*5+3.5}:.8);

        \draw[thick] (v1) -- (A1) -- (v2) -- (A2) -- (v4) -- (A3) -- (v6) -- (A4) -- (v8) -- (A5) -- (v9);
        \draw[thick] (v7) -- (A4) -- (B2) -- (A3) -- (v5) -- (C1) -- (v3) -- (A2) -- (B1) (C2) -- (v7);
        \draw[thick,double] (D) -- (A5);
        \draw[thick,red] (B1) -- (C1) -- (B2) -- (C2);
        \draw[very thick,violet] (C2) -- (B1); 
        \draw[thick,blue] (B1) -- (A1) -- (D) -- (C2);

        \foreach \x in {B1,B2,D}{
            \filldraw (\x) circle (2pt);
        }
        \foreach \x in {A1,A2,A3,A4,A5,C1,C2}{
            \filldraw[fill=white] (\x) circle (2pt);
        }

    \end{tikzpicture}
    \quad
    \begin{tikzpicture}[scale=0.8,baseline={(current bounding box.center)},xscale=-1]
        \def \cen{(0,0)};
        \def \mov{{-.2,0}};
        \def \rad{1.5};
        \draw {\cen} circle (\rad); 
        
        \foreach \x in {1,2,3,4,5,6,7,8,9}{
            \filldraw[fill=black,draw=black] \cen+(40*\x:\rad) circle (2pt);
            \draw[white] \cen+({40* (\x)}:{\rad+0.3}) circle (0.1) node[black]{\tikz{\node at (0,0) {$\x$}}};
            \coordinate(v\x) at (40*\x:1.5);}

        \foreach \x in {2,3,4,5}{
            \coordinate(A\x) at ({72*\x-15}:1.15);
        }
        \coordinate(A1) at ({72-3}:.9);

        \foreach \x in {3,4}{
            \coordinate(B\x) at ({7*40+30*\x}:{.7+0.07*\x});
        }
        \coordinate(B2) at ({7*40+25*2}:.6);
        \coordinate(B1) at ({7*40+10}:.7);

        \coordinate(C1) at (0,0);
        \coordinate(C2) at ({40*3}:0.5);
        \coordinate(C3) at ({40*4}:0.8);
        \coordinate(C4) at ({40*5.5}:0.7);

        \draw[thick] (v1) -- (A1) -- (v2) -- (A2) -- (v4) -- (A3) -- (v6) -- (A4) -- (v8) -- (A5) -- (v9);
        \draw[thick] (v3) -- (A2) -- (C2) -- (C3) -- (v5) -- (A3) -- (C4) -- (A4) -- (v7);
        \draw[thick,double] (B1) -- (B2) (B3) -- (B4);
        \draw[thick] (v6) -- (B1) -- (v7) (v1) -- (B4) -- (v2);
        \draw[thick,red] (B2) -- (A1) -- (C2)(C1) -- (B3) -- (A5) -- (B2);
        \draw[thick,blue] (C2) -- (C3) -- (C4) -- (C1);
        \draw[very thick,violet] (C1) -- (C2);

        \foreach \x in {B2,B3,C2,C4}{
            \filldraw (\x) circle (2pt);
        }
        \foreach \x in {A1,A2,A3,A4,A5,B1,B4,C1,C3}{
            \filldraw[fill=white] (\x) circle (2pt);
        }

    \end{tikzpicture}
    \quad
    \begin{tikzpicture}[scale=0.8,baseline={(current bounding box.center)},xscale=-1]
        \def \cen{(0,0)};
        \def \mov{{-.2,0}};
        \def \rad{1.5};
        \draw {\cen} circle (\rad); 
        
        \foreach \x in {1,2,3,4,5,6,7,8,9}{
            \filldraw[fill=black,draw=black] \cen+(40*\x:\rad) circle (2pt);
            \draw[white] \cen+({40* (\x)}:{\rad+0.3}) circle (0.1) node[black]{\tikz{\node at (0,0) {$\x$}}};
            \coordinate(v\x) at (40*\x:1.5);}

        \foreach \x in {2,5}{
            \coordinate(A\x) at ({72*\x-15}:1.15);
        }
        \coordinate(A1) at ({72-3}:.9);
        \coordinate(A3) at ({72*3-15}:1);
        \coordinate(A4) at ({72*4+10}:1.1);

        \coordinate(B4) at  ({7*40+30*4}:{.7+0.07*4});
        \coordinate(B3) at ({7*40+30*4}:{.3+0.07*4});
        \coordinate(B2) at ({72*2-15}:.8);
        \coordinate(B1) at ({72*2+12}:.8);
        
        \coordinate(C1) at (-.2,-.3);
        \coordinate(C2) at (.2:0.4);
        \coordinate(C3) at ({270*2-40*4-60}:0.7);
        \coordinate(C4) at ({270}:0.7);

        \draw[thick] (v1) -- (A1) -- (v2) (v3) -- (A2) -- (v4) -- (A3) -- (v6) -- (A4) -- (v8) -- (A5) -- (v1);
        \draw[thick,blue] (C1) -- (C4) -- (C3) -- (C2);
        \draw[very thick,violet] (C1) -- (C2);
        \draw[thick,double] (B4) -- (B3) (B2) -- (A2);
        \draw[thick,red] (C2) -- (B3) -- (B1) -- (B2) -- (A1) -- (C1);
        \draw[thick] (C1) -- (A3) -- (v5) -- (B1) -- (v6) (v5) -- (C4) -- (v7) -- (A4) -- (C3) -- (A5) -- (v9) -- (C2);
        \draw[thick] (v1) -- (B4) -- (v2);

        \foreach \x in {A1,A2,A3,A4,A5,C2,C4,B4,B1}{
            \filldraw[fill=white] (\x) circle (2pt);
        }
        \foreach \x in {C1,C3,B2,B3}{
            \filldraw (\x) circle (2pt);
        }

    \end{tikzpicture}
    \label{eq:typethreewebs}
\end{equation}
where we highlight the two loops in color.  Each of these is skein-equivalent to a valid web (that means, with no 2-cycles or triple edges; see footnote~\ref{nonelliptic}).

As already noted above, it would be very interesting to find a natural web series to associate to these more complicated webs; the corresponding invariants might evaluate to rational functions with higher (than quadratic) order polynomials in their denominators.  It is interesting to note that the approaches of~\cite{Drummond:2019cxm,Herderschee:2021dez} also seem to encounter some difficulty when passing from $\Gr(4,8)$ to $\Gr(4,9)$, for essentially the same reason: Whereas the $\Gr(4,8)$ cluster algebra has finite mutation type~\cite{felikson2012skew}, and all exceptional rays can be asymptotically approached by repeated mutation on some quiver containing an $A_{1,1}$ subalgebra, $\Gr(4,9)$ does not have finite mutation type and has arbitrarily complicated quivers.  It would be interesting to more precisely understand how (if at all) this fact relates to webs of the type shown in~(\ref{eq:typethreewebs}).

\acknowledgments

We are grateful to N.~Arkani-Hamed, S.~He, N.~Henke, A.~Herderschee, T.~Lam, J.~Mago, G.~Papathanasiou, A.~Schreiber and A.~Yelleshpur Srikant for encouraging discussions, correspondence, and collaboration on closely related work.  This work was supported in part by the US Department of Energy under contract {DE}-{SC}0010010 Task A and by Simons Investigator Award \#376208 (AV).

\appendix

\section{Conventions}
\label{sec:conventions}

\subsection{Web Variables}
\label{sec:webvariables}

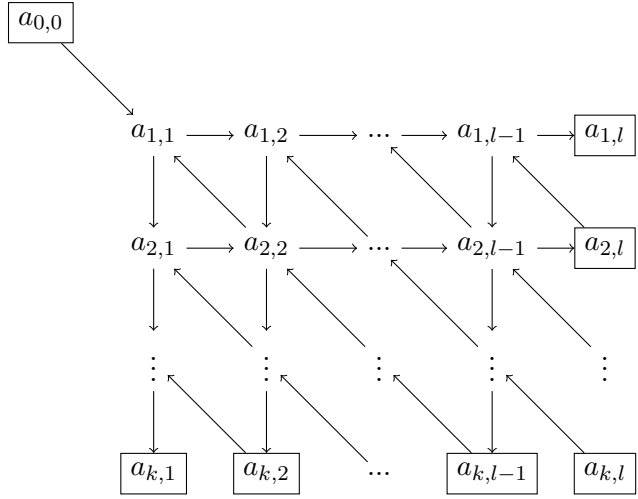
\begin{figure}[!htb]
     \centering
     \begin{tikzpicture}
     [frnode/.style={rectangle,draw=black}];
     \node[frnode](00) at (0,0) {$a_{0,0}$};
     \node(11) at (1.5,-1.5) {$a_{1,1}$};
     \node(12) at (3,-1.5) {$a_{1,2}$};
     \node(1dot) at (4.5,-1.5) {$...$};
     \node(1ll) at (6,-1.5) {$a_{1,l-1}$};
     \node[frnode](1l) at (7.5,-1.5) {$a_{1,l}$};
     \node(21) at (1.5,-3) {$a_{2,1}$};
     \node(22) at (3,-3) {$a_{2,2}$};
     \node(2dot) at (4.5,-3) {$...$};
     \node(2ll) at (6,-3) {$a_{2,l-1}$};
     \node[frnode](2l) at (7.5,-3) {$a_{2,l}$};
     \node(dot1) at (1.5,-4.5) {$\vdots$};
     \node(dot2) at (3,-4.5) {$\vdots$};
     \node(dotdot) at (4.5,-4.5) {$\vdots$};
     \node(dotll) at (6,-4.5) {$\vdots$};
     \node(dotl) at (7.5,-4.5) {$\vdots$};
     \node[frnode](k1) at (1.5,-6) {$a_{k,1}$};
     \node[frnode](k2) at (3,-6) {$a_{k,2}$};
     \node(kdot) at (4.5,-6) {$...$};
     \node[frnode](kll) at (6,-6) {$a_{k,l-1}$};
     \node[frnode](kl) at (7.5,-6) {$a_{k,l}$};
     \draw[->] (00) -- (11);
     \draw[->] (11) -- (12);
     \draw[->] (12) -- (1dot);
     \draw[->] (1dot) -- (1ll);
     \draw[->] (1ll) -- (1l);
     \draw[->] (21) -- (22);
     \draw[->] (22) -- (2dot);
     \draw[->] (2dot) -- (2ll);
     \draw[->] (2ll) -- (2l);
     \draw[->] (11) -- (21);
     \draw[->] (21) -- (dot1);
     \draw[->] (dot1) -- (k1);
     \draw[->] (12) -- (22);
     \draw[->] (22) -- (dot2);
     \draw[->] (dot2) -- (k2);
     \draw[->] (1ll) -- (2ll);
     \draw[->] (2ll) -- (dotll);
     \draw[->] (dotll) -- (kll);
     \draw[->] (22) -- (11);
     \draw[->] (2dot) -- (12);
     \draw[->] (2ll) -- (1dot);
     \draw[->] (2l) -- (1ll);
     \draw[->] (dot2) -- (21);
     \draw[->] (dotdot) -- (22);
     \draw[->] (dotll) -- (2dot);
     \draw[->] (dotl) -- (2ll);
     \draw[->] (k2) -- (dot1);
     \draw[->] (kdot) -- (dot2);
     \draw[->] (kll) -- (dotdot);
     \draw[->] (kl) -- (dotll);
     \end{tikzpicture}
     \caption{Initial seed for the $\Gr(k,n)$ cluster algebra~\cite{gekhtman2010cluster,keller2008cluster}, where $l=n-k$ and $a_{x,y}$ denotes the Pl\"ucker coordinate $\langle{1,\ldots,k{-}x, k{+}y{-}x{+}1,\ldots,k{+}y}\rangle$.  The arrows here are reversed (and the figure is transposed) with respect to that used in~\cite{chang2020quantum}; see the discussion in Sec.~\ref{sec:langlands}.}
     \label{figinigkn}
\end{figure}

We begin by reviewing the web variables of~\cite{speyer2005tropical}.  First consider the $k \times n$ \textit{web matrix}
\begin{align}
\left(
\begin{tabular}{cc}
${1}_{k \times k}$ & $\calW$
\end{tabular}
\right)
\label{eq:webmatrix}
\end{align}
where $\calW$ is the $k \times (n{-}k)$ matrix constructed as follows.  Draw a $(k{-}1) \times (n{-}k{-}1)$ array with faces labeled by \textit{web variables} $x_1$ through $x_d$ (reading down each column, from left to right).  Label the horizontal lines $1, \ldots, k$ from top to bottom and the vertical lines $k+1, \ldots, n$ from left to right.  Give each horizontal edge a rightward orientation and each vertical edge an upward orientation. To each path $p$ through the diagram we associate the product of all web variables above $p$, which we denote by $p(x)$.  Then the $i,j$ element of $\calW$ is given by
\begin{align}
\calW_{ij} = (-1)^{k-i} \sum_{p: i \to j + k} p(x)\,.
\end{align}
For example, for $k=3$ the array looks like
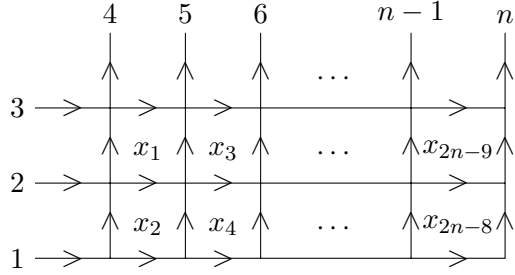
\begin{center}
 \tikz[scale=1,yscale=-1]{
 \draw (0,0) node[left] {1}
 --(1,0) node[pos=.5]{$>$}node[pos=.5,above=5pt]{}
 --(2,0) node[pos=.5]{$>$}node[pos=.5,above=5pt]{$x_2$}
  --(3,0) node[pos=.5]{$>$}node[pos=.5,above=5pt]{$x_4$}
  --(5,0)node[pos=.5,above=5pt]{$\cdots$}
   --(6.25,0) node[pos=.5]{$>$}  node[pos=.5,above=5pt]{$x_{2n-8}$}
  --(6.25,-1) node[pos=.5]{\rotatebox {90} {$>$}}
  --(6.25,-2) node[pos=.5]{\rotatebox {90} {$>$}}
  --(6.25,-3) node[pos=.5]{\rotatebox {90} {$>$}} node[above] {$n$}
  (0,-1) node[left] {2}
     --(1,-1) node[pos=.5]{$>$}node[pos=.5,above=5pt]{}
 --(2,-1) node[pos=.5]{$>$}node[pos=.5,above=5pt]{$x_1$}
  --(3,-1) node[pos=.5]{$>$}node[pos=.5,above=5pt]{$x_3$}
  --(5,-1)node[pos=.5,above=5pt]{$\cdots$}
  --(6.25,-1) node[pos=.5]{$>$}  node[pos=.5,above=5pt]{$x_{2n-9}$}
 
      (0,-2) node[left] {3}
     --(1,-2) node[pos=.5]{$>$}node[pos=.5,above=5pt]{}
 --(2,-2) node[pos=.5]{$>$}node[pos=.5,above=5pt]{}
  --(3,-2) node[pos=.5]{$>$}node[pos=.5,above=5pt]{}
  --(5,-2)node[pos=.5,above=5pt]{$\cdots$}
  --(6.25,-2) node[pos=.5]{$>$}  node[pos=.5,above=5pt]{}
  (1,0)
   --(1,-1) node[pos=.5]{\rotatebox {90} {$>$}}
  --(1,-2) node[pos=.5]{\rotatebox {90} {$>$}}
  --(1,-3) node[pos=.5]{\rotatebox {90} {$>$}} node[above] {4}
      (2,0)
   --(2,-1) node[pos=.5]{\rotatebox {90} {$>$}}
  --(2,-2) node[pos=.5]{\rotatebox {90} {$>$}}
  --(2,-3) node[pos=.5]{\rotatebox {90} {$>$}} node[above] {5}
          (3,0)
   --(3,-1) node[pos=.5]{\rotatebox {90} {$>$}}
  --(3,-2) node[pos=.5]{\rotatebox {90} {$>$}}
  --(3,-3) node[pos=.5]{\rotatebox {90} {$>$}} node[above] {6}
              (5,0)
   --(5,-1) node[pos=.5]{\rotatebox {90} {$>$}} 
  --(5,-2) node[pos=.5]{\rotatebox {90} {$>$}}
  --(5,-3) node[pos=.5]{\rotatebox {90} {$>$}} node[above] {$n-1$}
  ; }
\end{center}
from which we can read off the $\Gr(3,n)$ web matrix
\begin{align}
\left(
\begin{tabular}{ccccccc}
1 & 0 & 0 & 1 & $1{+}x_1{+}x_1 x_2$ & $1{+}x_1{+}x_1x_2{+}x_1x_3{+}x_1x_2x_3{+}x_1x_2x_3x_4$ & $\cdots$ \\
0 & 1 & 0 & $-1$ & $-1{-}x_1$ & $-1{-}x_1{-}x_1x_3$ & $\cdots$ \\
0 & 0 & 1 & 1 & 1 & 1 & $\cdots$
\end{tabular}
\right).
\end{align}
The web matrix associated to $\Gr(k,n)$ provides a parameterization of $\Gr_+(k,n)/T$ as the $d$ web variables range over $\mathbb{R}^d_{>0}$, and (importantly for our purposes) the web variables are precisely the cluster $\mathcal{X}$-coordinates associated to the initial seed of $\Gr(k,n)$ shown in Fig.~\ref{figinigkn}.  Specifically: when evaluated on the web matrix~(\ref{eq:webmatrix}), the cluster $\mathcal{X}$-coordinate (see Sec.~\ref{sec:gvectorappendix}) attached to any mutable node of the initial quiver is equal to the web variable $x$ that appears in the same position of the web array described above.

\subsection{Kinematic Space and Kinematic Functions}
\label{sec:kinematicspace}

Next we review the planar kinematic variables first employed for $k=2$ in the construction of~\cite{Arkani-Hamed:2017mur}.  We introduce $\binom{n}{k}$ (generalized) \textit{Mandelstam variables}~\cite{Cachazo:2019ngv} $s_{i_1,i_2,\ldots,i_k}$, fully symmetric in all indices, subject to the ``on-shell'' condition $s_{i,i,\ldots} = 0$ and the ``momentum conservation'' condition
\begin{align}
\label{fqfeq}
\sum_{\{ i_2,\cdots,i_k\} \subset \{1,\cdots,n\}\backslash \{i_1\} } s_{i_1,i_2,\cdots,i_k}=0 \quad \forall i_1\,.
\end{align}
The resulting $\binom{n}{k}-n$-dimensional space spanned by these variables is called the \textit{kinematic space} $\calK_{k,n}$.

Since the Mandelstam variables are not linearly independent, our next step is to define a particular basis for $\calK_{k,n}$~\cite{Arkani-Hamed:2019mrd}.  To that end we consider
\begin{align}
R_{k,n} = \prod_{1 \le i_1 < i_2 < \cdots < i_k \le n}
\langle i_1, i_2, \ldots, i_k\rangle^{\alpha' s_{i_1,i_2,\ldots,i_k}}
\end{align}
where $\alpha'$ is a positive constant that is irrelevant for our purposes.  Note that thanks to~(\ref{fqfeq}), $R_{k,n}$ is invariant under the torus action that rescales each $Z_i^a$ independently, and therefore is well-defined on $\Gr(k,n)/T$.

Something remarkable happens when $R_{k,n}$ is evaluated on the $\Gr(k,n)$ web matrix.  The $\binom{n}{k}$ minors fall into two categories.  First, there are $n+d$ \textit{trivial minors} that evaluate to 1 or to monomials in web variables; these include the $n$ frozen variables
\begin{align}
\langle 1,2,\ldots,k\rangle, \langle 2,3,\ldots,k{+}1\rangle, \quad \ldots, \quad \langle 1,2,\ldots,k{-}1,n\rangle
\end{align}
as well as the $d$ non-frozen variables of the form
\begin{align}
\langle 1,2,\ldots,j,l,l{+}1,\ldots\,l{+}k{-}j{-}1\rangle \quad 1\leq k,\ j+1\leq l\leq n-(k-j-1)\,.
\end{align}
Each of the remaining $\binom{n}{k}-n-d$ minors $\langle I \rangle$ factors into a monomial in web variables times a single polynomial $P_I(x)$ that is unique to each minor.  Moreover, each of these polynomials is subtraction free and has constant term 1.  By collecting all of the overall monomials from both the trivial and non-trivial minors, we can rewrite~\cite{Arkani-Hamed:2019mrd}
\begin{align}
R_{n,k}(x_1,\ldots,x_d) = \prod_{a=1}^d x_a^{\alpha' X_a} \prod_{I} P_I(x)^{
- \alpha' c_I}\,,
\label{fqe3fq}
\end{align}
where the product runs over all non-trivial minors, the power of each overall $x_a$ is $\alpha'$ times some linear combination of Mandelstam variables that we denote $X_a$, and we set $s_I = - c_I$.  Altogether the total number of $X_a$ and $c_I$ variables is $\binom{n}{k}-n$, and they provide the desired basis for $\calK_{k,n}$.

Here we explain the \emph{kinematic functions} that first appeared in Sec.~\ref{sec:warmup}.  We associate to any point ${\bf y} = (y_1, \ldots, y_d)$ in the integer lattice $\mathbb{Z}^d$ the function $F_{\bf y}$ on kinematic space defined by
\begin{align}
\label{eq:fromy}
F_{\bf y} = \frac{1}{\alpha'} \Res_{\epsilon = 0}\, d \log R_{k,n}(\epsilon^{-y_1}, \ldots, \epsilon^{-y_d})\,.
\end{align}
The properties of $R_{n,k}$ ensure that $F_{\bf y}$ is always an integer linear combination of the $X_a$ and $c_I$.  In fact the coefficient of $X_a$ is just $y_a$, so we have
\begin{align}
\label{eq:toy}
{\bf y} = - (\partial_{X_1} F_{\bf y}, \ldots,
\partial_{X_d} F_{\bf y})\,.
\end{align}
Using~(\ref{eq:fromy}) and~(\ref{eq:toy}) we can pass back and forth between ${\bf y}$ and $F_{\bf y}$ at ease.  The two vertical arrows on the right side of Fig.~\ref{fig:outline} apply this correspondence to the case when ${\bf y}$ is taken to be the generator ${\bf y}$ of an outward-pointing normal ray to a $\Gr(k,n)$-polytope.

For example, for $\Gr(3,5)$ a simple calculation reveals that
\begin{align}
R_{3,5}(x_1,x_2) = x_1^{\alpha' X_1} x_2^{\alpha' X_2}
(1 + x_1)^{-\alpha' c_{135}} (1+x_2)^{-\alpha' c_{245}}
(1+x_1+x_1 x_2)^{-\alpha' c_{235}}
\end{align}
where $X_1 = s_{123}$ and $X_2 = s_{345}$, and it is also easy to check that~(\ref{eq:fromy}) computes the first column of Tab.~\ref{tab:one} from the data given in the second column.

Note it is manifest (by homogeneity) that $F_{m {\bf y}} = m F_{\bf y}$ for any non-negative integer $m$.  This bears resemblance to the statement about cluster algebra bases that $\mathcal{B}(m {\bf y}) = \mathcal{B}({\bf y})^m$, but the former holds for any lattice point ${\bf y}$ while the latter holds only if ${\bf y}$ is inside the cluster fan.

\subsection{\texorpdfstring{$g$-Vectors}{g-Vectors}}
\label{sec:gvectorappendix}

Next we review the horizontal arrows at the bottom of Fig.~\ref{fig:outline}.  We order the $d+n$ cluster variables ($\mathcal{A}$-coordinates) appearing in the initial quiver (Fig.~\ref{figinigkn}) $a_1, a_2, \ldots, a_{d+n}$, first reading the mutable variables down each column from left to right, and then the frozen variables counterclockwise starting from $a_{0,0}$.  Next recall that the associated exchange matrix is given by $B_{ij} = (\# \text{arrows}\; i \to j) - (\# \text{arrows}\; j \to i)$ where $i,j$ run over the nodes, and the cluster $\mathcal{X}$-coordinate associated to node $i$ is related to the $\mathcal{A}$-coordinates of its neighbors by $x_i = \prod_j a_j^{B_{ji}}$.

To any monomial $\prod_i a_i^{g_i}$  we associate the vector of powers ${\bf g} = (g_1,\ldots,g_{d+n})$.  We introduce a partial order on such vectors by saying that ${\bf g}' \preceq {\bf g}$ iff ${\bf g}' -{\bf g}$ is a non-negative linear combination of the first $d$ columns of $B$ (the columns corresponding to mutable nodes).  If $a$ is a sum of monomials in the $a_i$ we define the ${\bf g}$-vector of $a$ to be that of the term whose ${\bf g}$-vector is largest with respect to $\preceq$ (if such a term exists).  It is always sufficient to truncate ${\bf g}$ to its first $d$ components.  If $a$ is a cluster variable of $\Gr(k,n)$, then the ${\bf g}$-vector of $a$ exists and it is said to be a ${\bf g}$-vector of the cluster algebra.

We have therefore explained the upward pointing arrow in Fig.~\ref{fig:outline}.  For example, for $\Gr(3,5)$ we have
\begin{align}
B^{\rm T} = \left(\begin{array}{cc;{2pt/2pt}ccccc}
0 & -1\ &\ 1 & 0 & 0 & 1 & -1 \\
1 & 0\ &\ 0 & -1 & 1 & -1 & 0
\end{array}\right),
\end{align}
the initial cluster variables (given in the caption) are
\begin{align}
(a_1,\ldots,a_7) =
(\langle 1\,2\,4 \rangle,
\langle 1\,3\,4 \rangle,
\langle 1\,2\,3 \rangle,
\langle 2\,3\,4 \rangle,
\langle 3\,4\,5 \rangle,
\langle 1\,4\,5 \rangle,
\langle 1\,2\,5 \rangle)\,,
\end{align}
and the remaining three cluster variables are given in terms of these by
\begin{align}
\begin{split}
\langle 2\,3\,5 \rangle &= \frac{a_3 a_5}{a_2} + \frac{a_3 a_4 a_6}{a_1 a_2} + \frac{a_4 a_7}{a_1}\,, \\
\langle 2\,4\,5 \rangle &= \frac{a_1 a_5}{a_2} + \frac{a_4 a_6}{a_2}\,, \\
\langle 1\,3\,5 \rangle &= \frac{a_3 a_6}{a_1} + \frac{a_2 a_7}{a_1}\,.
\end{split}
\label{eq:threegvectors}
\end{align}
Here we have written the terms in each sum in increasing order with respect to $\preceq$ so it is easy to read off, from the last term in each line, the ${\bf g}$-vectors $(-1,0)$, $(0,-1)$ and $(-1,1)$, as shown in Tab.~\ref{tab:one}.

Going the other way, down the dotted arrow in Fig.~\ref{fig:outline} to compute the cluster variable (or more general basis element) associated to a given lattice vector ${\bf x}$, is not so simple.  In practice one often resorts to a computer search by repeatedly mutating away from the initial seed until one has the fortune to chance upon a cluster variable whose ${\bf g}$-vector is ${\bf x}$.  Of course for infinite algebras this algorithm may take an indefinite amount of time.  Even worse, ${\bf x}$ may lie outside the cluster fan in which case one will never find a match.

One alternative, suggested in~\cite{Arkani-Hamed:2019rds}, is to read Corollary 7.3 of~\cite{chang2020quantum} as providing an explicit formula for an element of the canonical basis~\cite{lusztig1990canonical} associated to every ${\bf g}$ that agrees with the usual cluster algebraic definition when ${\bf g}$ lies inside the cluster fan.  Although the required computation is manifestly finite for any ${\bf g}$, its enormous computational complexity makes it impractical in many cases of interest.

\subsection{\texorpdfstring{$\Gr(k,n)$-Polytopes}{G(k,n)-Polytopes}}
\label{sec:polytopes}

Before ending this section, we are finally in a position to review the construction of the \emph{$\Gr(k,n)$-polytopes} of interest, which generalize the well-known Stasheff polytope~\cite{stasheff1,stasheff2} building on a construction introduced in~\cite{Arkani-Hamed:2017mur}.  These polytopes lie in the $d$-dimensional subspace $\calH_{k,n}$ of $\calK_{k,n}$ obtained by setting all of the $c_I$ to positive constants.  Here we see that the purpose of defining $X_1, \ldots, X_d$ in the previous subsection is that we can take these as coordinates on $\calH_{k,n}$.

The polytope called $\calC(k,n)$ in~\cite{Arkani-Hamed:2019rds} (called ${\cal P}(k,n)$ or dual $\Trop \Gr(k,n)$ in some other references) is defined by taking the Minkowski sum of the Newton polytopes (with respect to $x_1,\ldots,x_d$) associated to the polynomials $P_I$ appearing in~(\ref{fqe3fq}).  Other polytopes can be constructed by only including proper subsets of the $P_I$ in the Minkowski sum.  For example, of particular interest is the polytope called $\calC^\dagger(4,n)$ in~\cite{Arkani-Hamed:2019rds} (also studied in~\cite{Drummond:2019cxm,Henke:2019hve}).  It is defined as the polytope obtained by including only polynomials associated to $\langle I\rangle$'s of the form $\langle i\,i{+}1\,j\,j{+}1\rangle$ or $\langle i{-}1\,i\,i{+}1\,j\rangle$, and may be obtained from $\calC(4,n)$ by setting to zero all $c_I$ except those corresponding to these $I$'s.  Here we define $\calC^\dagger(3,n)$ to be the polytope obtained by keeping only $\langle I\rangle$'s of the form $\langle i\,i{+}1\,j\rangle$.

More general polytopes of the same basic type can be constructed by including other proper subsets of the $P_I$ in the Minkowski sum, or by including polynomials obtained by evaluating more complicated $\Gr(k,n)$ cluster variables on the web matrix.  An example of the latter was considered in~\cite{Drummond:2019cxm}.

\subsection{Langlands Dual Conventions}
\label{sec:langlands}

Because our work touches on a wide range of previous work in the physics and math literature, we find it helpful to clearly connect to two different choices of convention that are related to each other by what could be called ``Langlands duality'' (see Remark 7.15 of~\cite{fomin2007cluster}).  By this we mean performing the following compatible set of changes:
\begin{enumerate}
\item{inverting each $x_i \to 1/x_i$ in the web parameterization of Sec.~\ref{sec:webvariables},}
\item{reversing each arrow in Fig.~\ref{figinigkn},}
\item{and, correspondingly, changing the sign of the $B$-matrix with respect to which ${\bf g}$-vectors are computed as described in Sec.~\ref{sec:gvectorappendix}.}
\end{enumerate}

The conventions outlined in Sec.~\ref{sec:webvariables} through Sec.~\ref{sec:gvectorappendix} correspond to what we call the ``left'' convention starting in Sec.~\ref{sec:Xmapdefinition}.  To illustrate the different conventions we present in Tab.~\ref{tab:two} the ``right'' convention version of the $\Gr(3,5)$ data from Tab.~\ref{tab:one}.  Note that the form of the equations~(\ref{eq:threegvectors}) is the same for both choices, and while the ``left'' ${\bf g}$-vectors shown in Tab.~\ref{tab:one} can be read off from the last term in each line, we can similarly read off the ``right'' ${\bf g}$-vectors (shown in Tab.~\ref{tab:two}) from the first term on each line.  We explain a general relation between the two conventions, at the level of our $X$-map applied to general tensor diagrams, in Sec.~\ref{sec:generalrelation}.

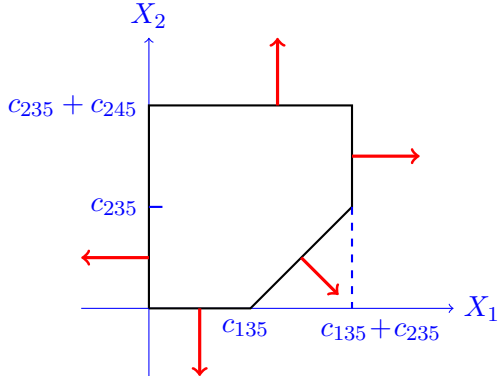
\begin{figure}
    \centering
    \begin{tikzpicture}[scale=0.9]
        \draw[->,blue] (-1,0) -- (4.5,0) node[right]{$X_1$};
        \draw[->,blue] (0,-1) -- (0,4) node[above]{$X_2$};
        \draw[thick] (0,0) -- (0,3) node[left,blue]{$c_{235} + c_{245}$} -- (3,3) -- (3,1.5) -- (1.5,0) node[below, blue]{$c_{135}\ $} -- (0,0);
        \draw[thick,dashed,blue] (3,1.5) -- (3,0) node[below, blue]{$\quad\quad c_{135}\!+\!c_{235}$};
        \draw[thick,blue] (0,1.5) node[left,blue]{$c_{235}$} -- (0.2,1.5);

        \draw[very thick,red,->] (0,0.75) -- (-1,0.75);
        \draw[very thick,red,->] (0.75,0) -- (0.75,-1);
        \draw[very thick,red,->] (2.25,0.75) -- (3-0.2,.2);
        \draw[very thick,red,->] (3,2.25) -- (4,2.25);
        \draw[very thick,red,->] (1.9,3) -- (1.9,4);
    \end{tikzpicture}
    \caption{The polytope ${\cal C}(3,5)$ according to the ``right'' conventions; see Tab.~\ref{tab:two} and contrast with Fig.~\ref{qiupfhq9}.}
    \label{qiupfhq8}
\end{figure}

\begin{table}
\begin{center}
\begin{tabular}{c|c|c}
kinematic function & generator & cluster variable \\
\hline\hline
$\color{brown} s_{123} = \color{black} c_{135} + c_{235} - X_1$ & $(1,0)$ & $\langle 1\,2\,4 \rangle$\\
\hline
$\color{brown} s_{345} = \color{black} c_{235} + c_{245} - X_2$ & $(0,1)$ & $\langle 1\,3\,4 \rangle$\\
\hline
$\color{brown} s_{125} = \color{black} X_1$ & $(-1,0)$ & $\langle 1\,3\,5 \rangle$\\
\hline
$\color{brown} s_{234} = \color{black} X_2$ & $(0,-1)$ & $\langle 2\,3\,5 \rangle$\\
\hline
$\color{brown} s_{145} = \color{black} c_{135} - X_1 + X_2$ & $(1,-1)$ & $\langle 2\,4\,5 \rangle$\\
\end{tabular}
\end{center}
\caption{The correspondence between kinematic functions, generators, and cluster variables for the $\mathcal{C}(3,5)$ polytope according to the ``right'' conventions, shown in Fig.~\ref{qiupfhq8}, in contrast to Tab.~\ref{tab:one} which shows the correspondence for the ``left'' conventions.}
\label{tab:two}
\end{table}

\section{Summary of Known Symbol Letters}
\label{sec:summaryofsymbolletters}

Here we summarize what is known about the symbol alphabet $S_n$ of $n$-particle amplitudes in SYM theory.  In this discussion we of course restrict our attention to those amplitudes which are of polylogarithmic type, and so have conventionally-defined symbols.

Let us begin with the rational letters.  All currently known rational letters are cluster coordinates of $\Gr(4,n)$, which (according to the FP conjectures) means that we can represent them as arborizable webs.  It expected that the $n$-particle symbol alphabet is a strict subset of the $n'$-particle symbol alphabet for all $n' > n$, which corresponds to the fact that we can always make a valid $n'$-particle web by adding $n' - n$ boundary vertices, with no edges attached, to an $n$-particle web.  Therefore it is convenient to categorize different types of symbol letters according to the smallest value of $n$ at which they first appear; we also categorize them by Pl\"ucker degree.  In this way we encounter five basic types of rational letters for $n \le 9$:

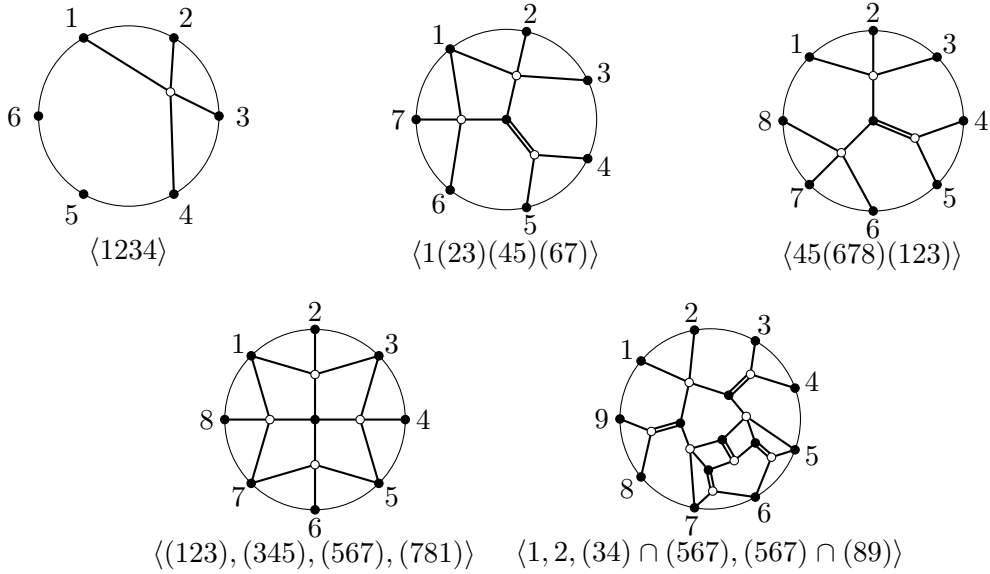
\begin{figure}
\centering
\begin{subfigure}{0.31\textwidth}
\centering
    \begin{tikzpicture}[xscale=-1,scale=.8]
        \def \cen{(0,0)};
        \def \rad{1.5};
        \draw {\cen} circle (\rad); 

        \foreach \x in {1,2,3,4,5,6}{
             \filldraw[fill=black,draw=black] \cen+(60*\x:\rad) circle (2pt);
             \draw[white] \cen+({60* (\x)}:{\rad+0.4}) circle (0.1) node[black]{\tikz{\node at (0,0) {$\x$}}};
             \coordinate(v\x) at (60*\x:1.5);}

        \coordinate(A0) at ({60*2.5}: .8);

        \draw[thick] (v1) -- (A0) -- (v2) (v3) -- (A0) -- (v4);

        \filldraw[fill=white] (A0) circle (2pt);

	\node at (0,-2.25) {$\langle 1234 \rangle$};
   \end{tikzpicture}
\end{subfigure}
\begin{subfigure}{0.31\textwidth}
\centering
    \begin{tikzpicture}[xscale=-1,scale=.8]
        \def \cen{(0,0)};
        \def \rad{1.5};
        \draw {\cen} circle (\rad); 
        \def \ang{360/7};
        
        \foreach \x in {1,2,3,4,5,6,7}{
             \filldraw[fill=black,draw=black] \cen+(\ang*\x:\rad) circle (2pt);
             \draw[white] \cen+({\ang* (\x)}:{\rad+0.3}) circle (0.1) node[black]{\tikz{\node at (0,0) {$\x$}}};
             \coordinate(v\x) at (\ang*\x:1.5);}

        \coordinate (A123) at ({\ang*2}:0.75);
        \coordinate (B45) at ({\ang*4.5}:0.75);
        \coordinate (A678) at ({\ang*7}:0.75);
        \coordinate (C) at (0,0);

        \draw[thick] (v1) -- (A123) -- (v2) (A123) -- (v3);
        \draw[thick] (v7) -- (A678) -- (v1) (A678) -- (v6);
        \draw[thick] (v4) -- (B45) -- (v5);
        \draw[thick] (A678) -- (C) -- (A123);
        \draw[thick,double] (B45) -- (C);

        \filldraw[fill=white] (A123) circle (2pt);
        \filldraw[fill=white] (B45) circle (2pt);
        \filldraw[fill=white] (A678) circle (2pt);
        \filldraw (C) circle (2pt);

        \node at (0,-2.25) {$\langle 1 (23) (45) (67) \rangle$};
   \end{tikzpicture}
\end{subfigure}
\begin{subfigure}{0.31\textwidth}
\centering
    \begin{tikzpicture}[xscale=-1,scale=0.8]
        \def \cen{(0,0)};
        \def \rad{1.5};
        \draw {\cen} circle (\rad); 
        
        \foreach \x in {1,2,3,4,5,6,7,8}{
             \filldraw[fill=black,draw=black] \cen+(45*\x:\rad) circle (2pt);
             \draw[white] \cen+({45* (\x)}:{\rad+0.3}) circle (0.1) node[black]{\tikz{\node at (0,0) {$\x$}}};
             \coordinate(v\x) at (45*\x:1.5);}

        \coordinate (A123) at (90:0.75);
        \coordinate (B45) at ({180+22.5}:0.75);
        \coordinate (A678) at ({270+45}:0.75);
        \coordinate (C) at (0,0);

        \draw[thick] (v1) -- (A123) -- (v2) (A123) -- (v3);
        \draw[thick] (v7) -- (A678) -- (v8) (A678) -- (v6);
        \draw[thick] (v4) -- (B45) -- (v5);
        \draw[thick] (A678) -- (C) -- (A123);
        \draw[thick,double] (B45) -- (C);

        \filldraw[fill=white] (A123) circle (2pt);
        \filldraw[fill=white] (B45) circle (2pt);
        \filldraw[fill=white] (A678) circle (2pt);
        \filldraw (C) circle (2pt);

        \node at (0,-2.25) {$\langle 45(678)(123) \rangle$};
   \end{tikzpicture}
\end{subfigure}
\begin{subfigure}{0.31\textwidth}
\centering
   \begin{tikzpicture}[xscale=-1,scale=0.8]
        \def \cen{(0,0)};
        \def \rad{1.5};
        \draw {\cen} circle (\rad); 
        
        \foreach \x in {1,2,3,4,5,6,7,8}{
            \filldraw[fill=black,draw=black] \cen+(45*\x:\rad) circle (2pt);
            \draw[white] \cen+({45* (\x)}:{\rad+0.3}) circle (0.1) node[black]{\tikz{\node at (0,0) {$\x$}}};
            \coordinate(v\x) at (45*\x:1.5);}

        \coordinate (A123) at ({45*2}:0.75);
        \coordinate (A345) at ({45*4}:0.75);
        \coordinate (A567) at ({45*6}:0.75);
        \coordinate (A781) at ({45*8}:0.75);
        \coordinate (C) at (0,0);

        \draw[thick] (v1) -- (A123) -- (v2) (A123) -- (v3);
        \draw[thick] (v3) -- (A345) -- (v4) (A345) -- (v5);
        \draw[thick] (v5) -- (A567) -- (v6) (A567) -- (v7);
        \draw[thick] (v7) -- (A781) -- (v8) (A781) -- (v1);
        \draw[thick] (A345) -- (C) -- (A123) (A567) -- (C) -- (A781);

        \foreach \x in {C}{
            \filldraw[fill=black,draw=black] (\x) circle (2pt);}
        \foreach \x in {A123, A345, A567, A781}
            \filldraw[fill=white,draw=black] (\x) circle (2pt);

        \node at (0,-2.25) {$\langle (123),(345),(567),(781) \rangle$};
    \end{tikzpicture}
\end{subfigure}
\begin{subfigure}{0.31\textwidth}
\centering
    \begin{tikzpicture}[baseline={([yshift=-.5ex]current bounding box.center)},xscale=-1,scale=0.8]
        \def \cen{(0,0)};
        \def \rad{1.5};
        \draw {\cen} circle (\rad); 
        \def \ang{360/9};

        \foreach \x in {1,2,3,4,5,6,7,8,9}{
                \filldraw[fill=black,draw=black] \cen+(40*\x:\rad) circle (2pt);
                \draw[white] \cen+({40* (\x)}:{\rad+0.3}) circle (0.1) node[black]{\tikz{\node at (0,0) {$\x$}}};
                \coordinate(v\x) at (40*\x:1.5);}

        \coordinate(A5671) at (4.4*\ang:.6);
        \coordinate(A5672) at (7.6*\ang:.6);
        \coordinate(A34) at (3.3*\ang:1);
        \coordinate(A89) at (8.7*\ang:1);
        \coordinate(B34567) at (3.2*\ang: 0.5);
        \coordinate(B56789) at (8.8*\ang:0.5);
        \coordinate(C12) at (1.5*\ang:0.7);
        \coordinate(A56) at (5.3*\ang:1.2);
        \coordinate(B56) at (5.2*\ang:.85);
        \coordinate(A67) at (6.7*\ang:1.2);
        \coordinate(B67) at (6.8*\ang:.85);
        \coordinate(A6) at (6*\ang:.8);
        \coordinate(B6) at (6*\ang:0.4);

        \draw[thick] (v7) -- (A5672) (v5) -- (A5671);
        \draw[thick] (v6) -- (A67) -- (v7) (v6) -- (A56) -- (v5);
        \draw[thick,double] (A67) -- (B67) (A56) -- (B56) (A6) -- (B6);
        \draw[thick] (A5672) -- (B67) -- (A6) -- (B56) -- (A5671) -- (B6) -- cycle;
        \draw[thick] (v3) -- (A34) -- (v4) (v8) -- (A89) -- (v9);
        \draw[thick,double] (A34) -- (B34567) (A89) -- (B56789);
        \draw[thick] (A5671) -- (B34567) (A5672) -- (B56789);
        \draw[thick] (v1) -- (C12) -- (v2) (B34567) -- (C12) -- (B56789);

        \foreach \x in {A5671,A5672,A34,A89,C12}
            \filldraw[fill=white] (\x) circle (2pt);
        \foreach \x in {B34567,B56789}
            \filldraw (\x) circle (2pt);

        \foreach \x in {A56,A6,A67}
            \filldraw[draw,fill=white] (\x) circle (2pt);
        \foreach \x in {B56,B6,B67}
            \filldraw[draw,fill=black] (\x) circle (2pt);
        \node at (0,-2.25) {$\langle 1,2, (34) \cap (567),(567)\cap(89) \rangle$};
    \end{tikzpicture}
\end{subfigure}
\caption{Examples of webs, and their corresponding invariants (cluster variables), for the five different basic types (defined in the text) of rational symbol letters that are known to appear in SYM theory.}
\label{fig:symbolwebs}
\end{figure}

\begin{table}
\begin{center}
\begin{tabular}{c|c|c|c|c|c|c}
$n$ & (1) & (2) & (3) & (4) & (5) & References \\
\hline\hline
6 & 15 & 0 & 0 & 0 & 0 & \cite{Bern:1994zx,Bern:1994cg,Goncharov:2010jf,Dixon:2011pw,Dixon:2011nj,CaronHuot:2011kk,Caron-Huot:2020bkp} \\
7 & 35 & 14 & 0 & 0 & 0 & \cite{Bern:1994zx,Bern:2004ky,CaronHuot:2011ky,Drummond:2014ffa,Dixon:2016nkn,Drummond:2018caf} \\
8 & 68 & 88 & 8 & 16 & 0 & \cite{Bern:1994zx,Britto:2004nc,Bern:2004bt,CaronHuot:2011ky,Zhang:2019vnm} \\
9 & 117 & 270 & 63 & 72 & 9 & \cite{Bern:1994zx,Britto:2004nc,Bern:2004bt,CaronHuot:2011ky,He:2020vob}
\end{tabular}
\end{center}
\caption{Enumeration of rational symbol letters of $n$-particle amplitudes in SYM, based on the results of all explicit computations available to date, and categorized according to the five types defined in the text.  The lines $n=6,7$ are believed to be complete (see~\cite{Prlina:2018ukf}), comprising the 15 and 49 cluster variables of $\Gr(4,6)$ and $\Gr(4,7)$, respectively.  For $n > 7$ the counts may grow as computations are pushed to higher order in the future; additional types of rational letters may also appear.}
\label{tab:letters}
\end{table}

(1) $S_{n \ge 6}$ contains the Pl\"ucker coordinates of the form $\langle 1 \, 2 \, a \, b\rangle$ for $3 \le a < b \le n$, and their cyclic images.  (For $n < 8$ all Pl\"ucker coordinates are of this type.)

(2) $S_{n \ge 7}$ contains letters that are quadratic in Pl\"ucker coordinates having the form $\langle a(b\,c)(d\,e)(f\,g)\rangle := \langle a\,b\,d\,e \rangle \langle a\,c\,f\,g\rangle-\langle a\,b\,f\,g\rangle \langle a\,c\,d\,e\rangle$.  Specifically, $S_7$ contains the 14 non-Pl\"ucker cluster variables of $\Gr(4,7)$: $\langle 1(23)(45)(67) \rangle$, $\langle 1(72)(34)(56)\rangle$ and their cyclic images.  The letters of this type for $n=8,9$ are listed in~\cite{Zhang:2019vnm,He:2020vob}.

(3) $S_{n \ge 8}$ contains additional quadratic letters having the form~(\ref{eq:capdef}); specifically $\langle 12(abc)\cap(def)\rangle$ for $3 \le a < b < c < d < e < f \le n$ and their cyclic images.

(4) $S_{n \ge 8}$ also contains certain cubic letters listed for $n=8,9$ in~\cite{Zhang:2019vnm,He:2020vob}.

(5) For $n \ge 9$ $S_n$ contains a second type of cubic letter; see~\cite{He:2020vob}.

In Tab.~\ref{tab:letters} we tabulate the number of cluster variables of each type that appear in the $n \le 9$-particle amplitudes whose symbols have been explicitly computed to date.  For $n\ge 8$ the numbers are expected to grow as higher-loop calculations are carried out, and there is no reason to expect that more complicated types will not be encountered.  However, some evidence suggests that for every fixed value of $n$, the total number of symbol letters (at arbitrary finite order in perturbation theory) might be finite~\cite{Prlina:2018ukf} (unlike the number of cluster variables, which is infinite).

Next we turn to the algebraic letters that are known to start appearing in $S_n$ for $n \ge 8$.  The two-loop NMHV amplitudes have 18, 99 multiplicatively independent algebraic letters respectively for $n=8,9$~\cite{Zhang:2019vnm,He:2020vob}.  As reviewed in Tab.~\ref{tab:symboldata}, these respectively involve 2, 9 distinct square roots of Pl\"ucker polynomials; all are of four-mass box type, having the form $\sqrt{A^2 - 4 B}$ in terms of~(\ref{eq:ABforfirstray}), or cyclic images thereof.  In our approach, as we found in Sec.~\ref{sec:series}, each of these arises from a web series associated to (a cyclic image of) the almost arborizable web shown in Fig.~\ref{fig:nonarb}(b) (or, for $n=9$, the same web but with a ninth boundary point added anywhere in the diagram).

\section{Some Notation for Kinematic Functions}
\label{sec:notation}

In this appendix we collect some notation, originally introduced in~\cite{He:2020ray}, to efficiently encode certain kinematic functions.  If $A$ is a subset of $\{1,\ldots,n\}$, we define
\begin{align}
S_A = \sum_{a_1 < a_2 < a_3 \in A} s_{a_1a_2a_3}\,.
\end{align}
If $A, B$ are two subsets, then we define
\begin{align}
S_{A(B)} = S_{(B)A} = S_A + S_{(B)} + S_{A|B} \quad
\text{where} \quad
S_{A|B} = \sum_{a_1<a_2 \in A, b \in B} s_{a_1a_2b}
\end{align}
and, on the right-hand side, $(B)$ means the complement of $B$ in $\{1,\ldots,n\}$.

For $k=4$ we require some additional notation.  If $A, B$ and $C$ are subsets of $\{1,\ldots,n\}$ then we define
\begin{align}
     {\bm S}_{A[B(C)]}= {\bm S}_{[(C)B]A}= {\bm S}_{A}+{\bm S}_{[B(C)]}+\!\!\sum_{\substack{a_1<a_2<a_3\in A\\i\in B\cup C}}\!\!s_{a_1a_2a_3i}
      +\!\!\sum_{\substack{a_1<a_2\in A\\b_1<b_2\in B}} \!\!s_{a_1a_2b_1b_2}
      +\!\!\sum_{\substack{a_1,a_2\in A\\b\in B,c\in C}}\!\!s_{a_1a_2bc}
\end{align}
where
\begin{equation}
     {\bm S}_{[B(C)]}=  {\bm S}_{[B]}+ {\bm S}_{(C)}+  \sum_{\substack{b_1<b_2\in B\\c\in C, i\notin B\cup C}} s_{b_1b_2ci}+  \sum_{\substack{b_1<b_2<b_3\in B\\c\in C}} s_{b_1b_2b_3c}+  \sum_{\substack{b_1<b_2\in B\\c_1<c_2\in C}} s_{b_1b_2c_1c_2}
\end{equation}
and
\begin{equation}
     \begin{split}
          \bms_A &=\sum\limits_{a_1<a_2<a_3<a_4\in A}s_{a_1a_2a_3a_4}\,,\\
          \bms_{[B]} &=\sum\limits_{b_1<b_2<b_3<b_4\in B}2s_{b_1b_2b_3b_4}+\sum\limits_{\substack{b_1<b_2<b_3\in B\\i\notin B}}s_{b_1b_2b_3i}\,,\\
          \bms_{(C)} &=\sum\limits_{\substack{c_1<c_2<c_3<c_4\in C}}3s_{c_1c_2c_3c_4}+\sum\limits_{\substack{c_1<c_2<c_3\in C\\i\notin C}}2s_{c_1c_2c_3i}+\sum\limits_{\substack{c_1<c_2\in C\\ i,j\notin C}}s_{c_1c_2ij}\,.
     \end{split}
\end{equation}

\bibliographystyle{JHEP}

\bibliography{reference}

\end{document}